\documentclass[a4paper,11pt]{article}
\usepackage{jheppub}
\usepackage{latexsym,amsmath,amsfonts,amssymb}
\usepackage{epsfig,graphics,graphicx,caption,subcaption}

\newcommand{\be}{\begin{equation}}
\newcommand{\ee}{\end{equation}}

\newcommand{\no}{{\nonumber}}

\newcommand{\Z}{{\mathbb Z}}

\newcommand{\al} {\alpha}
\newcommand{\N}{{\cal N}}

\newcommand{\ep}{{\epsilon}}

\newcommand{\Om} {\Omega}

\newcommand{\La}{{\Lambda}}

\newcommand{\cT}{{\cal T}}
\newcommand{\cA}{{\cal A}}
\newcommand{\cB}{{\cal B}}
\newcommand{\cC}{{\cal C}}
\newcommand{\cO}{{\cal O}}

\newcommand{\diag} {\operatorname{diag}}

\newcommand{\bra}[1] {\left<#1\right|}
\newcommand{\ket}[1] {\left|#1\right>}

\newcommand{\brai}[2] {{\vphantom{\bra{#2}}}_{#1}\!\bra{#2}}
\newcommand{\vev}[1] {\left<#1\right>}

\newcommand{\alpr} {\alpha^\prime}

\title{Lattice String Field Theory: The linear dilaton in one dimension}
\author{Francis Bursa$^{(1)}$, Michael Kroyter$^{(2)}$}
\affiliation{$^{(1)}$ School of Physics and Astronomy\\
University of Glasgow\\
Kelvin Building, University Avenue, Glasgow\\
G12 8QQ United Kingdom
\ \\
$^{(2)}$ School of Physics and Astronomy\\
The Raymond and Beverly Sackler Faculty of Exact Sciences\\
Tel Aviv University, Ramat Aviv, 69978, Israel}

\emailAdd{francis.bursa@glasgow.ac.uk}
\emailAdd{mikroyt@tau.ac.il}

\abstract{We propose the use of lattice field theory for the
study of string field theory at the non-perturbative quantum level.
We identify many potential obstacles and examine possible resolutions thereof.
We then experiment with our approach in the particularly simple case
of a one-dimensional linear dilaton and analyse the results.}

\keywords{String Field Theory, Lattice Gauge Field Theories}
\preprint{TAUP-2795-13}
\begin{document}

%=========
\maketitle
%=========

\section{Introduction}
\label{sec:intro}

String theory is currently the most promising candidate for the unification of all
forces. Unfortunately, it is neither clear what string theory is nor even how to define
it. The most common ``definition'' of string theory found in the literature uses
scattering amplitudes that are obtained from world-sheet perturbation theory.
However, this perturbative expansion cannot be considered as defining a theory,
since the series obtained is most probably an asymptotic one,
i.e. it has a vanishing radius of convergence.
This state of affairs is very similar to the one in field theory, where the
Feynman diagrams themselves cannot be considered as a definition of a theory, but
the field theory action, from which they are derived, does define a theory.
One could hope that something similar could be achieved for string theory, which
would be defined as a field theory of first quantized
strings\footnote{See e.g.~\cite{Polchinski:1998rq} for an introduction to string theory
and the reviews~\cite{Ohmori:2001am,Taylor:2003gn,Fuchs:2008cc}
for an introduction to string field theory.}.
The world-sheet expressions for the scattering amplitudes would then be derived from
the field theory action using standard perturbation theory methods.
This approach towards the definition of string theory goes under the name of
{\it string field theory}.

Furthermore, string field theory should, at least in principle, be good not only for
defining string theory, but also for studying string theory when the world-sheet tools
are less adequate. This is completely analogous to the case of standard field theory,
when one cannot rely on the standard perturbative approach at strong coupling,
high temperature or high density. Of course, some of the most interesting questions
one can pose relate to such regimes. In string theory one could hope that string field theory
would be useful for the study of many questions, among which we can find:
\begin{itemize}
\item The identification of phases of string theory at large coupling and temperature,
      the phase transitions of the theory and their type.
\item The examination of consistency and stability of string theory compactified to
      different dimensions and of more general, e.g. non-geometric string theory backgrounds.
\item The study of solitons, in particular D-branes, time-dependent solutions,
      and other classical objects.
\item The study of various quantum effects, such as the scale-dependence of masses and couplings,
      which are not protected by supersymmetry.
\item A particularly ambitious task would be the study of (portions of) the string theory landscape.
      A particular example could be the understanding of the landscape that is related to changes
			of the open string background.
\item The study of known string theory dualities and the identification of new ones.
\end{itemize}

While other approaches towards the non-perturbative definition of string theory also exist,
string field theory is very natural in principle and construction of such theories
was attempted already in the first days of string theory~\cite{Kaku:1974zz,Kaku:1974xu}.
We now have several such formulations. Among these formulations, the more
promising ones are those in which the theory is covariant and
universal\footnote{``Universality'' here refers to the property of having the
same functional definition regardless of the background. This is ``almost''
as good as background independence. Universal formulations usually depend on the BRST
world-sheet quantization of the string, e.g. in the bosonic case they depend on the $bc$-ghosts
in addition to
the usual
space-time degrees of freedom.}. Such formulations were introduced for
the bosonic open string~\cite{Witten:1986cc}, for the bosonic closed
string~\cite{Zwiebach:1992ie} and to some extent also for the open
superstring~\cite{Witten:1986qs,Preitschopf:1989fc,Arefeva:1989cp,Berkovits:1995ab,Berkovits:2005bt,Kroyter:2009rn,Erler:2013xta}
and the heterotic string~\cite{Saroja:1992vw,Okawa:2004ii,Berkovits:2004xh,Kunitomo:2013mqa}.
Interesting new ideas regarding closed superstring field theory were
presented in~\cite{Jurco:2013qra,Matsunaga:2013mba,Erler:2014eba}.

Initially, string field theories were put to the test by demonstrating that
they lead to the same scattering amplitudes as the world-sheet theory,
i.e. one attempted to demonstrate that a proper single covering of the relevant
moduli space is achieved and that the perturbative expansion for the amplitudes
is correctly reproduced~\cite{Giddings:1986wp,Zwiebach:1990az}.
For quite some time such perturbative studies formed the main
focus for research in the field. This state of affairs changed
following the realization that Sen's conjectures~\cite{Sen:1999mh,Sen:1999xm}
can be studied using string field theory, i.e. that a field theoretical approach
is a most adequate one for studying non-perturbative classical solutions,
in particular solutions describing the condensation of the open string tachyon.
Following the first attempts to address these questions in the cases of the
bosonic string~\cite{Sen:1999nx,Moeller:2000xv} and the superstring~\cite{Berkovits:2000hf},
the interest of the community drifted towards the study of such classical solutions.
This led to a large body of work,
which culminated with the construction of the first
non-trivial
analytical solution to
string field theoretical equations of motion by Schnabl~\cite{Schnabl:2005gv}
(see also~\cite{Okawa:2006vm,Fuchs:2006hw,Ellwood:2006ba}).
The new tools that were developed for the construction of this tachyon vacuum solution
were further used for the construction of other analytical solutions,
including the construction of simpler tachyon vacuum
solutions~\cite{Erler:2009uj}, similar superstring field theory solutions~\cite{Erler:2007xt,Fuchs:2008zx,Aref'eva:2008ad,Erler:2013wda}
and to solutions describing marginal
deformations~\cite{Schnabl:2007az,Kiermaier:2007ba,Fuchs:2007yy,Kiermaier:2007vu,Erler:2007rh,Okawa:2007ri,Fuchs:2007gw,Noumi:2011kn,Maccaferri:2014cpa},
as well as to much further advance in the field.

Despite all this progress, a non-perturbative quantum mechanical study of string field
theory was never performed. Such a study could be useful for addressing the important
question of distinguishing ``bare'' string field theories from ``effective''
ones~\cite{Polchinski:1994mb}.
The latter ones being theories that, while capable of reproducing the correct
scattering amplitudes, do not make sense as quantum theories at the non-perturbative level,
since there is no way to regularize or re-sum their perturbation series.
A related but simpler problem that might also benefit from such a study is that of
the gauge invariance and gauge fixing of string field theories:
While it was demonstrated that in a specific gauge Witten's open bosonic
string field theory reproduces the correct covering of moduli space, the quantum master
equation of this theory, which would ensure gauge invariance at the quantum level,
is singular~\cite{Thorn:1988hm}. The situation with many other open string field theories
seems to be similar. Many other issues that cannot currently be examined, such as
the fate of the closed string tachyon in open string field theories could also be
examined.

Another important motivation for such a study is that it could enable us to address
the big challenges that we listed in the beginning of this introduction.
Thus, such a study could be of use to the general string theory and high energy community,
as it would significantly extend the usefulness of
string field theory to the general research in the field. This could be particularly
important, as string field theoretical research tends much to concentrate around
string field theory itself.

However, the quantum non-perturbative study of string field theory is an enormous
endeavour. In this work we attempt a first small step towards this goal.
We consider the simplest possible string field theory, namely
Witten's bosonic open string field theory and examine the possibility of
studying this theory using lattice field theoretical methods.
Our aim is to provide a proof of concept for a lattice approach to string field theory
by identifying the many obstacles such an approach would have, suggesting
various ways to deal with these difficulties, and examining these issues
``experimentally'' by lattice simulations of a particularly simple setup.

The motivation for a lattice study is the complexity of the theories that
we are interested in. Even for regular field theories there is not much that
can be said analytically about the quantum non-perturbative regime
without, e.g. supersymmetry. Furthermore, even in the latter cases there are many
aspects of the theory that cannot be addressed analytically. Given the
limitations of analytical studies, it is only natural to consider a numerical
approach that is adequate for the study of non-perturbative physics of field
theories. Among the various possibilities, the lattice
approach~\cite{Wilson:1974sk,Creutz:1980zw,Creutz:1984mg,Rothe:1992nt,Montvay:1994cy,Smit:2002ug,DeGrand:2006zz,Gattringer:2010zz}
is probably the most established and the most useful one.

The choice of Witten's theory among the various possibilities is also
easily motivated, as it is the simplest and the most well understood among
the universal string field theories. Unlike the closed string field theories
it relies on a single product that can be explicitly expressed in terms of
known coefficients describing the coupling of various fields. Moreover,
studying the bosonic theory enables us to avoid the various complications related
to properly choosing picture numbers
of string fields, from which superstring field theories suffer.

As was already mentioned, there are several difficulties with the proposed approach.
Let us mention here a couple of obvious ones, and postpone the discussion
on the other ones to latter sections.
Witten's theory is cubic, implying that the action is unbounded from
below. While this is usually attributed to the unphysical nature of many of
the component fields, it is bound to lead to problems for a lattice simulation.
We attempt to resolve this problem using analytical continuation from a setup,
to be defined in what follows, in which the cubical terms are purely imaginary.
We thus trade the instability by a convergent oscillatory behaviour.

Another complication of Witten's theory comes from the fact that the bosonic
theory in the critical dimension, as well as in any other dimension above
two is presumably not well defined, due to the presence of the closed string tachyonic
instability.
Moreover, running lattice simulations in the critical $d=26$ dimensions seems hopeless
from a computational perspective.
A possible way to overcome both these problems is to study linear dilaton backgrounds
with $d\leq 2$. In this paper we focus mainly on the simplest one among all these models,
namely the $d=1$ case. While the $d\leq 2$ theories are not really
``stringy'' ones, it should be possible to generalize the (universal) string
field theoretical language used here to the more involved cases.

The rest of this paper is organised as follows:
In the remainder of this section we introduce the reader to the basics of
Witten's string field theory and to the formulation of string theory in a linear
dilaton background. In particular we define the ``level'' of a string field
and briefly introduce the notion of level truncation.
We also introduce four ``schemes'', which we study
in latter sections, for addressing the gauge symmetry of string field theory.
In section~\ref{sec:truncation} we
further dwell upon
the main tool used in the
numerical study of string field theory, namely level truncation, and explain
how to use it in the current context for the various schemes. We also remind
of the reality condition obeyed by the string field and perform an analytical
evaluation for the case of a single mode.
Then, in section~\ref{sec:setup} we explain how to put the string field theory
expressions on a lattice using the analytical continuation mentioned above.
We introduce observables whose dependence on various parameters can be studied,
and discuss some difficulties related to simulations involving a complex action.
The results of the lattice simulations are given in section~\ref{sec:results}.
Dependence of the observables on several parameters is studied and further
understanding of numerical and computational challenges in this approach
is achieved.
Some concluding remarks are offered in section~\ref{sec:conc}.

\subsection{String field theory}
\label{sec:SFTintro}

String field theory is a second quantization approach to string theory.
The classical string field
is identified with the quantum Fock
space of the first quantized (world-sheet) string theory. The world-sheet
theory is a two dimensional conformal field theory (CFT). Thus, its Fock
space is infinite dimensional. Hence, the string field is an infinite sum of
regular (component) fields.
The string field is assumed to be real. The reality condition for the string
field translates to reality conditions on the component
fields, to be described in~\ref{sec:RealityCond}.

The action of the string field is,
\begin{equation}
\label{action}
S\equiv S_2+S_3=-\int\Big(\frac{1}{2\al'}\Psi\star Q\Psi
   +\frac{g_o}{3}\Psi\star\Psi\star\Psi\Big),
\end{equation}
where $\Psi$ is the string field and the star product, the integral,
and $Q$ are defined below.
The constant $\al'$, related to the string tension,
defines the string length, a natural length scale for the string,
\begin{equation}
\label{ls}
l_s\equiv \sqrt{\al'}\,,
\end{equation}
and $g_o$ is the open string coupling constant.
A rescaling of $\Psi$ by $g_o$ would result in a global prefactor of
$g_o^{-2}$ in front of the action.
Note, however, that the way we define it here, $g_o$ is a dimensionful parameter.
Hence, we cannot expect to obtain canonical normalizations for
the component fields both in the way the action is written here and
after dividing by $g_o$. Jumping ahead to the equations that will follow upon
using level truncation we see that in the current form of the action,
canonical dimension for the scalar tachyon field is obtained
in~(\ref{tachQuad}). Then, we infer from~(\ref{S3T}) that
$g_o$ has a mass dimension of $\frac{6-d}{2}$.
Thus, we leave the action in the form~(\ref{action}).

In order to make sense out of the action~(\ref{action}),
the entities that appear in it should first be defined.
The bi-linear star product takes two string fields and gives back a single string
field. It has the geometric interpretation of gluing the right half of the
first string with the left half of the second string. Hence, it
is a non-commutative, associative product. The introduction of the star
product turns the space of string fields into an algebra.
The integral symbol represents ``integration over the space of string fields''.
It is performed by gluing of the left half of a single string to its right
half, followed by the evaluation of the CFT expectation value of the
resulting configuration.
The kinetic term is produced using the operator $Q$, which is the BRST
charge of the world-sheet theory. It is given by a contour integral of a
current $J$ around the state $\Psi$ in the CFT. An important property of $Q$
is that it is an odd derivation with respect to the other two operations,
\begin{align}
\int Q\Psi &=0\qquad \forall \Psi\,,\\
Q(\Psi_1\star\Psi_2) &=(Q\Psi_1)\star\Psi_2+(-)^{\Psi_1}\Psi_1\star(Q\Psi_2)\,.
\end{align}
Here $(-)^{\Psi_1}$ represents the parity of the string field $\Psi_1$.
The physical string
field is
odd and
leads
to a minus sign in the definition above,
while the string field that plays the role of a gauge parameter is even.

The fact that the string field includes an infinite amount of component
fields is a subtlety that any numerical method should address.
A common way to deal with the infinite amount of fields is to truncate the
string field to a finite sum by considering only fields whose ``level'' is
below some value $l_1$ and terms in the action integral, which are below some
$l_2$~\cite{Kostelecky:1988ta,Kostelecky:1990nt}. This is referred to as a
truncation to level $(l_1,l_2)$.
The level of a field is defined to be its conformal weight plus a constant
that sets the zero-momentum lowest level state to $l=0$.
Since the Virasoro operator $L_0$, which reads the conformal weight of
a state, serves as the (gauge fixed) kinetic term for the string field, the level
$l_0$ is essentially the on-shell mass of the string excitation considered.
Hence, level truncation has the natural physical interpretation of considering
only low-mass states.

The level is invariant under the action of $Q$. However, the star product mixes
different levels: the star product of string fields
that were truncated to a given level results in expressions that are not truncated
to this level. Hence, after the evaluation of the action in terms of component fields,
the action should also be truncated.
As $l_{1,2}$ are
sent to infinity one expects to obtain the result of using the full string
field. There is no proof that this should work and subtleties might well
arise. However, in the past it always did work.
The inclusion of a kinetic term implies $l_2\geq 2 l_1$ while the cubicity
of the action implies $l_2\leq 3 l_1$. In practice one always works either
in the $(l,2l)$ (which is simpler - has far fewer terms) or the
(more ``physical'') $(l,3l)$ level-truncation.

The conformal dimension of a field depends on its momentum. Most past
papers considered only the zero momentum sector. Those who did consider
non-zero momentum either considered the double limit of a truncation
in which the zero-momentum level and the momentum were considered separately,
or took the more physical choice of considering the total
conformal weight as a single level parameter.
In both cases, the only allowed momentum was along a
compactified space-like direction. Hence, this momentum was quantized and
its contribution to the level was always positive, which is sensible
from the perspective of level truncation.
This is not quite the case that we consider.
We do not have compactified directions and we consider the most general
space-time dependence of the fields.
However, the use of a lattice implies that we have to Wick-rotate the time
direction and to evaluate the action on a finite range of space-time, with
some arbitrary (Neumann/Dirichlet) boundary conditions.
These two restrictions turn the use of the more physical level truncation
into a sensible choice, which we adopt. Not only would that free us from
considering double limits, but it would also simplify and make more accurate
the consideration of the string-field-theory-inherent non-localities.
Thus, we define the total level of a string field to be
\begin{equation}
\label{l0lp}
l=l_0+l_p\,,
\end{equation}
where $l_0$ is defined by the mass of the specific excitation and $l_p$
includes the contribution of the momentum to the conformal weight.

Another important issue is that of the space of string fields,
a proper definition of which is still lacking.
Currently it is not even clear which mathematical concepts are needed in order to
properly define it. However, the general problems with the definition of
this space should not emerge in the context of level truncation. On the other hand,
we should decide whether the space of string fields should be restricted
to string fields of a given (first-quantization) ghost number and whether
dependence on the ghost zero mode should be allowed in its definition.

The ghost system that we refer to here is the $bc$ system
used to fix the conformal
gauge
symmetry on the world-sheet. This symmetry is
generated by the energy momentum tensor $T(z)$ which is an even object
of conformal dimension $h=2$. Thus, it is fixed by a system of two conjugate
odd bosons, $b$ whose conformal dimension is also $h=2$ and $c$ with $h=-1$.
The ghost number is defined as the number of $c$ insertions minus the number
of $b$ insertions.
These first-quantized ghosts manifest themselves in the second quantized
formulation by declaring that the string field $\Psi$ is a functional not
only of the space-time $X^\mu$ variables, but also of the $bc$ system.
In terms of modes these conformal fields can be expanded as\footnote{Our
conventions and world-sheet analysis follow Polchinski's
textbook~\cite{Polchinski:1998rq}.}
\begin{equation}
b(z)=\sum_{n=-\infty}^\infty \frac{b_n}{z^{n+2}}\,,\qquad
c(z)=\sum_{n=-\infty}^\infty \frac{c_n}{z^{n-1}}\,.
\end{equation}
The fact that these fields are canonically conjugate implies\footnote{In this
paper the brackets represent the graded commutator, which for the current case
is an $\mbox{anticommutator}$.}
\begin{equation}
[b_n,c_m]=\delta_{n,-m}\,.
\end{equation}

The world-sheet quantum space is defined in terms of vertex operators,
which are restricted to carry ghost number one.
The classical string field $\Psi$ generalizes the space of vertex operators
and should therefore also be restricted to carry ghost number one.
However,
a proper treatment of the gauge invariance of the classical action~(\ref{action}),
can modify this restriction.
The gauge transformation that leaves the action invariant is
\begin{equation}
\label{gaugeSym}
\delta\Psi=Q\La+\al'g_o\big(\Psi\star\La-\La\star\Psi\big)\,.
\end{equation}
A common way to classically fix the gauge is to impose the Siegel gauge,
which is a string field theoretical extension of the Feynman gauge for the
vector component field. This gauge choice is enforced by requiring that the
$b$-ghost zero mode annihilates the string field,
\begin{equation}
b_0\Psi=0\,.
\end{equation}
The Fock vacuum is annihilated by $b_0$. Hence, the Siegel gauge can also be
defined as the space of states built from the vacuum without using $c_0$.
A quantum treatment of the gauge symmetry should take into account the fact
that the gauge symmetry is redundant and ``uses the equations of motion''.
The most natural framework for addressing such a system covariantly
is the field-antifield BV formalism. This formalism was applied to Witten's
string field theory in~\cite{Bochicchio:1986zj,Bochicchio:1986bd,Thorn:1986qj}.
The result is very elegant:
prior
to gauge fixing
the action should be replaced by a ``master action'', which is identical in
form to the classical action~(\ref{action}), except that the string field
$\Psi$ should no longer be constrained to carry ghost number one\footnote{The
string field should still be an odd object. Hence, the component fields at
even ghost numbers have to be odd. This is also consistent with the general
parity assignments of the BV/BRST formalisms.}. One can
gauge fix the master action to obtain the Siegel gauge in the space of string
fields with unrestricted ghost number. A potential subtlety with the master
action is that according to the general BV formalism it should obey the
``quantum master equation''. However, it is still not clear if this is the
case and it is only certain that it obeys the ``classical master equation''.

While gauge fixing is necessary in perturbation theory, this is
not always the case in
a lattice approach, in which the infinite gauge orbits are replaced by finite ones.
Hence,
we can consider the following four options for our space of string fields,
to which we refer in the following as {\it schemes}:
\begin{enumerate}
\item Classical string fields without gauge fixing, i.e. $\Psi$ carries
      ghost number one but has no ghost zero-mode restrictions.
\item Classical gauge-fixed string fields.
			This is probably inconsistent, since there is no justification for
			gauge fixing without a proper treatment of the gauge symmetry.
\item BV string field without gauge fixing. Here, all ghost numbers are
      considered and the string field is allowed a $c_0$ dependence.
\item Gauge-fixed BV string field, i.e. $\Psi$ carries all possible ghost
      numbers but is $c_0$-independent.
\end{enumerate}
One of the important advantages of the lattice approach to field theory
is that it provides a regulator that does not break gauge symmetry.
In our case, on the other hand, the lattice does break the gauge symmetry,
since the star product mixes all levels. Hence, one should expect that
the gauge fixed schemes, in which the gauge symmetry was already taken care of,
would be better behaved. While we will experiment with all the four schemes,
we will see that, as we expect, scheme 4 seems to be the most promising one.

\subsection{CFT and the linear dilaton at $d\leq 2$}

The first step towards the explicit construction of the string field theory
is the choice of the background CFT. In string theory the background always
includes the $bc$ ghost system, which we already described.
Other than the ghost system the world-sheet theory depends also on a matter
system, which can be any CFT, as long as its central charge is $c_m=26$.
This value for the central charge is needed in order to cancel the central
charge of the ghost system which is $c_{gh}=-26$. The vanishing of the
total central charge is necessary for avoiding conformal anomalies.

In this paper, we work with a one dimensional linear dilaton theory.
But let us for now consider the more general case of a linear dilaton in
$d\leq 2$ dimensions\footnote{In
the two dimensional case, the world-sheet degrees of freedom are
the ``tachyon'' field and the ``discrete states'', which are
physical only for specific momenta. Scattering amplitudes
of both tachyons and discrete states are
known~\cite{Bershadsky:1991ay,Bershadsky:1992ub}.
They were considered from the string field theory perspective
in~\cite{Arefeva:1992yh,Urosevic:1993yx,Urosevic:1994yh}. Other relevant
papers include the identification of the ground ring/$W_\infty$
structure~\cite{Witten:1991zd}, the
introduction of ZZ~\cite{Zamolodchikov:2001ah} and FZZT
D-branes~\cite{Fateev:2000ik,Teschner:2001rv},
open/closed duality and relation to the supersymmetric
theories~\cite{McGreevy:2003kb} and the decay of ZZ-branes~\cite{Klebanov:2003km}.
}.
This case includes theories with $1<d<2$, where we identify the dimension $d$ with
the central charge of the non-linear-dilaton part of the matter sector plus one.
This identification stems from the fact, on which we elaborate below, that the
linear dilaton can be realized by a single non-homogenous direction,
while standard space-time directions correspond to one unit of central charge.
The $1<d<2$ theories can be realized using non-minimal
models, but we shall refrain in this work from going into details about these systems.
For the two dimensional case note that while we
Wick-rotate for enabling the lattice simulations, there is still a
distinction between the two dimensions, due to the linear dilaton
background. Also note that the dilaton gradient, $V^\mu$ to be defined below,
is of the order of the string scale,
\begin{equation}
\label{ms}
m_s\equiv l_s^{-1} = (\alpr)^{-\frac{1}{2}}\,.
\end{equation}
The string scale is usually identified (presumably up to a constant of order
one or around it) with the Planck mass.
Hence, these spaces differ
substantially from standard space-times.
We will further have to set two other scales, namely the lattice spacing and the lattice size.
In order to obtain a proper description of the physics, the first one
should be sent to zero and the second one should be sent to infinity,
in $l_s$ units, in order to approach the continuum limit.

In a theory of open strings, the matter fields can be expanded
as\footnote{In the literature it is
common to find the variables $\al_n^\mu\equiv \sqrt{|n|}a_n^\mu$.},
\begin{equation}
X^\mu(z,\bar z)=x^\mu-i\al'p^\mu \log|z|^2
 +i\sqrt{\frac{\al'}{2}}\sum_{n=1}^\infty \Big(
      \frac{a_n^\mu}{\sqrt{n}}(z^{-n}+\bar z^{-n})
      -\frac{a_n^{\dag \mu}}{\sqrt{n}}(z^n+\bar z^n)\Big)\,,
\end{equation}
where $\mu=1,2$ in the $2d$ case and
takes a single value and hence can be omitted in the $1d$ case.
Taking the derivative with respect to $z$ gives
\begin{equation}
\label{idx}
i\partial X^\mu(z)=\frac{\al'p^\mu}{z}
 +\sqrt{\frac{\al'}{2}}\sum_{n=1}^\infty \sqrt{n} \Big(
      a_n^\mu z^{-n-1}+a_n^{\dag \mu}z^{n-1}\Big)\,.
\end{equation}
These fields have the following operator product expansion (OPE)
\begin{equation}
\label{matterOPE}
X^\mu(z,\bar z)X^\nu(w,\bar w)\sim-\frac{\al'}{2}\eta^{\mu\nu}\log|z-w|^2\,,
\end{equation}
which translate into the standard commutation relations among the infinite
set of creation and annihilation operators,
\begin{equation}
[a_n^\mu,a_m^{\dag \nu}]=\eta^{\mu\nu}\delta_{m,n}\,.
\end{equation}
We assume a flat (space-time) metric,
\begin{align}
& \eta^{\mu\nu}=\diag(1,1) &d=2\,,\\
& \eta=1 &d=1\,.
\end{align}
Momentum states are represented as usual using the operators
$e^{ik\cdot X}$. The OPE~(\ref{matterOPE}) implies that these exponents
suffer from short distance singularities and should be normal-ordered.
We do not write the normal ordering symbol and assume that all fields that
we consider are implicitly normal ordered. In the standard flat background
with constant dilaton, the normal-ordered exponents
carry well-defined conformal weights,
\begin{equation}
(h,\bar h)=\Big(\frac{\al'k^2}{4},\frac{\al'k^2}{4}\Big).
\end{equation}

Since we consider open strings, our fields are given by insertions on the
boundary of the CFT theory, which we identify with the real axis,
\begin{equation}
\Im(z)=0 \quad \Longleftrightarrow \quad z=\bar z\,,
\end{equation}
and the CFT is defined on the upper half plane. One usually extends the CFT
to the whole complex plane using the doubling trick, according to which
$z$-dependence above the real line is identified with $\bar z$-dependence
below it. Still, the real line is special, e.g. conformal fields may suffer
from extra singularities when approaching it, originating from collisions of
$z$ and $\bar z$. In particular, momentum states should now be normal-ordered
according to the boundary-normal-ordering. Then, they carry a well defined
single conformal weight
\begin{equation}
h(e^{ik\cdot X})=\al'k^2\,.
\end{equation}
Another confusion that can arise from the presence of the boundary is the
relation between the derivative with respect to $z$ and that with respect to
the boundary parameter. We will try to avoid this issue by always
considering only the $z$ variable.

In order to fully describe the CFT one has to define the energy-momentum
tensor,
\begin{equation}
T(z)=T^g(z)+T^m(z)\,,
\end{equation}
where the superscripts stand for ``ghost'' and ``matter''.
The ghost part of the energy momentum tensor is fixed,
\begin{equation}
T^g(z)=-\partial b c (z)-2b\partial c (z)\,,
\end{equation}
but the matter part is theory-dependent. In our case it equals,
\begin{equation}
\label{Tm}
T^m(z)=-\frac{1}{\al'}\partial X^\mu \partial X_\mu+V^\mu\partial^2 X_\mu\,.
\end{equation}
The second ``improvement'' term encodes the linear dilaton nature of the
background.
This term changes the central charge to a new value,
\begin{equation}
c=d+6\al'V^2\,.
\end{equation}
In order to obtain the correct
value of 26 for the central charge, one has to impose
\begin{equation}
\label{genV}
V^2=\frac{26-d}{6\al'}\,.
\end{equation}
In particular, the cases we are interested in are,
\begin{align}
& V^\mu=(0,V)\,,\qquad V=-\frac{2}{\sqrt{\al'}} &d=2\,,\\
& V=-\sqrt{\frac{25}{6\al'}} &d=1\,,
\end{align}
where the minus sign is just a convention.

In a linear-dilaton background the following modifications occur as
compared to a standard flat-space background~\cite{Hellerman:2008wp,Beaujean:2009rb}:
\begin{itemize}
\item The momentum-conservation delta function is modified in order to
      reflect the breakdown of translational invariance,
      \begin{equation}
	    \delta^d\big(\sum_n k_n^\mu\big)\longrightarrow
	     \delta^d\big(\sum_n k_n^\mu+i V^\mu\big)\,.
      \end{equation}
      Here, the delta function is formally defined by,
      \begin{equation}
	    \delta^d\big(\sum_n k_n^\mu+i V^\mu\big)\equiv \frac{1}{(2\pi)^d}
	       \int d^d x\, e^{-ix\cdot\sum k_n+x\cdot V}\,.
	    \end{equation}
\item The field $X$, aligned with the linear dilaton direction,
      is no longer a (logarithmic) conformal tensor,
      \begin{equation}
      \label{Xtrans}
	    X^\mu(z,\bar z)\rightarrow f\circ X^\mu=X^\mu(f(z),\bar f(\bar z))+
	      \frac{\al'}{2}V^\mu\log|f'(z)|^2\,.
      \end{equation}
      The momentum operators built from $X^\mu$ remain conformal tensors.
      However, their conformal weights change,
      \begin{equation}
      \label{confWeight}
	    h(e^{ik\cdot X})=\al'(k^2+ik\cdot V)\,.
	    \end{equation}
\item The change in $T^m$ induces a change in $Q$, since the following identity holds in general
      \begin{equation}
	    J(z)=cT^m(z)+bc\partial c(z)+\frac{3}{2}\partial^2 c(z)
			\,,
	    \end{equation}
			where $J(z)$ is the BRST current, which should be integrated to obtain $Q$.
\end{itemize}

The introduction of a linear dilaton introduces complications also from the lattice
perspective.
Specifically, the fact that the target space is no longer homogenous implies
that it would not be enough to consider the length $L$ of the lattice, as a single
length parameter. Instead, we would have to study the dependence on $x_{min}$ and
$x_{max}$, the minimum and maximum points of the spatial direction\footnote{Here,
it is assumed that
we work in $d=1$.
For $d>1$, $x_{min}$ and $x_{max}$
are the minimum and maximum values in the direction
of the linear dilaton.} $X$,
or alternatively on $x_{min}$ and
the lattice interval
\begin{equation}
L\equiv x_{max}-x_{min}\,.
\end{equation}

\section{Level truncation}
\label{sec:truncation}

In this section we
implement level truncation in order to
describe various aspects, problems and resolutions
of our program from the string field theory perspective. We evaluate
most
expressions
in an arbitrary dimension and concentrate
on the $d=1$ case at the end. Later on, in section~\ref{sec:setup},
we discuss the computational lattice perspective.
Here we start by evaluating the action for $l_0=0$
in~\ref{sec:levelZero}. Next we discuss the way by which higher
levels can be added in~\ref{sec:HigherLevels} and explain how to
impose the reality condition in~\ref{sec:RealityCond}.
Then, we describe explicitly the first case of a higher level,
namely $l_0=1$, in~\ref{sec:l01}. In order to go to higher levels
one has to define a systematic way for the evaluation of the
very many terms that are present in the action.
We sketch a method
that can be used
for an automatization of the evaluation
of the action in~\ref{sec:Automat} and
utilize
it
in~\ref{sec:scheme3} for the evaluation of the action
of scheme 3 at $l_0=1$. We further explain there
that there is a problem with this action that prevents us from
using scheme 3 together with level truncation, after which we
truncate the scheme 3 action to a scheme 1 action in~\ref{sec:scheme1}.
We end this section by an analytical study of the simplest
possible case, that of a single
mode, in~\ref{sec:lowest_mode},
followed by a discussion of one more potential obstacle
towards using our
methods,
in~\ref{sec:trivial_terms}.

\subsection{Truncation to zero $l_0$-level}
\label{sec:levelZero}

We consider classical string fields that are built upon the
vacuum of the first quantized theory,
\begin{equation}
\ket{\Om}\equiv c(0)\ket{0}=c_1 \ket{0}\,,
\end{equation}
where $\ket{0}$ is the SL(2) invariant vacuum\footnote{The conformal symmetry
is generated by the Virasoro operators $L_n$, $n \in \Z$.
The vacuum $\ket{0}$ is invariant under an SL(2) subalgebra of the Virasoro algebra
generated by $L_0$ and $L_{\pm 1}$. This subalgebra becomes useful for us already
at $l_0=1$~(\ref{MatterVirasoro}). The other Virasoro operators are useful
at higher levels.}. The vacuum $\ket{\Om}$ is odd and carries ghost number one,
as
is proper for a classical string field.
General string field configurations are built from this vacuum by acting
on it with the operators $b_{-n}$ with $n>0$, $c_{-n}$
with $n\geq 0$ and $a^\mu_n$ with $n>0$. 
Note that the $b_{-n}$ and $c_{-n}$ ghost operators
are odd, so each one of them can act at most once, while the
$a^\mu_n$ are even and so each one of them can act indefinitely.
In order to retain the restriction to
ghost number one, which is needed for schemes 1 and 2, the number of $b$ and $c$
insertions must be the same. For the other two schemes there is no such restriction.
However, if the total number of $b$ and $c$ ghost insertions is odd,
the relevant component field must also have odd parity in order not to
change the parity of the string field.

The momentum of the state is quantized when we work with $0<X<L$.
Assuming Dirichlet boundary conditions, the momentum dependence would
come from $\sin\big(\frac{\pi k \cdot X}{L}\big)$, with $k$ an integer.
We can now define the level as the sum of the indices of all the operators
plus a contribution from the momentum, e.g. the level of
$c_{-3}b_{-2}(a^{\dag 1}_4)^2\ket{\Om,p}$ is $l=l_0+l_p$, with
$l_0=3+2+4+4=13$. However,~(\ref{confWeight}) implies that the
sine factor is not an eigenvalue of the level operator. Not only that,
but the eigenvalues of the two distinct exponents composing the sine
are complex. We resolve this problem below.

At the lowest (zero) $l_0$-level the string field contains only two component
fields. We impose the reality condition on the string field. This implies
that the following two component fields are real,
\begin{equation}
\label{Psi0}
\Psi_0=\int d^d p\, \big(T(p)e^{ip\cdot X}
  + \cT(p)e^{ip\cdot X}c_0\big)\ket{\Om}.
\end{equation}
The first of these fields is the ``tachyon'' $T$ (not to be confused with
the energy-momentum tensor). It carries no $c_0$ insertion and has ghost
number one and so contributes in all four schemes. The second field
$\cT$ is odd. It carries ghost number two and contains the $c$ zero mode
in its definition. Hence, it contributes only in scheme number three. In any
case, it cannot contribute to the action without the presence of string
fields of ghost number less than one. Thus, it does not contribute at all
at zero $l_0$-level and is set for now to zero.
Direct calculation shows that,
\begin{equation}
\label{QPsi0}
Q\Psi_0=\int d^d p\, T(p)\al'\Bigg(\Big(p+\frac{i V}{2}\Big)^2+m_0^2\Bigg)
  \partial c c e^{ip\cdot X}\ket{\Om}\,,
\end{equation}
where we defined,
\begin{equation}
\label{m0}
m_0^2\equiv \frac{V^2}{4}-\frac{1}{\al'}\,.
\end{equation}
We see that the constant term inside the parentheses vanishes only for the value of $V$~(\ref{genV})
at $d=2$. This constant fixes the mass of this mode. Hence, we see that the
tachyon becomes massless exactly in two dimensions, while for $d=1$ it is
massive.

The kinetic term of the tachyon reads,
\begin{equation}
\label{Psi0QPsi0}
S_2=-\frac{1}{2\al'}\int \Psi_0 Q\Psi_0=
   -\frac{1}{2\al'}\vev{(I\circ \Psi_0)(0)Q\Psi_0(0)}\,,
\end{equation}
where the expectation value is evaluated in the upper half plane
and $I$ is the conformal transformation,
\begin{equation}
I(z)=-\frac{1}{z}\,.
\end{equation}
For the evaluation of~(\ref{Psi0QPsi0}) we have to regularize,
$z\rightarrow\ep$ and continue as in~\cite{LeClair:1989sp}.
The expectation value factorizes into matter and ghost parts.
The ghost part gives,
\begin{equation}
\vev{c\big(-\frac{1}{\ep}\big)c'c(0)}=\frac{1}{\ep^2}\,,
\end{equation}
while for the matter part we have to evaluate
\begin{equation}
\vev{e^{ip\cdot X}\big(-\frac{1}{\ep}\big)e^{iq\cdot X}(0)}=
\frac{(2\pi)^d \delta^d(p+q+iV)}{\Big(\frac{1}{\ep}\Big)^{-2\al'p\cdot q}}
\,.
\end{equation}
Using this result for~(\ref{Psi0}) and~(\ref{QPsi0}) leaves us with integration
over the momenta, as well as with a
factor coming from the conformal transformation,
\begin{equation}
\Big(\frac{dI}{dz}\Big)^h=
   \Big(\frac{1}{\ep^2}\Big)^{\al'(p^2+ip\cdot V)-1}\,.
\end{equation}
Using the delta function one sees that all $\ep$-dependent factors cancel
out, regardless of the specific background and the final result is,
\begin{equation}
\begin{aligned}
S_2&=-\frac{1}{2}\int d^d p \, d^d q \,
  (2\pi)^d \delta^d(p+q+iV) T(p)T(q)
   \Bigg(\Big(p+\frac{i V}{2}\Big)^2+m_0^2\Bigg)
\\ \label{tachQuad}
 &=-\frac{1}{2}\int d^d p \, d^d q \,d^d x\, e^{-ix\cdot (p+q+iV)}
      T(p)T(q) \Bigg(\Big(\frac{p-q}{2}\Big)^2+m_0^2\Bigg)
\\& =-\frac{1}{2}\int d^d x\,e^{V \cdot x}
 \Bigg(\frac{\nabla T(x) \cdot \nabla T(x)-T(x)\nabla^2 T(x)}{2}
        +m_0^2 T(x)^2\Bigg),
\end{aligned}
\end{equation}
where we abuse the notation by using the same symbol for the Fourier
conjugate fields $T(p)$ and $T(x)$.

We now have to evaluate the cubic term,
\begin{equation}
S_3=-\frac{g_o}{3}\int \Psi_0\star\Psi_0\star\Psi_0=
  -\frac{g_o}{3}\int d^d p\, d^d q\, d^d k\,
   \vev{(f_{-1}\circ \Psi_0) (f_0\circ \Psi_0) (f_1\circ \Psi_0)}.
\end{equation}
The three conformal transformations are obtained by sending the upper half
plane to the unit disk using,
\begin{equation}
\label{UHPtoDisk}
w=\frac{1+iz}{1-i z}\,,
\end{equation}
then rescaling $w$ and relocating it to the three points of the
``rotated Mercedes-Benz logo'',
\begin{equation}
w\rightarrow e^{\frac{2\pi i n}{3}}w^{\frac{2}{3}}\,,
\end{equation}
and finally sending it back to the upper half plane using the inverse
of~(\ref{UHPtoDisk}),
\begin{equation}
z=i\frac{1-w}{1+w}\,.
\end{equation}
The only relevant information about these conformal transformations is,
\begin{equation}
\label{fns}
f_0(0)=0\,,\qquad f_{\pm 1}(0)=\pm\sqrt{3}\,,\qquad f'_0(0)=\frac{2}{3}\,,
\qquad f'_{\pm 1}(0)=\frac{8}{3}\,.
\end{equation}
Now, the ghost part contributes,
\begin{equation}
\vev{c(-\sqrt{3})c(0)c(\sqrt{3})}=2\cdot 3^{\frac{3}{2}}\,,
\end{equation}
the matter part is,
\begin{equation}
\vev{e^{ip\cdot X}(-\sqrt{3})e^{ik\cdot X}(0)
   e^{iq\cdot X}(\sqrt{3})}=
    \frac{(2\pi)^d \delta^d(p+k+q+iV)}
      {3^{-\al'p\cdot k}\cdot 3^{-\al'k\cdot q}\cdot 12^{-\al'p\cdot q}}\,.
\end{equation}
and the conformal weights contribute,
\begin{equation}
\begin{aligned}
\Big(&\frac{df_{-1}}{dz}\Big)^{h_{-1}}
\Big(\frac{df_{0}}{dz}\Big)^{h_0}
\Big(\frac{df_{1}}{dz}\Big)^{h_1}\\&=
\Big(\frac{8}{3}\Big)^{\al'(p^2+ip\cdot V)-1}
\Big(\frac{2}{3}\Big)^{\al'(k^2+ik\cdot V)-1}
\Big(\frac{8}{3}\Big)^{\al'(q^2+iq\cdot V)-1}\,.
\end{aligned}
\end{equation}
All in all we get,
\begin{align}
\nonumber
S_3 &=
-g_o
\int d^d p\, d^d q\, d^d k \frac{(2\pi)^d \delta^d(p+k+q+iV)}{3}
  T(p)T(k)T(q) K^{3-\al'(p^2+k^2+q^2+V^2)}\\
\label{S3T}
 &=-\frac{g_o K^{3-\al'V^2}}{3}\int d^d x\, e^{V \cdot x} \tilde T (x)^3 \,.
\end{align}
Here (as usual) we defined,
\begin{equation}
K\equiv \frac{3\sqrt{3}}{4}\,,
\end{equation}
and used the delta function in the first equality. In the second equality
we moved to $x$-space, as in~(\ref{tachQuad}) and defined (as usual) the
tilded variables,
\begin{equation}
\tilde T(x)=K^{\al' \nabla^2}T(x)\,.
\end{equation}
Similar relations are to be understood for other tilded variables in what
follows.
The pre-factor $K^{3-\al'V^2}$ can be absorbed into a redefinition of
the coupling constant.
Note also the linear dilaton factor $e^{V \cdot x}$, which is common to the
quadratic and cubic terms. A rescaling of $T$ by this factor leads
to a space-dependent effective coupling,
\begin{equation}
\label{effCoup}
g_o^{eff}\sim e^{-\frac{V \cdot x}{2}}\,.
\end{equation}
This coupling goes from zero to infinity along the linear dilaton direction,
which implies that the pre-factor $g_o K^{3-\al' V^2}$ can be set to unity by
an appropriate choice of the origin.
Note, that this effective coupling constant is still dimensionful. It is
possible to multiply it by a proper power of $\al'$ in order to obtain a
dimensionless coupling constant.

Let us now perform the advocated field redefinition,
\begin{equation}
\label{fieldReDef}
T(x)=e^{-\frac{V \cdot x}{2}}\tau(x)\quad \Longleftrightarrow \quad
 T(p)=\tau \Big(p+i\frac{V}{2}\Big)\,.
\end{equation}
The kinetic term takes now the standard form,
\begin{equation}
\label{S2Tau}
S_2=-\frac{1}{2}\int d^d x\, \big(m_0^2\tau^2+(\nabla \tau)^2\big)\,.
\end{equation}
This field redefinition is defined pointwise. Hence, the Jacobian is just a
number and can be ignored.
The interaction term transforms under the above field redefinitions into,
\begin{equation}
\label{S3Tau}
S_3=-\frac{g_o K^{3\big(1-\frac{\al' V^2}{4}\big)}}{3}\int d^d x\,
 e^{-\frac{V \cdot x}{2}}\tilde \tau (x)^3 \,.
\end{equation}
It is easiest to obtain this expression in momentum space, where one has to
use the delta function and deform the contour of integration.
Note that the resulting interaction term is both space-dependent and
non-local. In $p$-space the
spatial-dependence and non-locality reverse their roles.

The spatial dependence of the coupling constant implies that we cannot use
periodic boundary conditions on the lattice, since it would glue a strong
coupling region with a weak coupling region, which is unphysical. Instead we
can use, as mentioned above, Dirichlet or Neumann boundary conditions.
We choose the former and evaluate the action
in a box $x_{min}^\mu<x^\mu<x_{max}^\mu$, with $x_{max}^\mu-x_{min}^\mu=L^\mu$.
Specifying now to $d=1$, we can expand,
\begin{equation}
\label{tauExpansion}
\tau(x)=\sqrt{\frac{2}{L}}\sum_{n=1}^\infty
  \tau_n \sin\Big(\frac{\pi n (x-x_{min})}{L}\Big)\,.
\end{equation}
We can now recognize one more advantage of $\tau$ over $T$. We mentioned
above that the expansion of $T$ in terms of sine modes does not involve
conformal eigenmodes and its expansion in exponents leads to complex eigenvalues.
Contrary to that, one can see that the expansion in sine modes of $\tau$ is well
defined and real.
Furthermore, we shall see in section~\ref{sec:RealityCond} that working with the
$\tau$ variable is essential also in order to obey the string field reality
condition.

Another issue that we would like to mention is that of the variational principle.
While we do not derive here equations of motion, since we concentrate on the action
itself, it could still be interesting to examine their derivation and their form
in the case at hand.
String field theory includes an infinite number of derivatives. It is known that
there might be subtleties with the definition of the variational principle in
such theories~\cite{Eliezer:1989cr,Moeller:2002vx}.
One could wonder whether our choice of boundary conditions is consistent
with a variational principle. The quadratic term is the standard one and so is
its variation. The variation of the cubic term leads to
\begin{equation}
\delta S_3=-\frac{g_o K^{3\big(1-\frac{\al' V^2}{4}\big)}}{3}\int d^d x\,
 e^{-\frac{V \cdot x}{2}}\tilde \tau (x)^2 \delta \tilde \tau(x) \,.
\end{equation}
It seems that in order to obtain an equation of motion we need to change the
$\delta \tilde \tau(x)$ term into a $\delta \tau(x)$ term, i.e. to integrate by parts
the $K^{\al' \nabla^2}$ factor acting on $\delta \tau(x)$, such that it would act
on the $\tilde \tau^2(x)$ factor and on the dilaton factor $e^{-\frac{V \cdot x}{2}}$.
Such an integration by parts would lead to boundary terms
that include the variation of various higher order derivatives of $\tau(x)$.
These expressions should not a-priori vanish. Moreover,
setting all these infinitely many terms to zero could lead
to very strict functional restrictions on $\tau(x)$.
However, we can take another approach. Given the expansion~(\ref{tauExpansion})
we obtain,
\begin{equation}
\tilde\tau(x)=\sqrt{\frac{2}{L}}\sum_{n=1}^\infty
  K^{-\al' \big(\frac{\pi n}{L}\big)^2}\tau_n \sin\Big(\frac{\pi n (x-x_{min})}{L}\Big)\,.
\end{equation}
When expressed in this form it seems that $\tilde \tau(x)$ and $\delta \tilde \tau(x)$
vanish at the boundaries.
Nonetheless, this assertion relies on some convergence properties of the expansion,
which might not be well justified. This issue is related to the discussion
in~\cite{Eliezer:1989cr,Moeller:2002vx} and more generally to the problem of
properly defining the space of string fields. Attempting to analyse it would
take us too far away. Hence, we do not dwell on these questions further.

We can now assign a level to the modes in the expansion
above,
\begin{equation}
l(\tau_n)=\al' \Big(\frac{\pi n}{L}\Big)^2\,.
\end{equation}
We have to include all levels that are smaller than some $l<1$. The
restriction $l<1$ comes from the fact that it was assumed that we
include only the tachyon field. For $l\geq 1$ higher $l_0$-modes might also
contribute\footnote{For a Dirichlet expansion these modes actually
contribute starting at some $l>1$.}. The physical origin of this restriction
is that if we decide to probe lower than string-scale size modes, we should
also include the higher modes, which are also of this size.

An $l$ that allows for $N$ modes is equivalent to working on an $N$-site
lattice. However, the non-locality and space-dependence of the action
simplify if we perform the analysis directly in terms of the modes.
The free part of the action is now,
\begin{equation}
\label{level 0 S2}
S_2=-\frac{1}{2}\sum_{n=1}^N \Big(\frac{1}{24\al'}
  +\big(\frac{\pi n}{L}\big)^2\Big) \tau_n^2\,.
\end{equation}
Assuming that we work in the $(l,3l)$ scheme (that is, if all interaction
terms are to be included), the interaction term reads,
\begin{align}
\label{level 0 S3}
& S_3 =-\frac{g_o K^{3\big(1-\frac{\al' V^2}{4}\big)}}{3} \sum_{n_{1,2,3}=1}^N
   K^{-\al'\big(\frac{\pi}{L}\big)^2(n_1^2+n_2^2+n_3^2)}
     \tau_{n_1}\tau_{n_2}\tau_{n_3}f_{n_1,n_2,n_3}\,,\\
& f_{n_1,n_2,n_3} \equiv \Big(\frac{2}{L}\Big)^\frac{3}{2}\int_{x_{min}}^{x_{max}} dx
  e^{-\frac{V x}{2}}\cdot\\
	\nonumber
	& \qquad \cdot\sin\Big(\frac{\pi n_1 (x-x_{min})}{L}\Big)
    \sin\Big(\frac{\pi n_2 (x-x_{min})}{L}\Big)\sin\Big(\frac{\pi n_3 (x-x_{min})}{L}\Big).
\end{align}
We can substitute $-\frac{V x}{2}=\sqrt{\frac{25}{24\al'}}x$. We left the
$V$ dependence for enabling the evaluation of this expression with
unphysical values of $V$.
We can now evaluate this action on the lattice.
Note, that our Wick-rotation was performed in such a way that we have to
consider,
\begin{equation}
\label{pathInt}
Z=\int \Big(\prod_n d\tau_n\Big) e^S\,.
\end{equation}

\subsubsection{Moving the non-locality to the quadratic term}
\label{sec:move non-locality}

In previous studies of level-truncated string field theory
it was suggested to move the non-locality from the cubic term to the
quadratic term. The motivation was the simplification of the
(relatively more complicated) interaction term.
Furthermore, as it is moved to the quadratic term, the non-locality becomes completely
diagonal, which results in somewhat simplified expressions.

At the lowest $l_0$ level, the action is still given
by the sum of~(\ref{S2Tau}) and~(\ref{S3Tau}), only now the fundamental
field, which should be expanded in modes is $\tilde \tau$, while $\tau$
is defined in terms of it as
\begin{equation}
\tau = K^{
\alpr
p^2} \tilde \tau\,.
\end{equation}
For simplicity of notations, we drop the tilde from now on, with the understanding
that the correct variables are used.

Expanding in the modes for the tachyon field, the action is modified to
\begin{align}
& S_2=-\frac{1}{2}\sum_{n=1}^N K^{-2\al'\big(\frac{\pi n}{L}\big)^2}\Big(\frac{1}{24\al'}
  +\big(\frac{\pi n}{L}\big)^2\Big) \tau_n^2\,,\\
& S_3 =-\frac{g_o K^{3\big(1-\frac{\al' V^2}{4}\big)}}{3} \sum_{n_{1,2,3}=1}^N
    \tau_{n_1}\tau_{n_2}\tau_{n_3}f_{n_1,n_2,n_3}\,,
\end{align}
where the definition of $f_{n_1,n_2,n_3}$ did not change.

As higher level fields are added, one can similarly decide whether the better
representation is the one in which the untilded fields
are the fundamental ones, or the one with the tilded fields.
The needed manipulations are completely analogous to what is done here.

\subsection{Higher levels}
\label{sec:HigherLevels}

Evaluating conformal transformations can be tedious, especially at higher
levels, where the coefficient fields are not represented by primary
conformal fields. A way to simplify calculations was actually devised even
before the CFT formulation of~\cite{LeClair:1989sp}. In this formulation
the action is given by,
\begin{equation}
S=-\frac{1}{2\al'}\,
  \brai{12}{V_2}\ket{\Psi}_1\ket{Q\Psi}_2
   -\frac{g_o}{3}\,\brai{123}{V_3}\ket{\Psi}_1\ket{\Psi}_2\ket{\Psi}_3\,,
\end{equation}
where the subscripts represent an index of a copy of the Hilbert space.
The two-vertex $V_2$ and three-vertex $V_3$ live in the spaces $H^2$ and
$H^3$ respectively. They are squeezed states and their form for flat
background was found
in~\cite{Gross:1987ia,Gross:1987fk,Gross:1987pp,Samuel:1986wp,Ohta:1986wn,Cremmer:1986if}.
The modification of these works to a linear dilaton background is relatively
simple, due to the similarity with the bosonized ghost sector, studied in
these papers. Explicit expressions for $d=2$ were given
in~\cite{Urosevic:1993yx}. Both factorize into matter and ghost sectors
(the matter sector further factorizes into $d$ independent sectors).

The two-vertex is explicitly given by,
\begin{subequations}
\label{V2}
\begin{align}
\label{V2m}
& \begin{aligned}
\brai{12}{V_2^m}=\int d^d p\,d^d q\,& \brai{1}{p}\brai{2}{q}\delta^d(p+q+iV)
\cdot \\
\cdot & \exp \!\bigg(\sum_{n=1}^\infty
 (-1)^{n+1}(a^\mu_n)^1(a^\mu_n)^2\Bigg),
\end{aligned} \\ &
\brai{12}{V_2^g}=\brai{12}{\Om}
 (c_0^1+c_0^2)\exp \!\bigg(\sum_{n=1}^\infty (-1)^n \Big
  (b^{1}_{n} c^{2}_{n}+b^{2}_{n} c^{1}_{n}
 \Big)\bigg).
\end{align}
\end{subequations}
The superscripts $m$ and $g$ in these expressions stand for ``matter'' and
``ghost''. The superscripts 1 and 2 over the oscillators represent the two
spaces.

In order to evaluate the kinetic term we also need to write down the
oscillator form of $Q$,
\begin{equation}
\label{Qmode}
Q=\sum_{n\in \Z} c_n (L_{-n}^m-\delta_{n,0})
 +\sum_{m,n\in \Z}\frac{m-n}{2}:c_m c_n b_{-m-n}:\,,
\end{equation}
where we indicated that the second term is normal ordered.
The matter Virasoro operators $L_n^m$ appearing in~(\ref{Qmode}) are
obtained from expending the energy momentum tensor $T^m$~(\ref{Tm}).
Explicitly, the relevant ones at $l_0=1$ are,
\begin{subequations}
\label{MatterVirasoro}
\begin{align}
L_0=& \, \sum_{n=1}^\infty n a_n^\dag a_n+\al'(p^2+ip V)\,,\\
L_{-1}=& \, \sqrt{2\al'}p\, a_1^\dag
  +\sum_{n=1}^\infty \sqrt{n(n+1)} a_{n+1}^\dag a_n\,,\\
L_1=& \, \sqrt{2\al'}(p+i V) a_1
  +\sum_{n=1}^\infty \sqrt{n(n+1)} a_n^\dag a_{n+1} \,.
\end{align}
\end{subequations}

For the evaluation of the cubic term we need to know the three-vertex,
which is unfortunately more complicated than the two-vertex.
\begin{subequations}
\label{V3}
\begin{align}
\label{V3m}
& \begin{aligned}
& \brai{123}{V_3^m} =\N \int d^d p_1\,d^d p_2\,d^d p_3\,
\brai{1}{p_1}\brai{2}{p_2}\brai{3}{p_3}
\delta^d(p_1+p_2+p_3+iV)\cdot \\
& \quad \cdot \exp \!\bigg(-\sum_{r,s=1}^3\!\bigg(\sum_{n,m=1}^\infty
 \frac{1}{2} a_n^r V_{nm}^{rs} a_m^s
  +\sum_{n=1}^\infty
 a_n^r V_{n0}^{rs}p_s +\frac{1}{2}
p_r V_{00}^{rs}p_s
 \Bigg)\Bigg),
\end{aligned} \\
& \brai{123}{V_3^g}=\brai{123}{\Om}
 c_0^3 c_0^2 c_0^1 \exp \!\bigg(\sum_{r,s=1}^3\sum_{m=0}^\infty \sum_{n=1}^\infty
    b^r_m X_{mn}^{rs} c^s_n\bigg).
\end{align}
\end{subequations}
Here, we suppressed the index
$\mu$,
on which the oscillators and some of the
coefficients depend, for clarity.
The $V_{nm}^{rs}$ and
$X_{mn}^{rs}$
coefficients are independent of the
linear dilaton. They are found, e.g. in~\cite{Taylor:2003gn},
\begin{align}
\label{Vnm}
V_{nm}^{rs}= & -\frac{1}{\sqrt{n m}}
  \oint \frac{dw}{2\pi i}\oint \frac{dz}{2\pi i}
  \frac{1}{z^m w^n}\frac{f_r'(z)f_s'(w)}{\big(f_r(z)-f_s(w)\big)^2}\,,\\
\label{Xmn}
X_{mn}^{rs}= & \oint \frac{dw}{2\pi i}\oint \frac{dz}{2\pi i}
  \frac{1}{z^{n-1} w^{m+2}}\frac{f'_s(z)^2}{f'_r(w)
\big(f_s(z)-f_r(w)\big)
}
    \prod_{I=1}^3 \frac{f_r(w)-f_I(0)}{f_s(z)-f_I(0)}\,,
\end{align}
where $f_r$ are the conformal transformations defining the
three-vertex~(\ref{UHPtoDisk})-(\ref{fns}).
The normalization factor $\N$ and the momentum dependence can be read by
comparing to the expressions obtained for the tachyon using CFT methods.
The result is
\begin{align}
\N & =K^{3-\al' V^2}\,,\\
V_{00}^{rs} & =2\al' \log K \delta^{rs}\,.
\end{align}

For evaluating $V_{n0}^{rs}$ we again compare the expressions obtained
using oscillator methods and CFT methods. The oscillator representation is,
\begin{equation}
\brai{123}{V_3^m} a^{\dag\,1}_n\ket{p_1}\ket{p_2}\ket{p_3}=
 -\sum_{s=1}^3 V_{n0}^{1s}p_s \big<\mbox{tachyon}\big>\,,
\end{equation}
where the value of the expression without the $a^\dag$ insertion, which
equals the expectation value for three tachyon interaction is written as
$\big<\mbox{tachyon}\big>$.
On the CFT side we obtain,
\begin{align}
\nonumber
&\!\!\!\!\!\!\brai{123}{V_3^m} a^{\dag\,r}_n \ket{p_1}\ket{p_2}\ket{p_3}\\
\nonumber
 &= \oint \frac{dz}{2\pi i}\sqrt{\frac{2}{\al'}}\frac{1}{\sqrt{n}z^n}
 \big< f_1 \circ (i \partial X)(z)
   f_1 \circ e^{i p_1 X}(0) f_2 \circ e^{i p_2 X}(0)
    f_3 \circ e^{i p_3 X}(0)\big>\\
 &= \oint \frac{dz}{2\pi i}\sqrt{\frac{2}{\al'}}\frac{\al'}{\sqrt{n}z^n}
  \Bigg(\sum_{s=1}^3 \frac{f_1'(z) p_s}{f_1(z)-f_s(0)}
   +\frac{i V}{2}\frac{f_1''(z)}{f_1'(z)}\Bigg)
  \big<\mbox{tachyon}\big>\\
\nonumber
 &= \oint \frac{dz}{2\pi i z^n}\sqrt{\frac{2\al'}{n}}
  \sum_{s=1}^3 p_s \Bigg(\frac{f_1'(z)}{f_1(z)-f_s(0)}
   -\frac{f_1''(z)}{2f_1'(z)}\Bigg)
  \big<\mbox{tachyon}\big>.
\end{align}
Here, we used the CFT definition of the expression and~(\ref{idx}) in the
first equality. Then, we used the non-tensor transformation
rule~(\ref{Xtrans}) in the second equality and the anomalous momentum
conservation in the last equality. We can now infer,
\begin{equation}
\label{Vn0}
V_{n0}^{1s}=-\oint \frac{dz}{2\pi i z^n}\sqrt{\frac{2\al'}{n}}
  \Bigg(\frac{f_1'(z)}{f_1(z)-f_s(0)}-\frac{f_1''(z)}{2f_1'(z)}\Bigg)\,.
\end{equation}
Note, that there is no lose of generality from choosing $r=1$, due to
the cyclicity property of the three-vertex,
\begin{equation}
V^{rs}= V^{(r+n)(s+n)}\quad \forall n\,,
\end{equation}
where the indices are added modulo 3.
Also note, that the expression we obtained does not agree with the literature
even in the limit $V\rightarrow 0$ (which is a trivial limit, since the
final expression is $V$-independent). The reason is that without a linear
dilaton $V^{rs}_{n0}$ is only defined up to
$V^{rs}_{n0}\rightarrow V^{rs}_{n0}+K^s_n$, for arbitrary constants $K^s_n$,
due to the non-anomalous momentum conservation. This freedom is used, e.g.
in~\cite{Rastelli:2001jb,Bonora:2003xp} in order to set
$\sum_s V^{rs}_{n0}=0$. If we do not wish to have a term proportional to
$iV a^\dag$ in the definition of the three-vertex, then we have no
redefinition freedom and we are forced to use~(\ref{Vn0}).
One can verify that this result indeed makes sense, by noticing that, unlike
other expressions for the three-vertex, it is SL(2) invariant.
We are almost ready now to address higher levels. The only issue that
should still be clarified is the form of the reality condition,
to which we turn next.

\subsection{The reality condition}
\label{sec:RealityCond}

Let us recall the reality condition of the string field.
This condition states that the string field is
left invariant under the combined action of two involutions, Hermitian
conjugation $\cO\rightarrow \cO^\dag$ and BPZ conjugation
$\cO\rightarrow \cO^\flat$. The former is the more familiar one.
It sends $\ket{0}$ to $\bra{0}$, $\cO_n$ to $\cO_{-n}$, where $\cO$ stands
for either $a$, $b$ or $c$, while reversing the order of operators. It also
induces complex conjugation. BPZ conjugation is performed by the action of
the two-vertex $\bra{V_2}$~(\ref{V2}). It also sends $\ket{0}$ to
$\bra{0}$. However, it does not induce complex conjugation, nor does it
change the order of operators. It also acts differently on the various
operators,
\begin{equation}
a_n^\flat=(-1)^{n+1} a_{-n}\,,\qquad
c_n^\flat=(-1)^{n+1} c_{-n}\,,\qquad
b_n^\flat=(-1)^n b_{-n}\,.
\end{equation}
The different signs originate from the odd conformal dimension of
$\partial X(z)$ and $c(z)$ versus the even one of $b(z)$.
Combining the two involutions leaves us with
$(-1)^{\#a_{even}+\#c_{even}+\#b_{odd}}$ times the original
operators inversely ordered. It is important to note that the coefficient
fields also change their order relative to the other expressions.
This is important when the Grassmann odd coefficient fields of
even-ghost-number string fields are considered. The
coefficient fields are also complex conjugated. We prefer to work
with coefficient fields which are defined to be real.
Thus, matching the signs translates into the
choice of putting an extra $i$ factor in front of some of the coefficients.
Note, that we do not have to separate the $c_1$ factor from the vacuum
$\ket{\Om}$, since it does not contribute a sign and also commutes with the rest
of the operators, which are Grassmann even when coefficient fields are included.

We write the action in terms of momentum modes.
The rules for settling the reality of the component fields, when
applied to the explicit momentum dependence lead to the (almost) standard
reality in momentum space,
\begin{equation}
\hat \phi(p)=\hat \phi(-p-iV)^*\,,
\end{equation}
where $\hat \phi$ is an arbitrary component field and $p$ is the momentum.
To get from this expression a genuine standard reality condition we have
to impose the same transformation that we imposed in~(\ref{fieldReDef}),
\begin{equation}
\label{GenFieldRedef}
\hat \phi(p)=\phi\Big(p+i \frac{V}{2}\Big).
\end{equation}
With this definition the reality condition takes the familiar form,
\begin{equation}
\label{realComp}
\phi(p)=\phi(-p)^* \quad \Longleftrightarrow \quad \phi(x)=\phi(x)^*\,.
\end{equation}
We would like to work from now on only with the real fields. This can be
achieved by redefining $Q$, the matter two-vertex
$\bra{V_2^m}$~(\ref{V2m}) and the matter three-vertex
$\bra{V_3^m}$~(\ref{V3m}) in a way that compensates for the
transformation~(\ref{GenFieldRedef}).
The redefinition of $\ket{V_2}$ is nothing but the replacement,
\begin{equation}
\label{redefDelta}
\delta^d(p+q+iV) \rightarrow \delta^d(p+q)\,,
\end{equation}
in~(\ref{V2m}).
The redefined $Q$ is the same as the old one, only with the matter Virasoro
operators redefined from~(\ref{MatterVirasoro})
to the more symmetric form,
\begin{subequations}
\label{NewMatterVirasoro}
\begin{align}
L_0=& \sum_{n=1}^\infty n a_n^\dag a_n+\al'\Big(p^2+\frac{V^2}{4}\Big),\\
L_{-1}=& \sqrt{2\al'}\Big(p-i \frac{V}{2}\Big)\, a_1^\dag
  +\sum_{n=1}^\infty \sqrt{n(n+1)} a_{n+1}^\dag a_n\,,\\
L_1=& \sqrt{2\al'}\Big(p+i \frac{V}{2}\Big) a_1
  +\sum_{n=1}^\infty \sqrt{n(n+1)} a_n^\dag a_{n+1} \,,
\end{align}
\end{subequations}
and similarly for the other modes. We see that not only the string fields,
but also the Virasoro modes obey now the standard reality condition,
\begin{equation}
L_n^\dag=L_{-n}\,.
\end{equation}
It is straightforward to see that with the new definition one
recovers~(\ref{S2Tau}).

Transforming the cubic interaction according to~(\ref{GenFieldRedef}) is
nothing but the replacement of $p_r$ by $p_r-\frac{i V}{2}$ everywhere in
the three vertex. This results in,
\begin{align}
\label{MomConserv}
\delta^d(p_1+p_2+p_3+iV) & \rightarrow
 \delta^d\Big(p_1+p_2+p_3-\frac{iV}{2}\Big)\,,\\
\label{newN}
\N & \rightarrow K^{3\big(1-\frac{\al' V^2}{4}\big)}\,,\\
\label{V0rule}
V_{n0}^{1s} &\rightarrow V_{n0}^{1s} - \sum_{r=1}^3 V_{n0}^{1r}\,.
\end{align}

\subsection{Truncation of the action to $l_0=1$ in scheme 4}
\label{sec:l01}

We now have all the ingredients needed for defining the $l_0=1$ action,
which we evaluate for a general dimension $d$.
The $l_0=1$ component of the string field can be written in terms
of six real component fields,
\begin{equation}
\begin{aligned}
\Psi_1=\int d^d p \bigg(& A_\mu(p) a_1^{\mu\,\dag}
  +B(p)b_{-1}+i C(p)c_{-1}+\\
     \Big(& \cA_\mu(p) a_1^{\mu\,\dag} + i \cB(p) b_{-1} +\cC(p) c_{-1}\Big) c_0
       \bigg) e^{ip\cdot X}\ket{\Om}.
\end{aligned}
\end{equation}
Of the new six fields, the second line includes the ones which are outside
the Siegel gauge. These fields do not contribute to our schemes 2 and 4.
The only fields with ghost number one are the ``photon'' $A$ and the
auxiliary field $\cB$. These are the fields that contribute to scheme
number 1. Of these, only the photon contributes to scheme 2.
Scheme 4 carries all the fields of the first line, while scheme 3 carries
not only all the new fields, but also the field $\cT$, from $l_0=0$, which
did not contribute to the action previously.
It is important to remember that the fields $B$, $C$, $\cT$ and
$\cA$ are Grassmann odd fields.

We now want to evaluate the kinetic term of the new fields. Since we assume
that the fields in $\Psi_1$ are real, we should work with the modified
$\bra{V_2}$~(\ref{redefDelta}) and $L_n$~(\ref{NewMatterVirasoro}). 
At this level the BRST charge $Q$ is truncated to
\begin{equation}
Q=c_0(L_0^m-1)+c_1 L_{-1}^m+c_{-1} L_1^m-b_{-1}c_0 c_1-c_{-1}c_0 b_1
   +2c_{-1}b_0 c_1\,.
\end{equation}
Assume for now that we work with scheme 4. Then, we have the even fields
$T$ and $A$ and the odd fields $B$ and $C$. Note, that due to our treatment
of the reality condition, $T$ now is what we called $\tau$ in
section~\ref{sec:levelZero}. The form of the BRST charge $Q$ can now
be further simplified by disregarding all terms that do not include $c_0$,
\begin{equation}
Q=c_0(L_0^m-1)-b_{-1}c_0 c_1-c_{-1}c_0 b_1\,.
\end{equation}
Direct evaluation now gives,
\begin{equation}
\label{S2l1}
S_2=-\int d^d x\, \Big(\frac{m_0^2 T^2+(\nabla T)^2}{2}+
   \frac{m_1^2 A^2+(\partial_\nu A_\mu)^2}{2}
    + i (m_1^2 B C+\nabla B\cdot \nabla C)\Big)\,.
\end{equation}
Here, we used the generalization of~(\ref{m0}),
\begin{equation}
\label{ml}
m_{l_0}^2\equiv \frac{V^2}{4}+\frac{l_0-1}{\al'}\,,
\end{equation}
for $l_0=1$. Note, that the kinetic term of the vector takes the
standard form of a vector in the Feynman gauge.

We now turn to evaluating the cubic terms. Ghost number conservation
dictates that the only possible interactions include $T^3$, $AT^2$,
$A^2 T$, $A^3$, $TBC$ and $ABC$. We have to evaluate all these terms.
To that end we need the coefficients $V^{rs}_{10}$,
$V^{rs}_{11}$ and $X^{rs}_{11}$.
Using~(\ref{Vn0}) we obtain,
\begin{equation}
V^{11}_{10}=0\,,\qquad V^{12}_{10}=\sqrt{\frac{8\al'}{27}}\,,
 \qquad V^{13}_{10}=-\sqrt{\frac{8\al'}{27}}\,.
\end{equation}
These values should have been modified according to~(\ref{V0rule}).
However, they do not change, since they sum up to zero. In fact, this
is the case for all odd values of $n$.
For terms at least quadratic in $A$ we also have to use~(\ref{Vnm}) for
evaluating,
\begin{equation}
V_{11}^{12}=V_{11}^{13}=-\frac{16}{27}\,,
\end{equation}
while for the terms involving the ghost fields we need~(\ref{Xmn}),
\begin{equation}
X_{11}^{12}=X_{11}^{21}=-\frac{8}{27}\,.
\end{equation}

The evaluation of the $T^3$ term is straightforward and leads to the
result already obtained~(\ref{S3Tau}),
\begin{equation}
\label{S31}
S_3^1=-\frac{g_o \N}{3}\int d^d x\,\tilde T^3 e^{-\frac{V \cdot x}{2}}\,.
\end{equation}
Here and in what follows we leave the $x$ argument (of $\tilde T(x)^3$)
implicit.
Next, we get the $T^2 A$ term,
\begin{align}
\nonumber
S_3^2 &= -\frac{g_o}{3}3\bra{V_3}\int d^d p_1 d^d p_2 d^d p_3
  \big(a_1^\dag A(p_1)\ket{p_1,\Om}\!\big)\big(T(p_2)\ket{p_2,\Om}\!\big)
    \big(T(p_3)\ket{p_3,\Om}\!\big)\\
  \nonumber
 & =-g_o \N\int d^{3d}p\,\delta\Big(\sum p_i-\frac{iV}{2}\Big)
         \tilde A(p_1)\tilde T(p_2)\tilde T(p_3)
          (-V_{10}^{11}p_1-V_{10}^{12}p_2-V_{10}^{13}p_3)\\
\label{S32}
 & =0\,,
\end{align}
where in the last equality we used the fact that we obtain in the integrand
an expression which is anti-symmetric with respect to $p_2\leftrightarrow p_3$.
The
$T A^2$ term, for which we have to write the space-time indices
explicitly, is
\begin{align}
\nonumber
S_3^3 & = -\frac{g_o}{3}3\bra{V_3}\int d^{3d}
  \big(a_1^{\mu\,\dag} A_\mu(p_1)\ket{p_1,\Om}\!\big)
  \big(a_1^{\nu\,\dag} A_\nu(p_2)\ket{p_2,\Om}\!\big)
    \big(T(p_3)\ket{p_3,\Om}\!\big)\\
  \nonumber
 & =-g_o \N\int d^{3d}p\,\delta\Big(\sum p_i-\frac{iV}{2}\Big)
       \tilde A_\mu(p_1)\tilde A_\nu(p_2)\tilde T(p_3)\Big(
   -V_{11}^{12}\eta^{\mu\nu}+\\
 \nonumber
      & \quad (-V_{10}^{11}p_1-V_{10}^{12}p_2-V_{10}^{13}p_3)^\mu
        (-V_{10}^{11}p_2-V_{10}^{12}p_3-V_{10}^{13}p_1)^\nu\Big)\\
\label{S33}
 & =-g_o \N\int d^{3d}p\,\delta\Big(\sum p_i-\frac{iV}{2}\Big)
        \tilde A_\mu(p_1)\tilde A_\nu(p_2)\tilde T(p_3)
				\cdot\\
     & \quad \cdot  \frac{8}{27}\Big(\al'\big(p_3^\mu p_1^\nu+p_3^\nu p_2^\mu
       -(p_3^\mu p_3^\nu+p_1^\mu p_2^\nu)\big)+2\eta^{\mu\nu}\Big)
        \nonumber
        \\ \nonumber
 & = -\frac{8g_o \N}{27}
 \int d^d x
  \Big(2\tilde A^\mu\tilde A_\mu\tilde T+\al'\big(
   \tilde A_\mu\tilde A_\nu\partial^\mu \partial^\nu \tilde T+
   \partial^\mu \tilde A_\nu\partial^\nu \tilde A_\mu \tilde T
     -2\tilde A_\nu \partial^\nu \tilde A_\mu \partial^\mu \tilde T\big)\!
  \Big)e^{-\frac{V \cdot x}{2}}
.
\end{align}
Then, we evaluate
the $A^3$ term,
\begin{align}
\nonumber
S_3^4 & = -\frac{g_o}{3}\bra{V_3}\int d^{3d}
  \big(a_1^\dag A(p_1)\ket{p_1,\Om}\!\big)
  \big(a_1^\dag A(p_2)\ket{p_2,\Om}\!\big)
  \big(a_1^\dag A(p_3)\ket{p_3,\Om}\!\big)\\
  \nonumber
 & =-\frac{g_o\N}{3}\int d^{3d}p\,\delta\Big(\sum p_i-\frac{iV}{2}\Big)\Big(
   V_{11}^{12}
\sum_{r,s} V_{10}^{rs}
p_s +
\sum_{r,s,t}
(V_{10}^{1r}p_r)(V_{10}^{2s}p_s)(V_{10}^{3t}p_t)
 \nonumber
\Big)\cdot\\
       & \quad \cdot \tilde A(p_1)\tilde A(p_2)\tilde A(p_3)\\
       \nonumber
 & =0\,.
\end{align}
Here,
we should have paid attention to the
Lorentz indices. The result, however, does not change by doing so.
We can now notice that, in the expressions above, all terms with an odd
number of (vector) $A$ fields vanish, as expected.
Similarly, the $ABC$ term vanishes. The calculation is the same
as in~(\ref{S32}), except that $T$ should be replaced by
$B b_{-1}+i C c_{-1}$.
Hence, we are left with the evaluation of the $TBC$ term,
\begin{align}
\nonumber
& S_3^5= -\frac{g_o}{3}3i\bra{V_3}\int d^{3d}
  \big(\tilde T(p_1)\ket{p_1,\Om}\!\big)
  \Big(\big(\tilde B(p_2)b_{-1}\ket{p_2,\Om}\!\big)
  \big(\tilde C(p_3)c_{-1}\ket{p_3,\Om}\!\big)\\
\label{S35}
  &\quad +
  \big(\tilde C(p_2)c_{-1}\ket{p_2,\Om}\!\big)
  \big(\tilde B(p_3)b_{-1}\ket{p_3,\Om}\!\big)\Big)\\
  \nonumber
 & =-\frac{16ig_o\N}{27}\int d^d x
  \tilde B\tilde C \tilde T e^{-\frac{V \cdot x}{2}}
\end{align}
The complete action up to $l_0=1$ (for scheme 4) is the sum
of $S_2$~(\ref{S2l1}) and $S_3$~(\ref{S31}), (\ref{S33}) and~(\ref{S35}).

\subsection{Automatization using conservation laws}
\label{sec:Automat}

So far we considered scheme 4 at $l_0=1$. If we remain at $l_0=1$, but switch to
scheme 3, we already have 8 component fields. This results in over a hundred possible
interaction terms. While many of those trivially vanish in light of, e.g. ghost
number conservation, many others have to be explicitly evaluated. Furthermore,
the number of terms grows fast as we increase the level $l_0$, which is essential
in order to obtain reliable results. Explicit evaluation of all terms would soon
become hopeless. The resolution of this difficulty is to automate the evaluation
of the various coefficients that appear in the action. The quadratic terms are
easily calculated. For the evaluation of the cubic terms, an efficient method
should be used. As in previous works that used level-truncation, we find that
the most efficient method for the evaluation of these terms is using conservation
laws of the cubic vertex~\cite{Rastelli:2000iu}.

Conservation laws are obtained by evaluating the expectation values of currents
in the geometry of the three-vertex. These currents are built from products of
the conformal
fields,
for which we want to derive the conservation
laws,
by conformal tensors of functions. The weight of these conformal tensors is properly
chosen in order to obtain a current, and the functions are constrained in order
to prevent singularities at any point other than the punctures, including infinity.
Closing such a current around the three
punctures leads to a linear combination of modes of the current, while deforming
the current to infinity leads to zero, as long as the functions were properly
constrained. Actually, some more terms can be obtained, both at infinity and
around the punctures, if the current is anomalous, as is often the case
(e.g. Virasoro operators in the case of a non-zero central charge $c$,
ghost current, and $\partial X$ in the case of a linear dilaton system).
However, these terms are also explicitly derived in~\cite{Rastelli:2000iu}.

Here, we need the conservation laws for the $b$ and $c$ ghosts and for the
$\partial X$ (matter) modes. The lowest order conservation laws are,
\begin{align}
\bra{V_3} c_0^{2} &=\bra{V_3}\Big(\frac{4}{3\sqrt{3}}\big(c_1^{1}-c_1^{3}\big)+\ldots\Big),\\
\bra{V_3} c_{-1}^{2} &=\bra{V_3}\Big(\frac{1}{27}\big(8c_1^{1}+8c_1^{3}+11c_1^{2}\big)+\ldots\Big),\\
\bra{V_3} b_{-1}^{2} &=
  \bra{V_3}\Big(\frac{4}{3\sqrt{3}}\big(b_0^{1}-b_0^{3}\big)-\frac{1}{27}\big(8b_1^{1}+8b_1^{3}+11b_1^{2}\big)+\ldots\Big),\\
\bra{V_3} a_{-1}^{2} &=\bra{V_3}\Big(\sqrt{\frac{8\al'}{27}}\big(p_3-p_1\big)+
  \frac{1}{27}\big(16a_1^{1}+16a_1^{3}-5a_1^{2}\big)+\ldots\Big),
\end{align}
where
again,
the superscript refers to the space in which the mode is defined and the
ellipses indicate higher level modes.
Note, that the matter conservation law includes the momentum explicitly.
In principle, the dilaton slope $V$ could also occur. However, we can
always eliminate it in favour of the momenta using the anomalous
momentum conservation~(\ref{MomConserv}). The result then holds in any
dimension. It
might differ
from the familiar flat space expressions by terms
proportional to $p_1+p_2+p_3$.

\subsection{The problem with scheme 3}
\label{sec:scheme3}

Since conservation laws are given in the momentum representation, it is easier
to write down the action in this representation.
For now we consider the one-dimensional case at $l_0=1$ in scheme 3.
Hence, Lorentz indices, when they appear, can obtain only a single value
and are therefore omitted.
The quadratic term is given by
\begin{align}
\nonumber
S_2^{(3)}=-\int d p\, \Big(& \frac{m_0^2 +p^2}{2}T(p)T(-p)+
   \frac{m_1^2 +p^2}{2}\big(A(p)A(-p)+2i B(p) C(-p)\big)\\
\label{Scheme4QuadAction}
& +\frac{\cB(p)\big(\cB(-p)+\sqrt{\frac{\al'}{2}}V A(-p)\big)}{\al'}
-\frac{i V}{\sqrt{2\al'}}B(p)\cA(-p) \Big).
\end{align}
Here, the first line is the expression that we had before and the second line
includes the new fields. It is seen that all these fields are auxiliary
fields, since there are no new kinetic terms. Reality of the action is a
consequence of the fact that products of even fields carry real coefficients,
while products of odd fields carry imaginary coefficients.

For the evaluation of the cubic terms we use the conservation rules,
which reduce the general cubic term to that of the elementary tachyon vertex
\begin{equation}
\bra{V_3}
\ket{\Psi_1}_1 \ket{\Psi_2}_2 \ket{\Psi_3}_3
 \propto \bra{V_3}\ket{\Om,p_1}_1 \ket{\Om,p_2}_2 \ket{\Om,p_3}_3\,.
\end{equation}
The conservation laws give the proportionality coefficients, which can be zero
and are momentum-dependent.
We already evaluated the fundamental (three tachyon) term,
\begin{equation}
\begin{aligned}
\label{S3TTT}
S_3^{(TTT)} &=-\frac{g_o}{3}\bra{V_3}\ket{\Om}_1 \ket{\Om}_2 \ket{\Om}_3
\\ &=-\frac{g_o \N}{3} \int dp_1 dp_2 dp_3\,
  \delta\Big(\sum_{j=1}^3 p_j-\frac{iV}{2}\Big) K^{-\al'\sum_{k=1}^3 p_k^2}\,.
\end{aligned}
\end{equation}
Here, we wrote $\ket{\Om}_k$ instead of $\ket{p_k, \Om}_k$ for short.
Also, recall that the coefficient $\N$ is given by~(\ref{newN}).

Even before the use of the conservation laws there are several terms that can be discarded
due to ghost number. The total ghost number of any three coefficient fields should equal
three. In our treatment, where we build the states over the ghost number one $\ket{\Om}$
vacuum, it means that the total ghost number other than that of the vacua should equal
zero. From the correlation between ghost number and statistics of the component fields
we can also infer that odd fields either do not appear, or appear as a pair, as should
be the case for obtaining an even action. That means that we would be able to continue
integrating those fields out, before commencing the simulations. All in all, there are
only 19 possible terms that
have to be evaluated.

In the evaluation of $S_3$ there are six contributions to a generic coefficient, which
come from the six possible orderings of the three coefficient fields involved. The
properties of the three-vertex implies that these coefficients can only depend on the
cyclic order of the fields. Hence, the term in the action that involves the component
fields $\Psi_1 \Psi_2 \Psi_3$ is given by,
\begin{equation}
\label{oneS3Term}
-g_o \bra{V_3}\Big(\ket{\Psi_1}_1 \ket{\Psi_2}_2 \ket{\Psi_3}_3+
  \ket{\Psi_3}_1 \ket{\Psi_2}_2 \ket{\Psi_1}_3\Big).
\end{equation}
It turns out that in several cases the two orderings produce expressions that
cancel out, after relabeling the three spaces, in particular, due to the momentum
dependence of the result.
Another issue, which we have to notice, is that of symmetry factors, i.e.
if two component fields are the same, e.g. $\Psi_1=\Psi_2$, the result should
be divided by two, while in the case $\Psi_1=\Psi_2=\Psi_3$, the result should
be divided by six. Even better (computationally) is to divide the result by
one and by three respectively, but to evaluate only one of the terms in~(\ref{oneS3Term}),
since in these cases there is no issue of different orderings.

We are now ready to write down the full expression in terms of~(\ref{S3TTT}),
\begin{align}
\no
S_3^{(3)}=& -\frac{g_o \N}{3} \int dp_1\, dp_2\, dp_3\,
  \delta\Big(\sum_{j=1}^3 p_j-\frac{iV}{2}\Big) K^{-\al'\sum_{k=1}^3 p_k^2}\Big(T(p_1)T(p_2)T(p_3)
\\ \no & + \frac{8}{9}A(p_1)A(p_2)T(p_3)\big(2-\al'(p_3-p_2)(p_3-p_1)\big)
	+\frac{16i}{9}T(p_1)B(p_2)C(p_3)
\\ 
\label{Scheme4CubicAction}
 & -\frac{16}{9}T(p_1)\cB(p_2)\cB(p_3)
\\ \no & +\frac{16\sqrt{2\al'}}{9}B(p_1)\big(\cA(p_2)T(p_3)-A(p_2)\cT(p_3)\big)(p_3-p_1)
\\ \no & +\frac{32i}{9}B(p_1)\cB(p_2) \cT(p_3)
\Big).
\end{align}
Here, the first two lines are the expression that we had for scheme 4, the third line
includes a new bosonic interaction and the last two lines include two new interaction
terms involving odd fields.

The path integral~(\ref{pathInt}) now contains also integration over the
various new modes. In particular one expects it to contain integration over
the odd variables included, namely, $B$, $C$, $\cT$ and $\cA$,
\begin{equation}
Z=\int \Big(\prod_j dT_j d\cT_j \Big)
\Big(\prod_n d\cB_n dB_n d\cC_n dC_n dA_n d\cA_n \Big) e^S\,.
\end{equation}
Here and in the rest of the paper $T_j$ (denoted $\tau_n$ above),
$\cT_j$, $A_n$, etc., represent the modes of the various fields.
The fields appear in the measure in pairs of an even and an odd
field, with the even ones written first. We need two different indices for
the products since the number of modes of a given field depends on its $l_0$.
We would also like, if possible, to integrate out the
bosonic auxiliary field $\cB$. Since at higher levels it would be quite impossible
to eliminate all the auxiliary fields, it could be nice to compare the results
with and without the elimination of $\cB$. The auxiliary bosonic field $\cC$ does
not appear in the action at all.

Inspecting the action~(\ref{Scheme4QuadAction}) and~(\ref{Scheme4CubicAction})
we recognize that it suffers from a major problem:
A Grassmann integral can be non-zero only if the integrand has a term, which
is linear with respect to all the Grassmann modes.
However, a term linear in all the odd fields is absent in the path integral.
Since odd terms enter the various terms in the action either quadratically
or not at all, the problem of saturating all the modes is that of the regularity
of the (bosonic-field-dependent) matrix of coefficients of the terms
quadratic with respect to the odd variables in the action.
The problem then is that this matrix turns out to be singular.

This problem occurs since level truncation does not commute with
Grassmann integration. Actually, we faced this problem already at level zero,
where we noticed that the field $\cT$, which is present in~(\ref{Psi0})
is absent from the action altogether.
There, we decided to ignore this field temporarily and it indeed enters
the action now. However, it is not clear which
fields
should we retain now
and which ones should be postponed to the next level. Inspecting the action
we see that the field $B$ is present in all the relevant expressions and is
saturated in each term by one of the other fermionic fields, namely $\cT$, $C$
and $\cA$. This is not particularly surprising, due to the ghost number of the
states that these component fields multiply. However, it is not clear
which modes should we keep now. The most ``natural'' choice would be to keep
$\cT$, since it already ``enters too late to the game''. However, one could
object to the idea of adding high $\cT$ modes before adding the first $C$ modes,
since it would make our cutoff $l_0$-dependent instead of $l$-dependent.
Furthermore, since the modes of $\cT$ and $B$ enter the level truncation
at different cut-off values, we would generally have a different number of
such modes and it would be impossible to saturate the Grassmann integral.

One could worry that such problems could occur also for scheme 4, which
also includes odd modes. This is not the case.
The source of the problem we
face
here is the fact that the fields
$\cT$ and $\cA$ are auxiliary and hence do not have kinetic terms.
The kinetic terms provide regular parts for the matrix. Hence,
the integral over the odd fields is regular for scheme 4, except perhaps
for some specific values for the bosonic fields.

An additional potential difficulty with scheme 3, is that it
is likely to inherit from scheme 1 the problem, to be described in~\ref{scheme1sec},
of a nearly-massless mode leading to large instabilities.
In light of all that we do not dwell further into scheme 3.

\subsection{The action in scheme 1}
\label{sec:scheme1}

We also would like to check scheme 1, in which we only keep the fields $T$, $A$ and $\cB$.
The action is just the truncation of the scheme-3 action to include only these fields.
The quadratic part of the action is
\begin{equation}
\begin{aligned}
\label{Scheme2QuadAction}
S_2^{(1)}=-\int d p\, \Big(& \frac{m_0^2 +p^2}{2}T(p)T(-p)+
   \frac{m_1^2 +p^2}{2}A(p)A(-p)
 \\& +\frac{\cB(p)\big(\cB(-p)+\sqrt{\frac{\al'}{2}}V A(-p)\big)}{\al'}\Big),
\end{aligned}
\end{equation}
and the cubic part is
\begin{align}
\no
S_3^{(1)}=& -\frac{g_o \N}{3} \int dp_1\, dp_2\, dp_3\,
  \delta\Big(\sum_{j=1}^3 p_j-\frac{iV}{2}\Big) K^{-\al'\sum_{k=1}^3 p_k^2}\Big(T(p_1)T(p_2)T(p_3)
\\ \no & + \frac{8}{9}A(p_1)A(p_2)T(p_3)\big(2-\al'(p_3-p_2)(p_3-p_1)\big)
	\\
\label{Scheme2CubicAction}
		& -\frac{16}{9}T(p_1)\cB(p_2)\cB(p_3)
\Big).
\end{align}
The explicit integration of the $\cB$ field should be much easier in this
scheme as compared to schemes 3 and 4.

\subsection{Analytical study of the lowest mode}
\label{sec:lowest_mode}

Before we attempt a numerical study of the case with many modes, we would
like to examine analytically the simplest possibility of retaining a single
mode. Hopefully, we could get some feeling about what should be expected
from this simple example. The lowest lying mode would be the first mode
of the tachyon field. Its level depends on the length $L$ of the range which we consider
for $X$
and it
is
given by
\begin{equation}
T(x)=\sqrt{\frac{2}{L}} \sin\Big(\frac{\pi n (x-x_{min})}{L}\Big)T\,,
\end{equation}
where $T$ is the only variable in the theory.
The action is
\begin{equation}
S=-\frac{1}{2} \Big(\frac{1}{24\al'}
  +\big(\frac{\pi}{L}\big)^2\Big) T^2-\frac{\tilde g_o f}{3} K^{-\al'\big(\frac{\pi}{L}\big)^2} T^3\,,
\end{equation}
where we absorbed some constants into the coupling constant and the single
coupling constant
of the theory
is found to be
\begin{equation}
f=\Big(\frac{2}{L}\Big)^{\frac{3}{2}}\int_{-\frac{L}{2}}^{\frac{L}{2}} dx
  e^{-\frac{V x}{2}}\sin^3\Big(\frac{\pi (x-\frac{L}{2})}{L}\Big)=\Big(\frac{2}{L}\Big)^{\frac{3}{2}}
	\frac{192 \pi^3 L \cosh \left(\frac{L V}{4}\right)}{L^4 V^4+40 \pi^2
   L^2 V^2+144 \pi^4}.
\end{equation}
Here for simplicity we take the range of integration to be symmetric with respect to the origin.
It is easy to see that, as one should expect, $f$ approaches infinity as $L\rightarrow \infty$.

Performing the advocated analytical continuation $T \to e^{i \pi/6} T$
(see section~\ref{sec:setup} for details) the action becomes
\begin{equation}
S=-\frac{e^{\frac{i\pi}{3}}}{2} \Big(\frac{1}{24\al'}
  +\big(\frac{\pi}{L}\big)^2\Big) T^2-\frac{i\tilde g_o f}{3} K^{-\al'\big(\frac{\pi}{L}\big)^2} T^3\,.
\end{equation}
The simplicity of this expression makes it possible to evaluate the partition function
analytically.
Write,
\begin{equation}
\label{SingleModeAction}
S=-a(L) T^2-i b(L,V) T^3\,.
\end{equation}
Then, the partition function is given by,
\begin{equation}
\label{Z single}
\begin{aligned}
Z=&\int_{-\infty}^\infty dT e^S=\int_{-\infty}^\infty dT e^{-a T^2-i b T^3}=
b^{-\frac{1}{3}}\int_{-\infty}^\infty dT e^{-a b^{-\frac{2}{3}} T^2-i T^3}\\=&
\frac{2 \pi  e^{\frac{2 a^3}{27 b^2}}}{(3b)^{\frac{1}{3}}}
  \text{Ai}\Big(\frac{a^2}{(3 b)^{4/3}}\Big)\,,
\end{aligned}
\end{equation}
where Ai is the Airy function. We know that the integral converges, since our $a$ has
a positive real part. However, the result is not real, as expected. Nonetheless,
when we take the limit $L\rightarrow \infty$, the partition function approaches a
real value. In this limit (setting $\al'=1$) we have (regardless of the value of $V$),
\begin{equation}
a\rightarrow \frac{e^{\frac{i\pi}{3}}}{48}\,,\qquad b\rightarrow \infty\,,
\end{equation}
where the approach of $b$ to infinity is along the positive real line.
The factor in front of the Airy function is real and approaches zero as $L\rightarrow \infty$.
The Airy function, on the other hand, is complex. However, it has a real limit,
\begin{equation}
\text{Ai}\Big(\frac{a^2}{(3 b)^{4/3}}\Big)\rightarrow \frac{1}{3^{\frac{2}{3}}\Gamma\big(\frac{2}{3}\big)}\,.
\end{equation}

In principle, we were not supposed to expect a real limit here, since we are truncating
to the lowest single mode. Nonetheless, it is encouraging to see that the wild oscillations
conspire to produce a real value already at this stage. Also, we see that reality is really
obtained only as we take the limit $L\rightarrow \infty$. Thus, comparing finite values
does not necessarily make sense.

Using~(\ref{Z single}), we can study the dependence of various expectation
values as a function of $a$ and $b$. One can obtain different values for these
coefficients in many ways, by using symmetric as well as non-symmetric limits
for $x_{min}$ and $x_{max}$.
The limit $b\rightarrow 0$ gives a free theory, in which
$\langle S \rangle =-\frac{1}{2}$, while
$\langle T \rangle $ approaches zero from the direction of $e^{\frac{7 i \pi}{6}}$ and
$\langle T^2 \rangle$ approaches zero from the direction $e^{-\frac{i \pi}{3}}$.
Conversely, in the mentioned above limit $b \rightarrow \infty$,
we find that $\langle S \rangle = -\frac{1}{3}$,
while $\langle T \rangle$ approaches zero from the direction $e^{-\frac{i \pi}{6}}$ and
$\langle T^2 \rangle$ approaches zero from the direction
$e^{\frac{i \pi}{3}}$.
We will compare these results to the lattice simulation of section~\ref{sec:results}\footnote{Note,
that in section~\ref{sec:results} we use a slightly different convention, in which the factor
of $e^\frac{i \pi}{6}$ explicitly multiplies the fields. This leads to different constant phases
as compared to what we did here.}.

Another important remark regarding the fact that the normalization factor approaches
zero: on the one hand, the normalization factor should be renormalized as we change our
parameters. Thus, from this perspective, there is nothing here to discuss.
On the other hand, keeping only the lowest level amounts to truncating more and more
modes as $L$ approaches infinity. This is not a natural limit and we took it only
for the purpose of verifying that we can reproduce on the lattice the analytical
expression that we obtain using it. The natural limit that we would have to consider
is taking $L$ to infinity while keeping $l$ fixed. This leads to more and more
modes, up to infinity at the limit. This is the physical limit. However, the increase
in the parameters, as well as the introduction of an ever growing number of modes,
imply that a non-trivial renormalization would be needed.

\subsection{Adding trivial terms to the action}
\label{sec:trivial_terms}

Another problem that a lattice simulation in a linear dilaton background faces
comes from the possible presence in the action of trivial terms.
By that we mean the presence in the definition of the cubic interaction of
terms that vanish due to the anomalous momentum conservation.
Such terms can be added to the definition of the vertex also in the
standard case of a constant dilaton. In any case they take the form
of conformal fields inserted at the three interaction points times
a momentum dependent function of the form
\begin{equation}
F(p_1,p_2,p_3)=p_1^{n_1} p_2^{n_2} p_3^{n_3}\Big(p_1+p_2+p_3 - \frac{iV}{2}\Big)
   \,\qquad n_{1,2,3}\geq 0\,.
\end{equation}
The expression in the brackets is identical to the argument of the
momentum conservation delta function and thus leads to zero contribution
of these terms, which can, therefore, be added to the definition of the interaction
at will. In previous works use was made of such terms in order to simplify
the form of the interaction in various contexts, e.g. in~\cite{Bonora:2003xp}.

While the ambiguity in these terms is usually harmless and could even be useful,
in our case new complications emerge. The momentum conservation is broken
by the introduction of the lattice. Thus, while the introduction of these terms
does not change the action before the introduction of the lattice, it does influence
the results when a lattice is used.
A simple idea for a resolution would be to avoid these terms altogether.
However, it is not clear how to distinguish the ``genuine'' interaction from
the trivial terms. The definition of the action is really ambiguous. This
is somewhat similar to the case of a gauge symmetry:
there
is no canonical
way to gauge fix. The analogue of gauge fixing in our case is the decision of which
is the correct form of the action. However, on the lattice different ``gauge fixings''
lead to different results. One would like to be able to show that as the lattice
cutoff is removed, the results tend to the same values regardless of the ``gauge choice''.
Unfortunately, this seems to be quite unlikely, since the coefficients of the
trivial terms can be arbitrary and more and more such terms pop up as the level is
being increased. One could try to fix the ambiguity by demanding that the form
of the interaction be ``as simple as possible''. While this statement makes sense
at low levels, it becomes ambiguous at higher levels.
Another possibility for a ``gauge fixing'' is to avoid the appearance of $V$ in
the action other than in the exponent. While this option does not necessarily
lead to the simplest possible expressions, as we have already seen in our $l_0=1$
example above, the expressions are unambiguous and are formally independent of
the dimension $d$. Unfortunately, it is not clear that the expressions obtained
in this way are more correct than those of any other ``gauge choice''.

It is important to stress that the problem could have been avoided had we been working
in a constant dilaton background. In such a case we would have chosen to work with
periodic boundary conditions that do not make sense in the case at hand. Then, momentum
conservation would not have been broken by the lattice.
Moreover, the presence of the linear dilaton leads to yet another problem due to the
anomalous form of the conservation law. The issue is not so much the fact that the sum
of momenta is non-zero, as with the fact that it is imaginary. This is not a problem
before the introduction of the lattice, since it only results in the exponential
term in coordinate space. However, with the introduction of the lattice actual
imaginary terms pop-up. These terms produce further problems:
as
we mentioned above,
the action being cubic is not bounded from below, a problem that we resolve
using a change of the contour of integration followed by an analytical continuation.
This procedure turns the real cubic terms to purely imaginary terms,
which results in convergence of the expressions. However, the imaginary terms become
real now, which brings us back to the starting point, in which no numerical analysis
is possible. One could hope that the ambiguity in the form of the interaction term can
be used in order to set to zero the imaginary part of the interaction.
We examine the consequences of adding trivial parts to the action
in section~\ref{sec:Adding_triv_terms}.

\section{Lattice setup}
\label{sec:setup}

We now want to use lattice simulations to calculate observables.
The degrees of freedom are the fields found above using level truncation,
up to some maximum total level $l_{max}$, not necessarily an integer.
Explicitly,~(\ref{l0lp}) can be written as
\begin{equation}
\label{l_of_l0_p}
l=l_0+\al' p^2
\,.
\end{equation}
For our sine-expansion,
$p=\frac{\pi n}{L}$, and since we have only evaluated the level truncation
up to $l_0=1$ we must choose $l < 2$. The number of modes for the $l_0=0$ fields is then
\begin{equation}
n_0=\lfloor \frac{L}{\pi}\sqrt{\frac{l_{max}}{\al'}} \rfloor
\end{equation}
and if $l_{max}>1$ the number of modes for the $l_0=1$ fields is 
\begin{equation}
n_1=\lfloor \frac{L}{\pi}\sqrt{\frac{l_{max}-1}{\al'}} \rfloor.
\end{equation}

Thus given a lattice size $L$ and a choice of `scheme', our degrees of
freedom will be a finite number of modes of one or more fields.
We can read off the action from the appropriate expressions above, e.g.
for scheme 4 and $l<1$ we would need the terms~(\ref{level 0 S2})
and~(\ref{level 0 S3}).
We remind
that the weight of a
configuration in the path integral is $e^S$ rather than $e^{-S}$ due
to the way we Wick-rotated.
In addition to the various `schemes' described above, we have also carried out additional
runs where we have removed the level-1 fields from the action.
This is to try to assess whether the higher level fields are helping to tame the instabilities.

Looking at the action we see an immediate problem:
The action has a
cubic instability. To proceed, we consider the integral over each mode
as a complex integral, and deform the
integration contour to be a straight line at an angle $\gamma$ to the
real axis.
If we choose $\gamma=\pi/6$, the cubic part of the action becomes pure
imaginary and so the action is no longer unstable.
In principle we could have chosen different phases, i.e. $\gamma=\pm \pi/6$
for different components of the string field. However, we refrain from
doing so in order to treat the string field as a uniform physical entity.
This is in accord with our strategy of using a single expression for
the level~(\ref{l_of_l0_p}), instead of considering separately $l_0$ and
the momentum.

However, taking the modes to be complex introduces another problem;
the action also becomes complex and so cannot be interpreted as a
weight for a Markov chain. Instead we simulate in the phase-quenched
ensemble and reweight. That is, we split $e^{S}$ into an amplitude and
a phase:
\begin{equation}
e^{S}=|e^{S}| e^{i\theta},
\end{equation}
and calculate the expectation value of an observable $\mathcal{O}$
using the identity
\begin{eqnarray}
\langle \mathcal{O} \rangle & = & \frac{\int \mathcal{O}|e^{S}|e^{i \theta}}{\int
  |e^{S}|e^{i \theta}} \\
& = & \frac{\langle \mathcal{O}e^{i \theta} \rangle_\mathrm{PQ}}{\langle e^{i
    \theta} \rangle_\mathrm{PQ}},
\end{eqnarray}
where the label $\mathrm{PQ}$ means the expectation value is evaluated in
the phase-quenched ensemble, i.e. with the weight $|e^{S}|$. This
is a real, positive weight, so
it can be used in a Monte Carlo simulation.

We generate configurations in the phase-quenched ensemble using a Metropolis algorithm,
chosen since it is simple to implement and to alter
for the variety of different field contents and actions we are concerned with.
In the cases where we have Grassmann-odd fields we include their contribution
by calculating the fermion determinant directly. This would be expensive for a
large number of modes (the cost scales as $n_1^3$) but is reasonable for the small
number of modes in our simulations (we have at most $n_1=9$). In any case since
the action is non-local the cost of evaluating it scales as $n_{0,1}^3$ even for the bosonic part.

Due to the phase-quenching, our errors increase as the imaginary part of the action increases,
i.e. as we move to larger $x$. To some extent it is possible to compensate for this by
increasing the number of configurations in our simulations, but the number of
configurations required increases exponentially with $x$ so eventually this becomes impossible.
The practical effect of this is that it gives an upper limit on the values of $x$ at which we can simulate;
it will be very difficult to go much beyond this in future work.

There is no general
method
known to avoid the exponentially large cost associated with complex actions.
In some specific cases the complex Langevin method (see~\cite{Aarts:2013uxa} for a recent review),
which does not have an exponential cost, can be used to bypass this `sign problem'.
The complex Langevin method is not a panacea, however; in some cases it converges to the
wrong limit~\cite{Aarts:2011ax}. We attempted to bypass the sign problem by implementing
the complex Langevin method for our system. We found results in agreement with the
conventional Monte Carlo simulations at weak coupling, but disagreement at strong coupling,
indicating that the complex Langevin method was converging to the wrong limit.
Thus we did not pursue this method further. 

As discussed in section~\ref{sec:move non-locality}, the action
can be reformulated so that the quadratic terms are non-local
while the cubic terms are local. The two formulations are equivalent
and therefore
should give identical results. Confirming that this is the
case is a useful additional check of the correctness of our code.
We have carried out this check for several sets of parameters and
indeed found good agreement. The run time and statistical errors
are similar for both formulations, so there is no particular benefit
from using either case. We have chosen to use the formulation with the
non-locality in the cubic term, and all our results below are for that case.

\subsection{Observables}

The observables we measure are the action $\langle S \rangle$, and the
expectation values of the
Grassmann even
fields and their
squares.
The specific field content is dictated by the choice of scheme and level.
For example, for
scheme 4,
we measure $\langle T_n \rangle$
and $\langle T_n^2 \rangle$ for all $l_{max}$, and also $\langle A_n \rangle$
and $\langle A_n^2 \rangle$ if $l_{max} > 1$. Here the subscript $n$
refers to the mode number, and we measure all
modes present.
We find that the $\langle A_n \rangle$ are always consistent with zero,
in some cases with very small errors, of order $10^{-4}$. This is
because $A$ only appears quadratically in the action --- there are no
terms linear or cubic in $A$. Hence we will not discuss $\langle A_n \rangle$ further.

One issue to be considered is whether or not to include the logarithm of the
fermion determinant in the action when Grassmann-odd fields are
present. (Here we refer to the action considered
as an observable, not to the action used for the update algorithm,
where of course the fermion determinant must be included.)
The statistical weight used when Grassmann-odd fields are present is
\begin{equation}
\mathrm{det} M e^{S_B},
\end{equation}
where $\mathrm{det} M$ is the fermion determinant and $S_B$ is the
bosonic part of the action. This can be rewritten as
\begin{equation}
e^{\mathrm{ln}\ \mathrm{det} M + S_B}.
\end{equation}
The question is whether to take $S_B$ or $(\mathrm{ln}\ \mathrm{det} M
 + S_B)$ as the action. Neither of these is obviously more
 physical than the other, but in the weak-coupling limit,
 $S_B$ will simply be $-\frac{1}{2}$ per bosonic degree of freedom,
 whereas $(\mathrm{ln}\ \mathrm{det} M
 + S_B)$ will contain additional $L$-dependent terms coming from $\mathrm{det} M$.
Because of this we have chosen to use $S_B$ as the action observable.

\subsection{Independence of the analytical continuation on the rotation angle}

As described above, we define the theory by an analytical continuation
of the integration contour, which is implemented by a rigid rotation
in the complex plane. So far we considered this rotation to be by
an angle of $\frac{\pi}{6}$, which is exactly what is needed in order
to make the cubic part of the action purely imaginary.
We define $\gamma$ as the angle of rotation, i.e. $\gamma=0$
is the original theory and $\gamma=\frac{\pi}{6}$ is the angle
that is needed for our analytical continuation.

Taking $\gamma=\frac{\pi}{6}$ has a large numerical cost, since then
the action has a large imaginary part. Because of this, we use
$\gamma=0$ when this is possible, i.e. when the cubic term is
small. In some cases we found that it is
possible to use intermediate values of
$\gamma$ when the cubic term is not too large; this is worth it
because even a small decrease in $\gamma$ from $\frac{\pi}{6}$ gives a
large saving in
computational cost.

When the instability is small it is possible to compare results for
different values of $\gamma$ in the range $0\leq \gamma \leq \frac{\pi}{6}$
in order to establish $\gamma$-independence.
We have done this for several sets of parameters and obtained good agreement. For example, at $\alpr=1$,
$V=-\sqrt\frac{25}{6\alpr}$, $L=20$, $x_{min}=-20$, and
$l_{max}=1.6$, we obtained
\begin{subequations}
\begin{align}
& \langle T_1 \rangle = -0.379(27)-0.140(18)i
   & \gamma=\frac{7\pi}{48}\,, \\
& \langle T_1 \rangle = -0.376(3)-0.146(4)i
   & \gamma=\frac{15\pi}{96}\,, \\
& \langle T_1 \rangle = -0.357(22)-0.164(19)i
   & \gamma=\frac{\pi}{6}\,.
\end{align}
\end{subequations}
These results are clearly in good agreement.
In this case, we found out that the metastability at $\gamma=\frac{7\pi}{48}$
is manifested only around $1.7\times10^7$ updates (the result above was obtained from $10^7$ updates),
while we did not observe the metastability at $\gamma=\frac{15\pi}{96}$.
Our results suggest that as long as the metastability does not manifest itself
the results are almost $\gamma$-independent. 

\subsection{$\alpha'$-independence}

$\alpr$, or equivalently
$l_s$~(\ref{ls}), or $m_s$~(\ref{ms}),
sets the scale for our
simulations --- for example the physical box size is $l_s L$. A
useful check on the code is that it gives the same results for
different $\alpr$ when all physical quantities (box size, $l_{max}$,
$g_o$,~\ldots) are the same\footnote{It is important to remember
that, as described in~\ref{sec:SFTintro},
$g_o$ has dimension
$\frac{5}{2}$.}.
We have carried out this check explicitly for the case $\alpr=1$, $L=20$,
$x_{min}=-20$, $l_{max}=0.9$ and rescaled versions thereof, indeed obtaining identical results.
Apart from this test, all the simulations have been
carried out
with $\alpr=1$;
thus the lattice units are equivalent to string units.

\subsection{Estimate of statistical errors}

It is important to have reliable estimates of the statistical errors
on our results, which we estimate using the jackknife method. This
should provide accurate estimates, provided that there are no large
auto-correlation
times in the data, i.e. provided that configurations
with large separation of lattice times are uncorrelated.

In a typical simulation we evaluate about $10^9$ updates, which we split
into about $100$ bins for analysis. The jackknife analysis is supposed to
work well provided these bins are uncorrelated. Thus, the
question is whether configurations $10^7$ updates apart are
correlated. Since we have only about 20 degrees of freedom or less, it
would be very surprising if this would have been the case.

As an additional check we have analysed how accurate the error
estimates are. First, we generated high statistics data ($2\times
10^{10}$ configurations) for a particular set of parameters
($\alpr=1$, $V=-\sqrt\frac{25}{6\alpr}$, $L=20$, $x_{min}=-16$,
$l_{max}=0.9$). This gave an accurate measurement of $\langle T_1
\rangle$ with very small errors:
$\langle T_1 \rangle = -0.0749(8)-1.0317(8)i$.
We then carried out ten independent low statistics runs ($2\times
10^{6}$ configurations each) with the same parameters. Each of these
gives an independent estimate of $\langle T_1
\rangle$ with errors, and if the error estimate is correct these
should all be consistent with the high-statistics result. For example,
the first low-statistics run gave
$\langle T_1 \rangle = -0.123(92)-1.043(112)i$,
which is indeed consistent.

Including both real and imaginary parts, this procedure gives 20
estimates with errors. 11 of these are within $1 \sigma$, 19 are within
$2 \sigma$, and all are within $3 \sigma$ of the high-statistics result.
This is completely consistent with the errors being estimated
correctly, and indeed shows that any bias in the errors must be quite small.

\section{Results}
\label{sec:results}

We focus our simulations on the issue of whether the theory becomes
stable (or at least less unstable) in the limit where $l_{max}$ and/or
$L$ go to infinity.
Recall, that we interpret stability as the vanishing of the imaginary parts
of the various expectation values. We have no reason to expect that those
will vanish already at the low level we work with here, but we would like to
observe a tendency of decreasing the imaginary parts as compared to the real
parts at least of some of the observables.
In principle we would also like to take $x_{max}$
to infinity, but as explained above this will not be possible and we
will have to be content with taking it as far into the strong coupling
region as we can. We expect that a sign of stability will be that the
imaginary parts of observables go to zero, or at least decrease.
Of course, it may be that this will work better for some observables
than others. In particular, we might expect that it will work best for
the lowest modes, which should be less affected by the missing higher
level modes and fields.

The results below are mainly for scheme 4, where we have the most
detailed results, organised roughly into sections dealing with the
effect of varying
a single parameter
(e.g. $L$, $V$, $\ldots$) at a time.
This is followed
by briefer overviews of our results for the other schemes.

\subsection{A single mode}
\label{singlemodesec}

A simple test of our code is to look at the case of a single mode of
the field $T$,
where we can compare to
the analytical results of
section~\ref{sec:lowest_mode}.

We can choose several sets of parameters that will give a single mode,
for example $L=20$, $l_{max}=0.05$, or $L=6$, $l_{max}=0.9$, both with
$\alpr=1$ and $V=-\sqrt\frac{25}{6\alpr}$. These should give identical
results, apart from an overall scale and shift in $x$ due to the
different values of $a$ and $b$. We have found that this is indeed the
case.

Another check is whether we obtain the correct limiting values. The
case $\langle S \rangle =
-\frac{1}{2}$ for $x_{min} \rightarrow -\infty$ is easy to check: for
example, for $L=6$ and $l_{max}=0.9$ we obtain
\begin{eqnarray}
& S=-0.534(2) - 0.0068(19)i\ ; & x_{min}=-6 \\
& S=-0.509(5) + 0.0001(2)i\ \ ; & x_{min}=-6.5,
\end{eqnarray}
which is already very close.

The case $x_{min} \rightarrow \infty$ is harder since the simulations
become expensive in this limit. However, it is still doable, and for
the same parameters we get
\begin{eqnarray}
& S=-0.334(2) - 0.005(2)i\ ; & x_{min}=-2 \\
& S=-0.333(2) + 0.001(2)i\ ; & x_{min}=-1.
\end{eqnarray}
This is clearly going to the correct limit of $-\frac{1}{3}$.

\begin{figure}
  \centering
  \epsfig{file=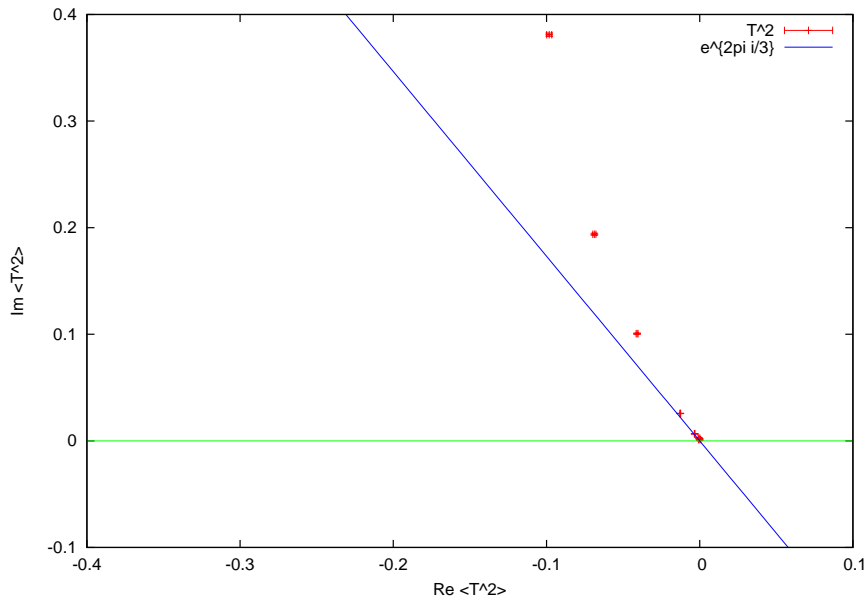,scale=0.45,angle=-90,clip}  
  \caption{$\langle T^2 \rangle$ in the complex plane for $L=6$, $l_{max}=0.9$
	         for $x_{min}$ in the range $-3$ to 2 (red), and the expected asymptotic
					 line (blue).}
 \label{tsquaredanglefig}
\end{figure}

Finally, we have looked at whether $T^2$ approaches the origin from the
correct direction, i.e. at an angle of $\frac{2 \pi}{3}$ to the real
axis. This is shown in Fig.~\ref{tsquaredanglefig}, again for the case
$L=6$ and $l_{max}=0.9$. We see that indeed the results appear to be going towards the right asymptotic line.
All other observables approach zero as well,
except for the action,
which as described above goes to $-\frac{1}{3}$. (The reason the action
is different is the factor of $b$ in eq.~(\ref{SingleModeAction}),
which diverges in this limit.)

\subsection{Changing $l_{max}$}
\label{l_maxsec}
$l_{max}$ is the highest level
allowed for fields
in the simulation.
Since we only include fields up to $l_0=1$, it must be less than 2.
Increasing $l_{max}$ means both
allowing more fields (e.g. the $A$ field only appears for $l_{max} >
1$) and increasing the number of momentum modes of each field.

Since increasing $l_{max}$ is like increasing the cutoff, we might
hope that for high enough $l_{max}$ the results will become
independent of $l_{max}$ (at least for some quantities). With this in
mind we have done some scans in $l_{max}$, keeping all other parameters fixed.

Generally we have concentrated on the observables $\langle S \rangle$
and $\langle T_1 \rangle$ to simplify the presentation. However results
for the other observables are similar. Throughout this section we
fix $\alpr=1$ and $V=-\sqrt\frac{25}{6\alpr}$.

\subsubsection{Extensive study at $L=20$}
\label{sec:l=20}

We begin by describing our results for $L=20$, where we have the most extensive results.
This value of $L$ is quite large, so there
is a reasonable number of modes --- 9 for $T$ and 6 for $A$
up to $l=2$.
We have done runs for every $l_{max}$ between 0 and 2 which gives a
different number of modes. We have also added runs where the level-1
fields $A$, $B$ and $C$ are not included in the simulation, to give
some idea of the effect the level-1 fields have on the physics.

\begin{figure}
  \centering
\hspace{-9 ex}
        \begin{subfigure}{0.45\textwidth}
                \centering
                \epsfig{file=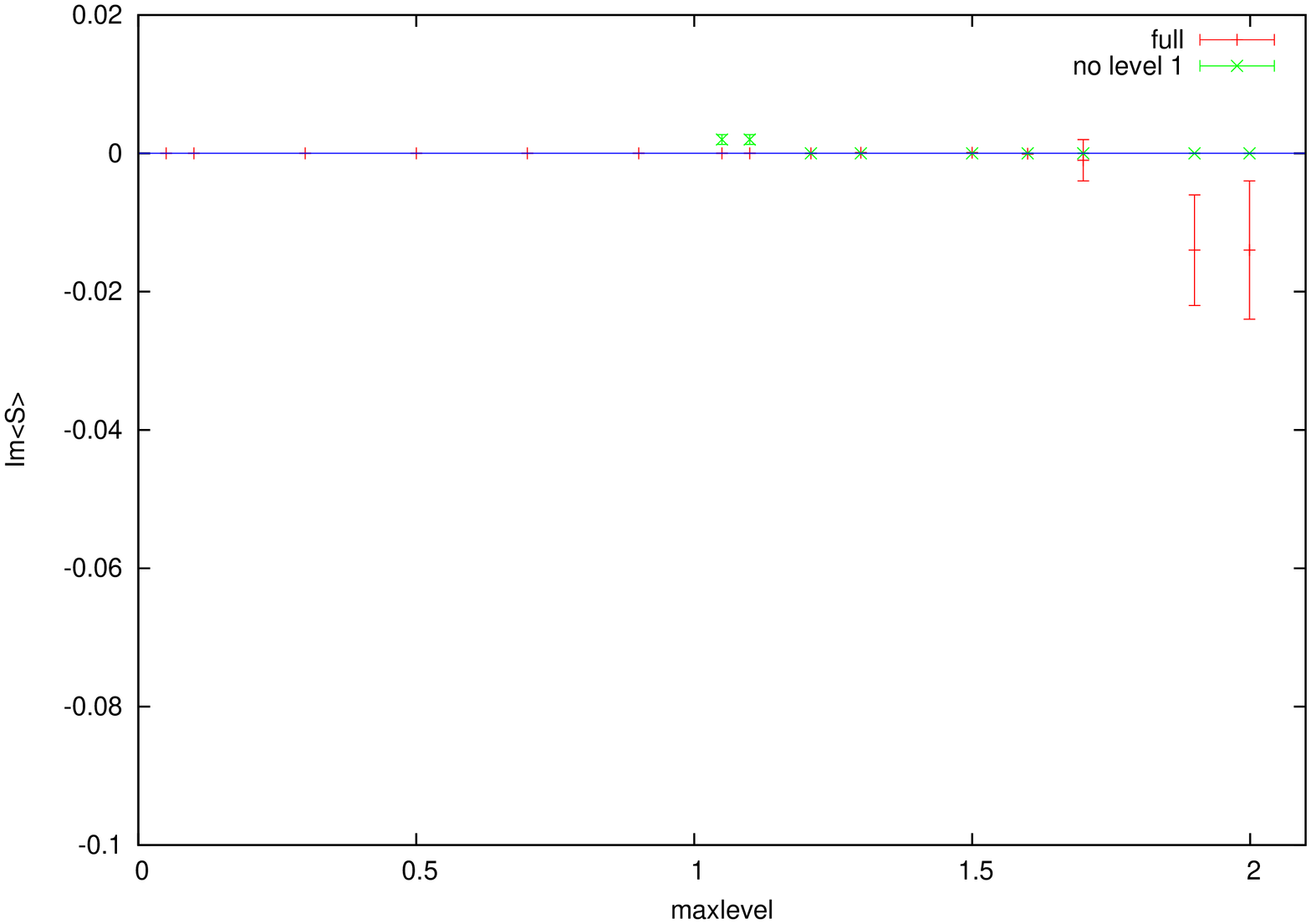,
								scale=0.3,
								angle=-90,clip}  
                \caption{$x_{min}=-21$}
								\vspace{1 ex}
        \end{subfigure}
	\hspace{4 ex}
        \begin{subfigure}{0.45\textwidth}
                \centering
                \epsfig{file=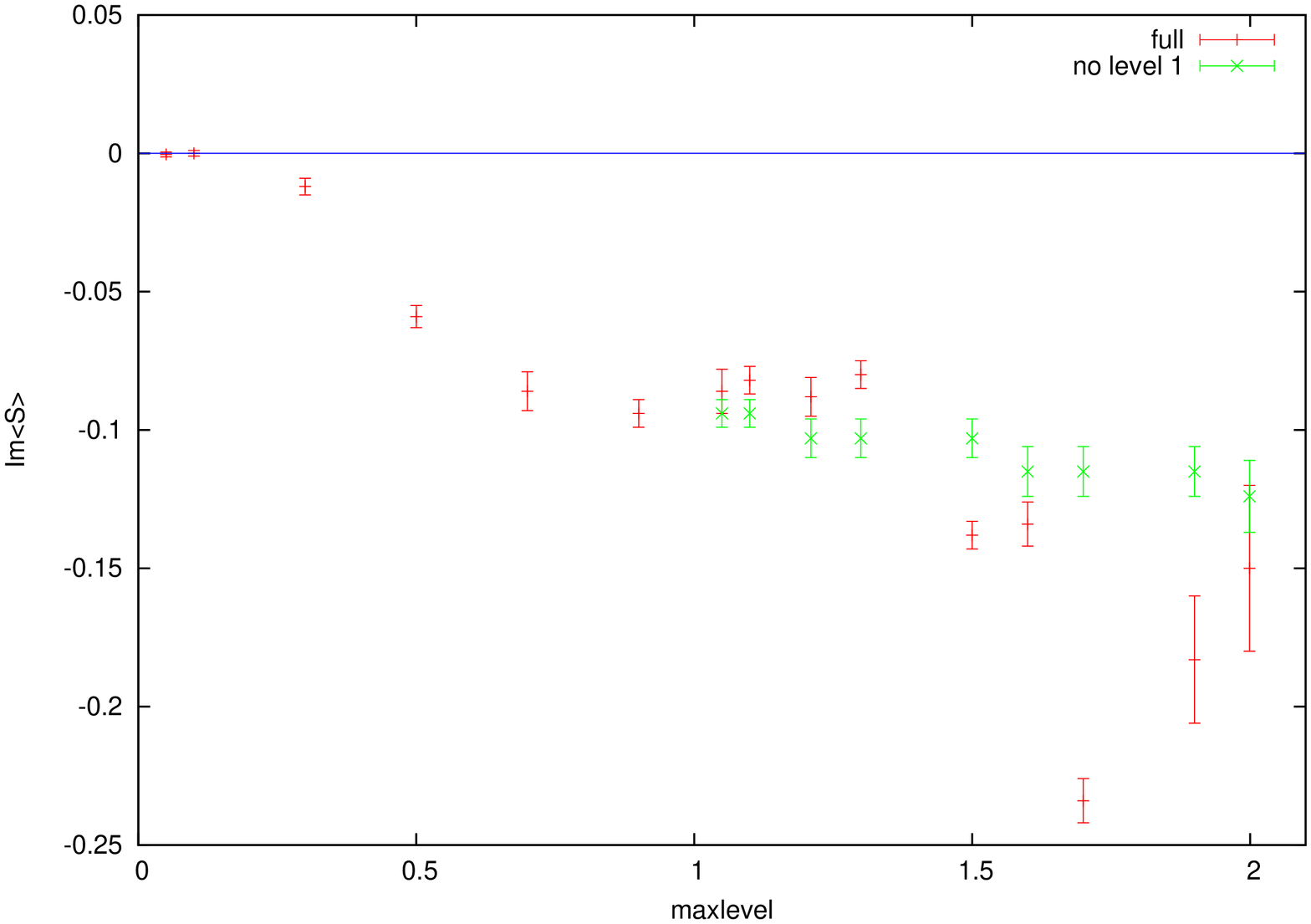,
								scale=0.3,
								angle=-90,clip}  
                \caption{$x_{min}=-20$}
								\vspace{1 ex}
        \end{subfigure}\\
				\hspace{-9 ex}
        \begin{subfigure}{0.45\textwidth}
                \centering
                \epsfig{file=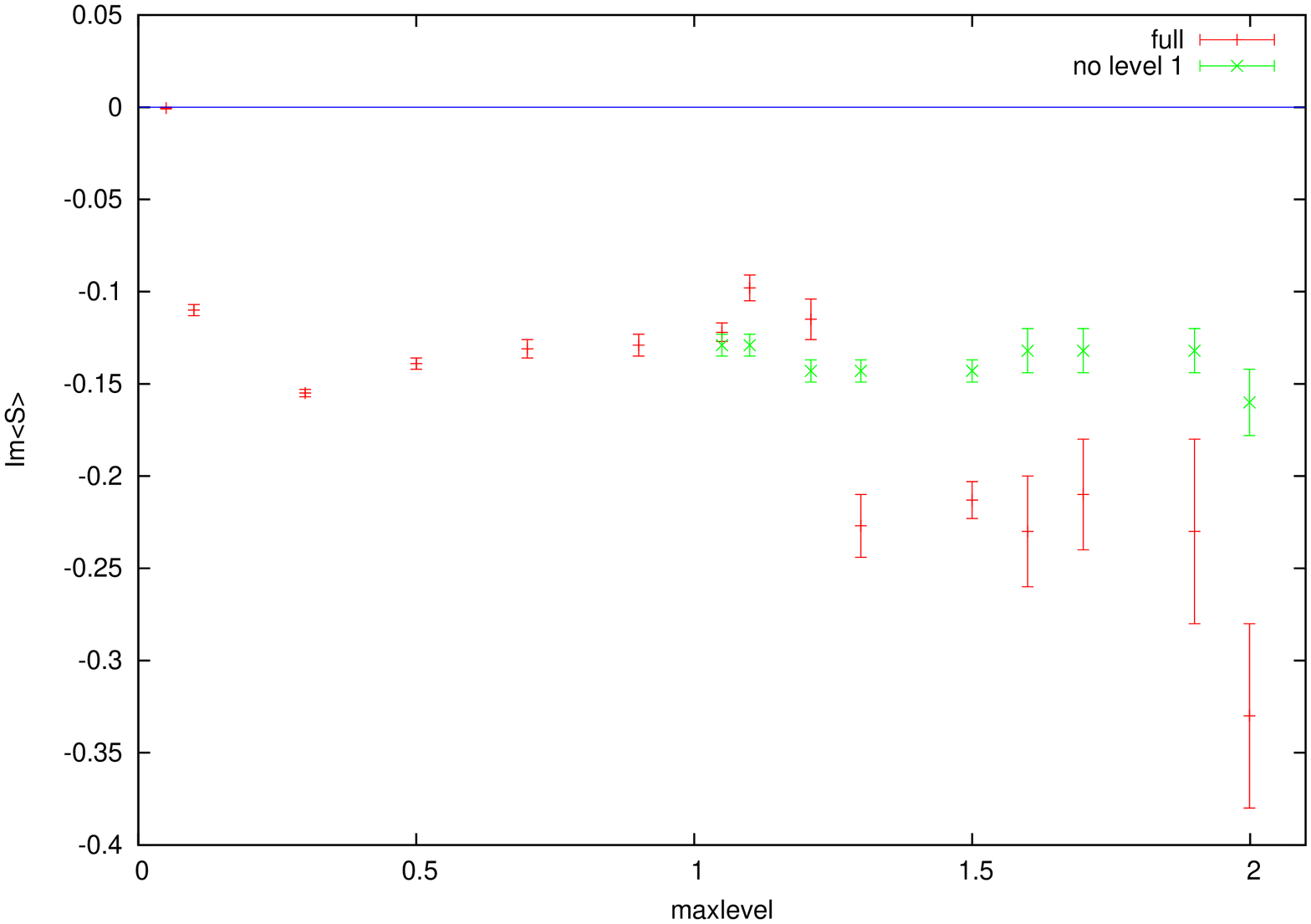,
								scale=0.3,
								angle=-90,clip}  
                \caption{$x_{min}=-19$}
        \end{subfigure}        
			\hspace{4 ex}		
        \begin{subfigure}{0.45\textwidth}
                \centering
                \epsfig{file=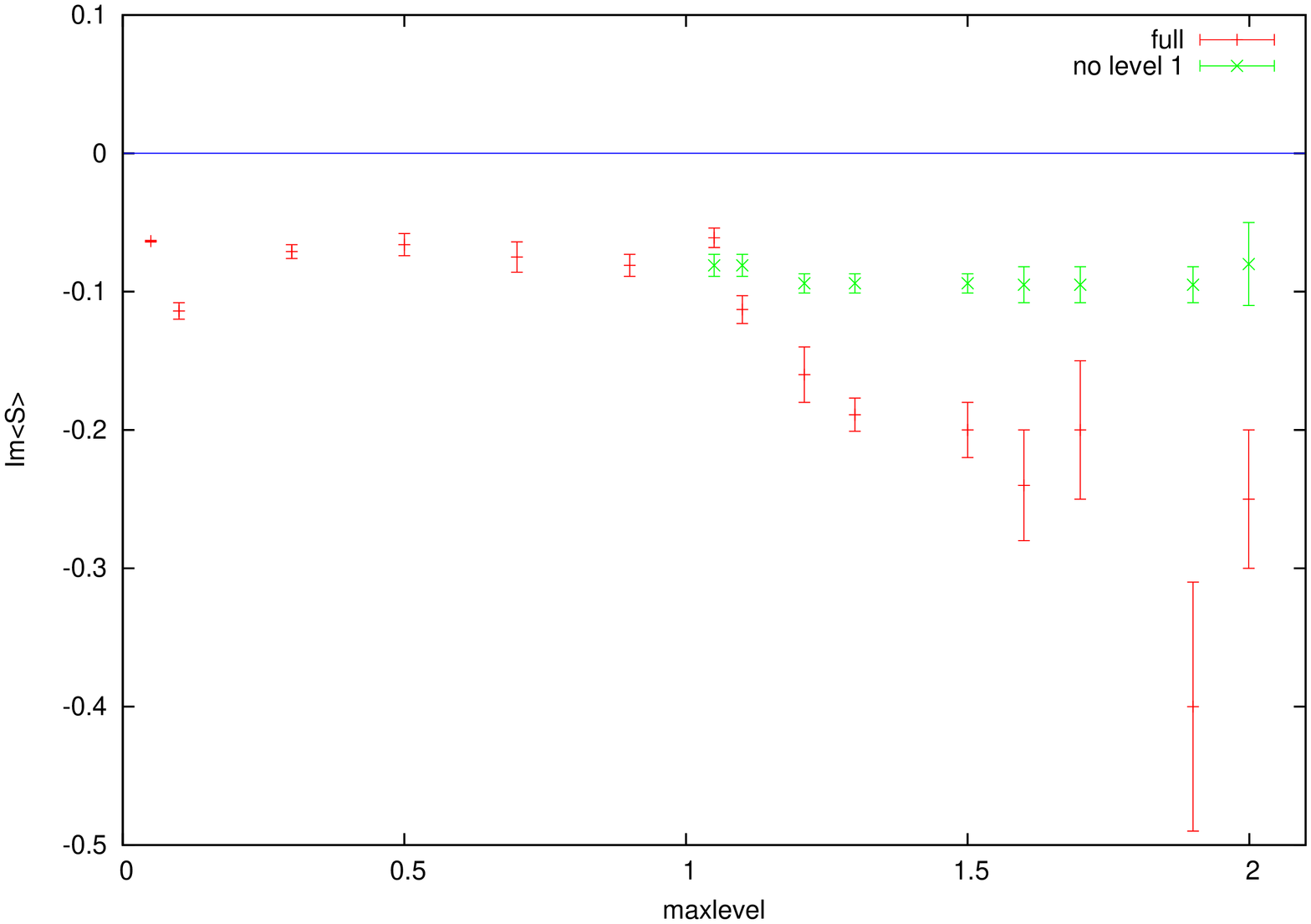,
								scale=0.3,
								angle=-90,clip}  
                \caption{$x_{min}=-18$}
        \end{subfigure}
        \caption{$\Im\langle S \rangle$ for $L=20$ for
the
full theory (red),
                 and without level-1 fields (green), for $x_{min}=-21,-20,-19,-18$.}
\label{l20actfig}
\end{figure}

\begin{figure}
  \centering
        \hspace{-9 ex}
        \begin{subfigure}{0.45\textwidth}
                \centering
                \epsfig{file=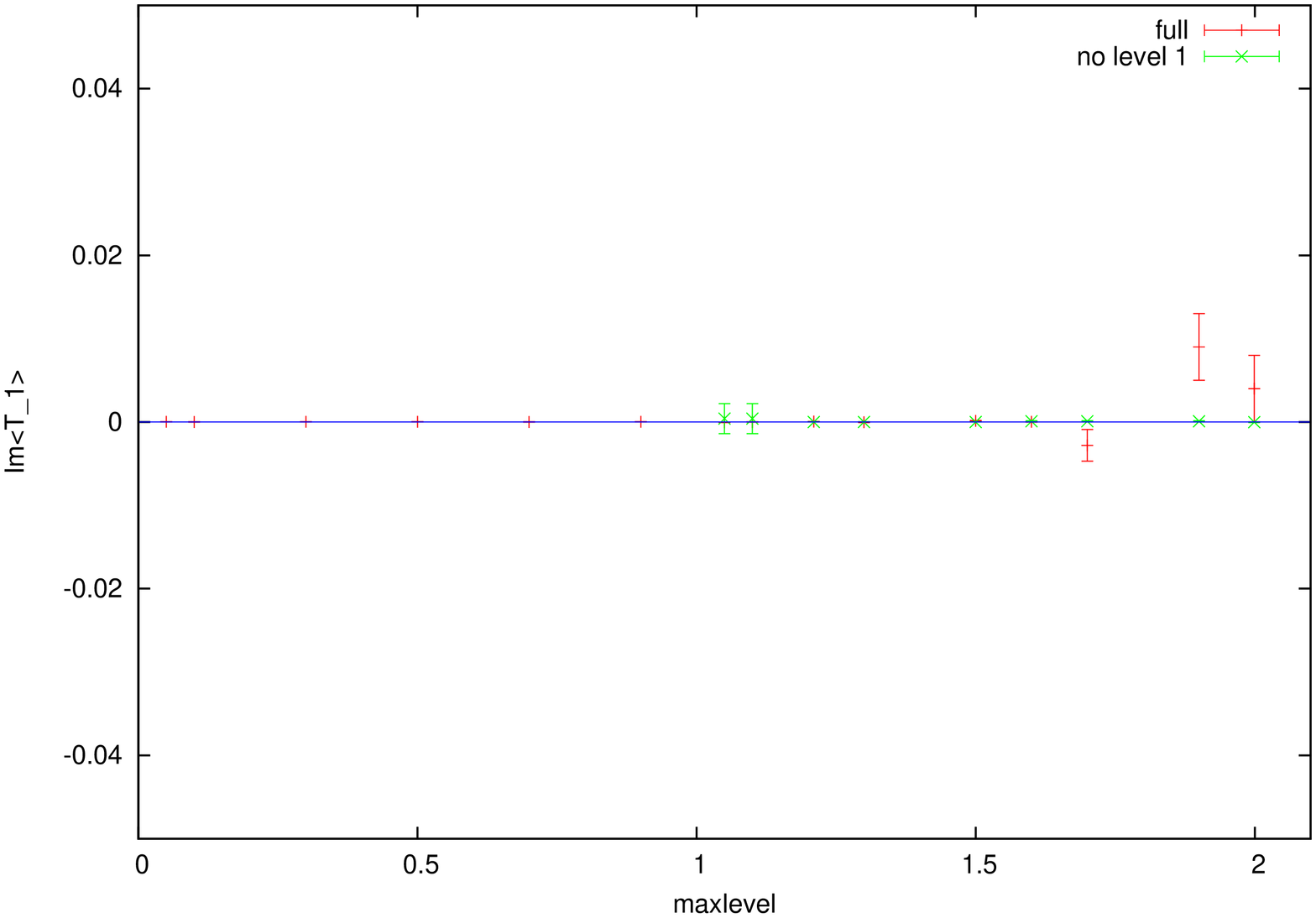,scale=0.31,angle=-90,clip}  
                \caption{$x_{min}=-21$}
								\vspace{1 ex}
        \end{subfigure}
        \hspace{4 ex}
        \begin{subfigure}{0.45\textwidth}
                \centering
                \epsfig{file=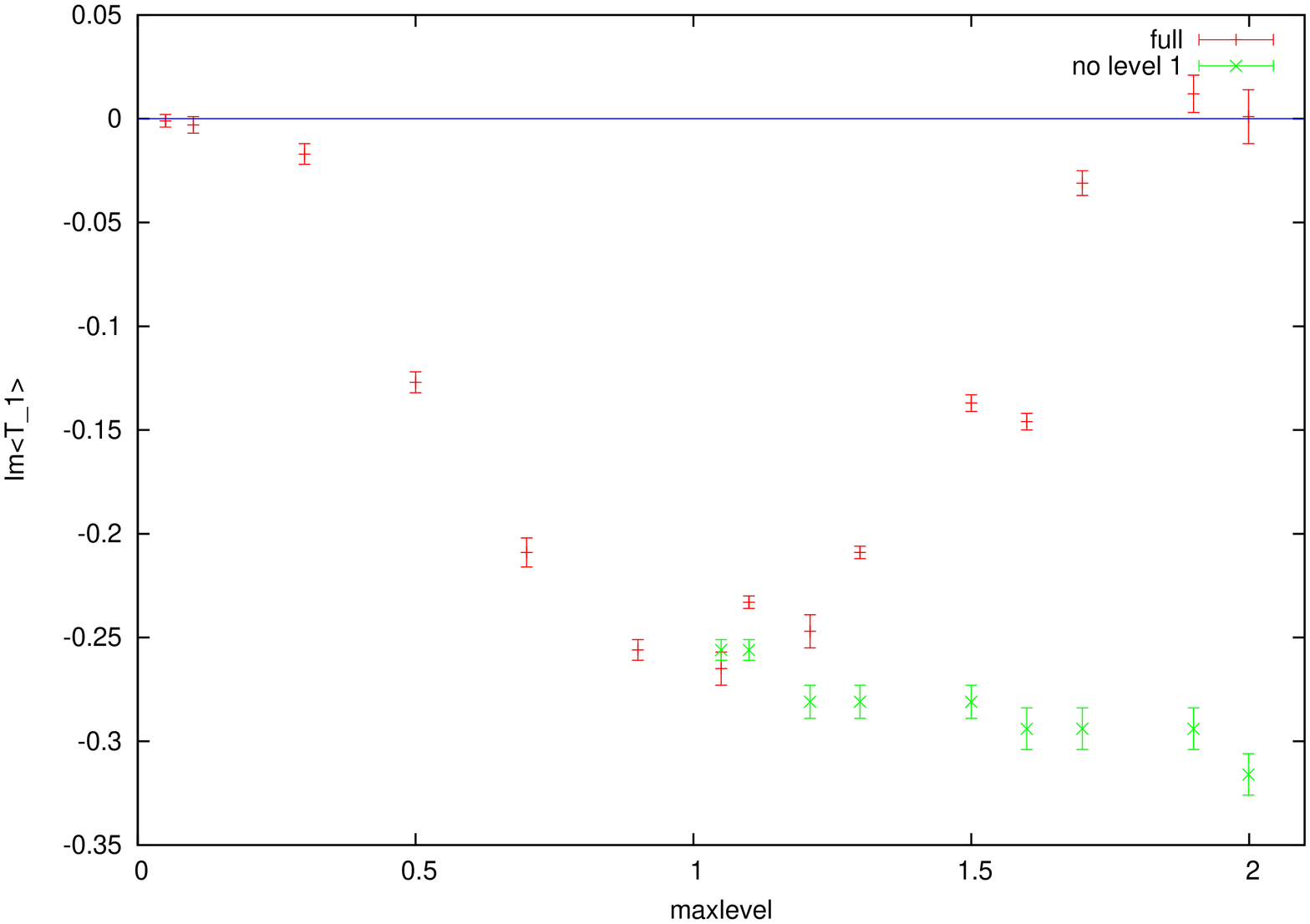,scale=0.31,angle=-90,clip}  
                \caption{$x_{min}=-20$}
								\vspace{1 ex}
        \end{subfigure}
				\\
        \vspace{1 ex}
        \hspace{-9 ex}
        \begin{subfigure}{0.45\textwidth}
                \centering
                \epsfig{file=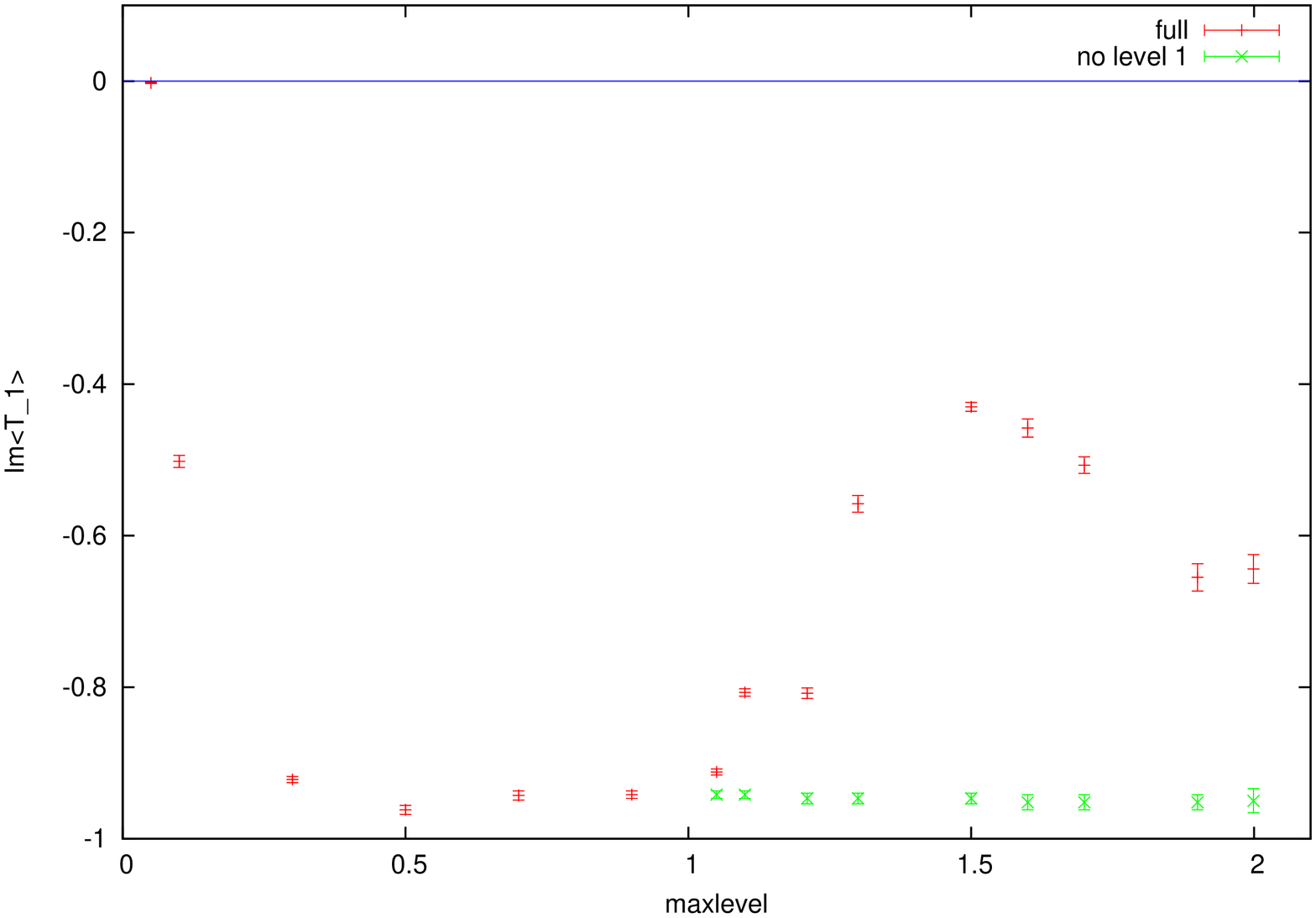,scale=0.31,angle=-90,clip}  
                \caption{$x_{min}=-19$}
        \end{subfigure}        
        \hspace{4 ex}
        \begin{subfigure}{0.45\textwidth}
                \centering
                \epsfig{file=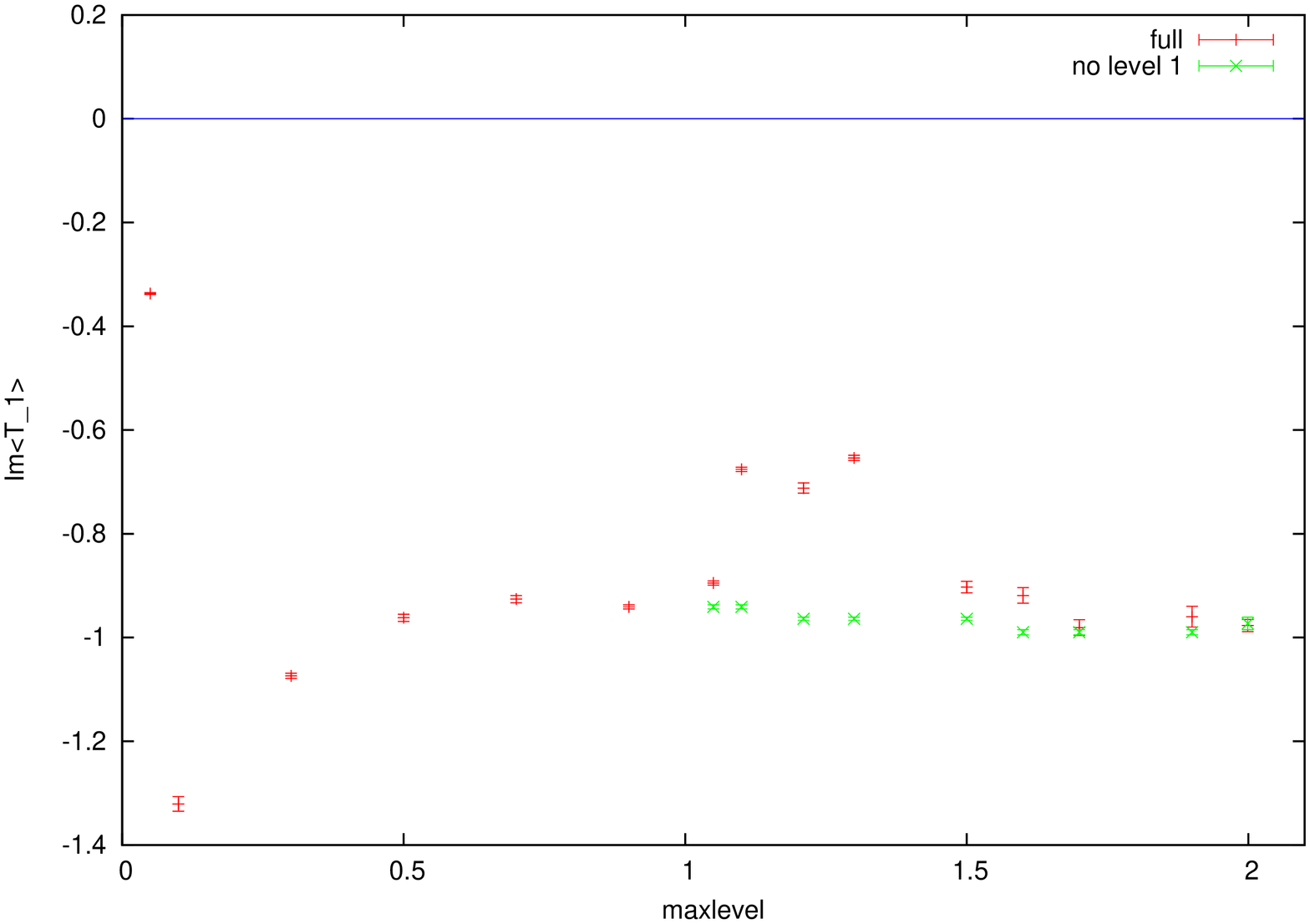,scale=0.31,angle=-90,clip}  
                \caption{$x_{min}=-18$}
        \end{subfigure}
        \caption{$\Im\langle T_1 \rangle$ for $L=20$ for
the
full theory (red),
and without level-1 fields (green), for $x_{min}=-21,-20,-19,-18$.}
\label{l20t1fig}
\end{figure}

The strong coupling region begins around $x_{min}=-20$, and we are able to get results with
reasonably small errors up to $x_{min}=-18$. We show results for the imaginary parts
of the action and the $T_1$ mode in Figs.~\ref{l20actfig} and~\ref{l20t1fig}.
Other observables show similar behaviour.
The general trends we see are quite clear. First, increasing
$l_{max}$ increases
$\Im \langle S \rangle$, and to a lesser extent  $\Im \langle T_1 \rangle$ as well.
In particular, there is no evidence that the imaginary parts are going to zero
as $l_{max}$ is increased. There is not a great difference between the runs with
and without the level-1 fields; in some cases they make the imaginary parts smaller,
and in other cases larger. We also see a general trend for the imaginary parts to be
larger at larger $x_{min}$; this is not surprising as the destabilising cubic terms
are becoming larger. Another point is that the results do not appear to become independent
of $l_{max}$ as it is increased, so we are not (yet?) seeing cutoff-independence.

From all of these, we also see that generally the results are
smooth in $l_{max}$. This means we can focus on a few key values
of $l_{max}$ to see the trends, which we often do from now on,
especially in section~\ref{xminsec}.

\subsubsection{Results at $L=10$}

This is a much smaller interval, so there are fewer modes,
specifically 4 for $T$ and 3 for $A$ up to $l=2$. There may
potentially be a problem with having too few modes --- e.g. one might
imagine that one needs many modes to see ``continuum" physics. On the
other hand having fewer modes makes the simulations faster, so we can
achieve smaller errors or go to stronger couplings.
Note that because the numbers of level-0 and
level-1 modes scale differently, there is no simple mapping between
$L=10$ and $L=20$.

Again, we have carried out simulations for every $l_{max}$ between 0 and 2 which gives a
different number of modes. We have also added runs where the level-1
fields $A$, $B$ and $C$ are not included in the simulation, to give
some idea of the effect the level-1 fields have on the physics.

\begin{figure}
  \centering
        \hspace{-9 ex}
        \begin{subfigure}{0.45\textwidth}
                \centering
                \epsfig{file=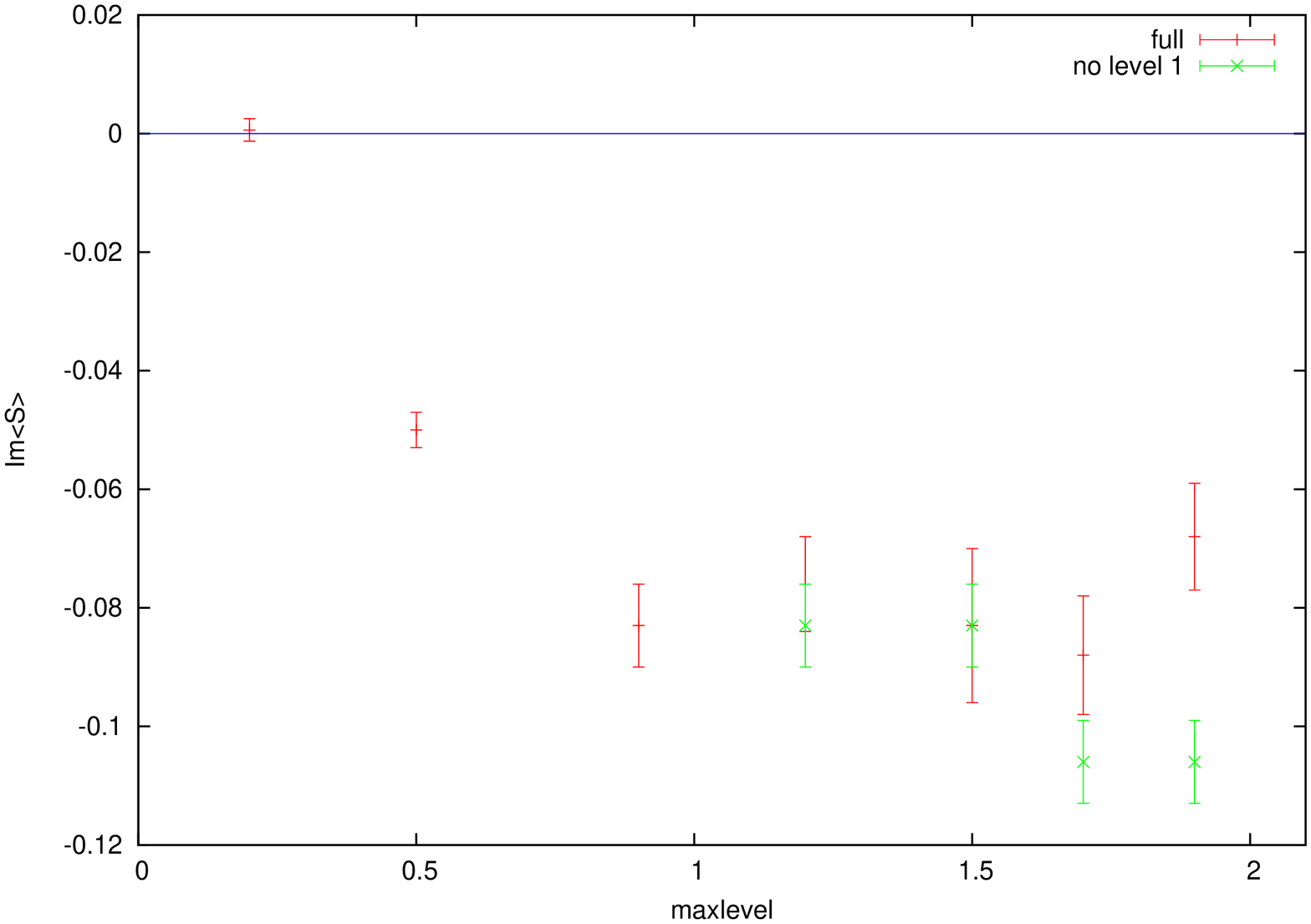,scale=0.31,angle=-90,clip}  
                \caption{$x_{min}=-10$}
        \end{subfigure}
        \hspace{4 ex}
        \begin{subfigure}{0.45\textwidth}
                \centering
                \epsfig{file=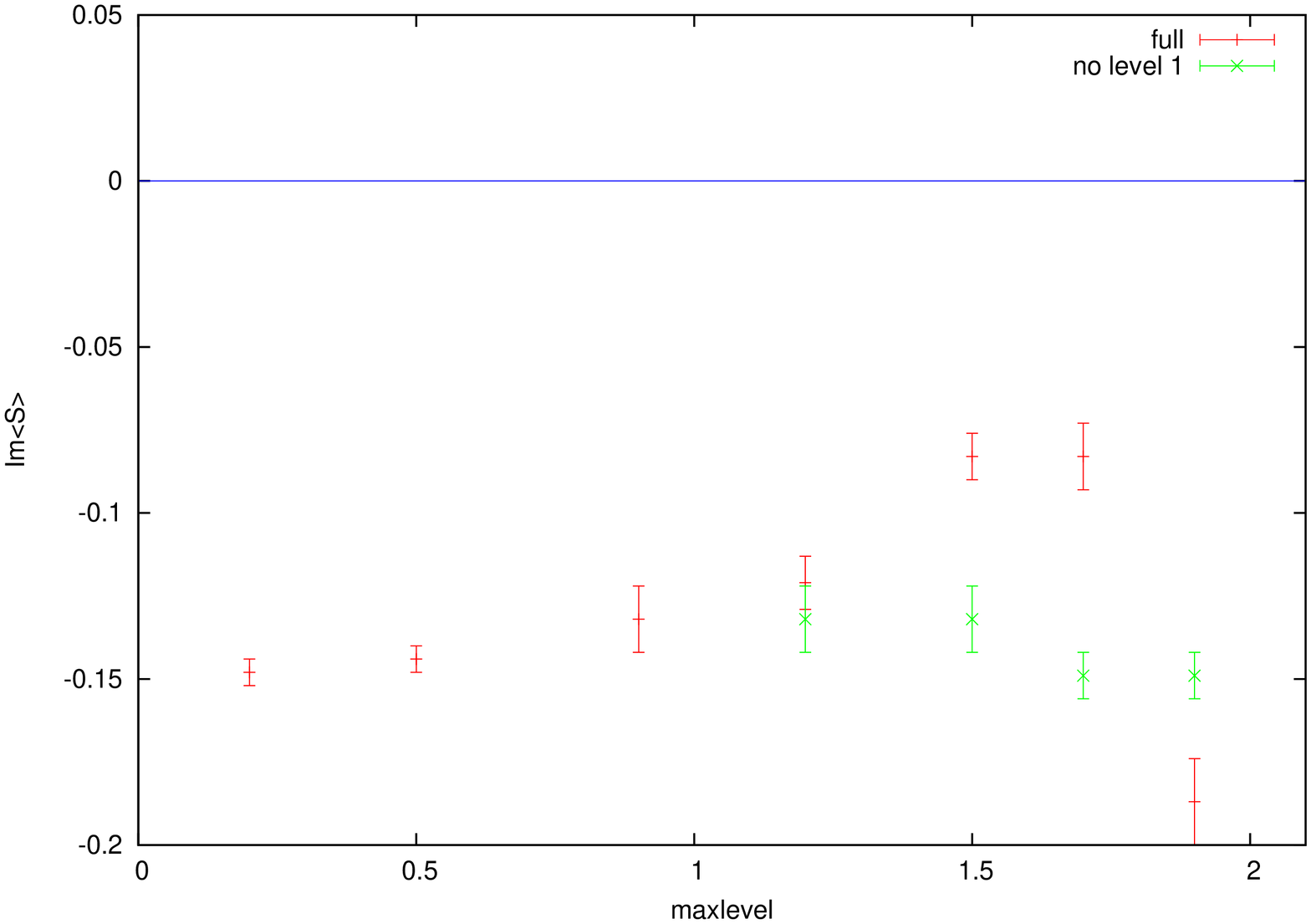,scale=0.31,angle=-90,clip}  
                \caption{$x_{min}=-9$}
        \end{subfigure}\ \\
        \vspace{2 ex}
        \hspace{-10 ex}
        \begin{subfigure}{0.45\textwidth}
                \centering
                \epsfig{file=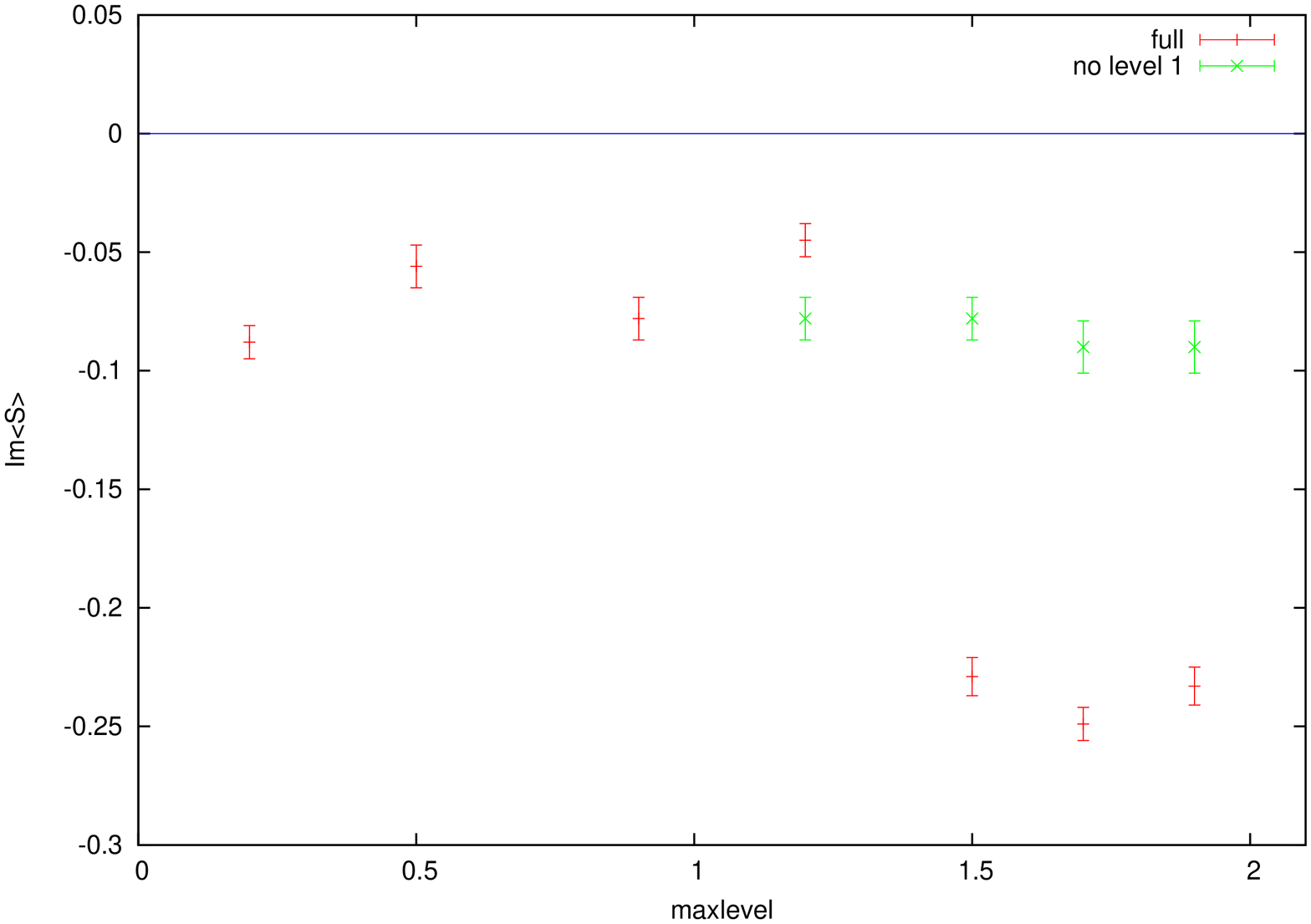,scale=0.31,angle=-90,clip}  
                \caption{$x_{min}=-8$}
        \end{subfigure}        
        \caption{$\Im\langle S \rangle$ for $L=10$ for
the
full theory (red),
and without level-1 fields (green), for $x_{min}=-10,-9,-8$.}
\label{l10actfig}
\end{figure}

\begin{figure}
  \centering
        \hspace{-9 ex}
        \begin{subfigure}{0.45\textwidth}
                \centering
                \epsfig{file=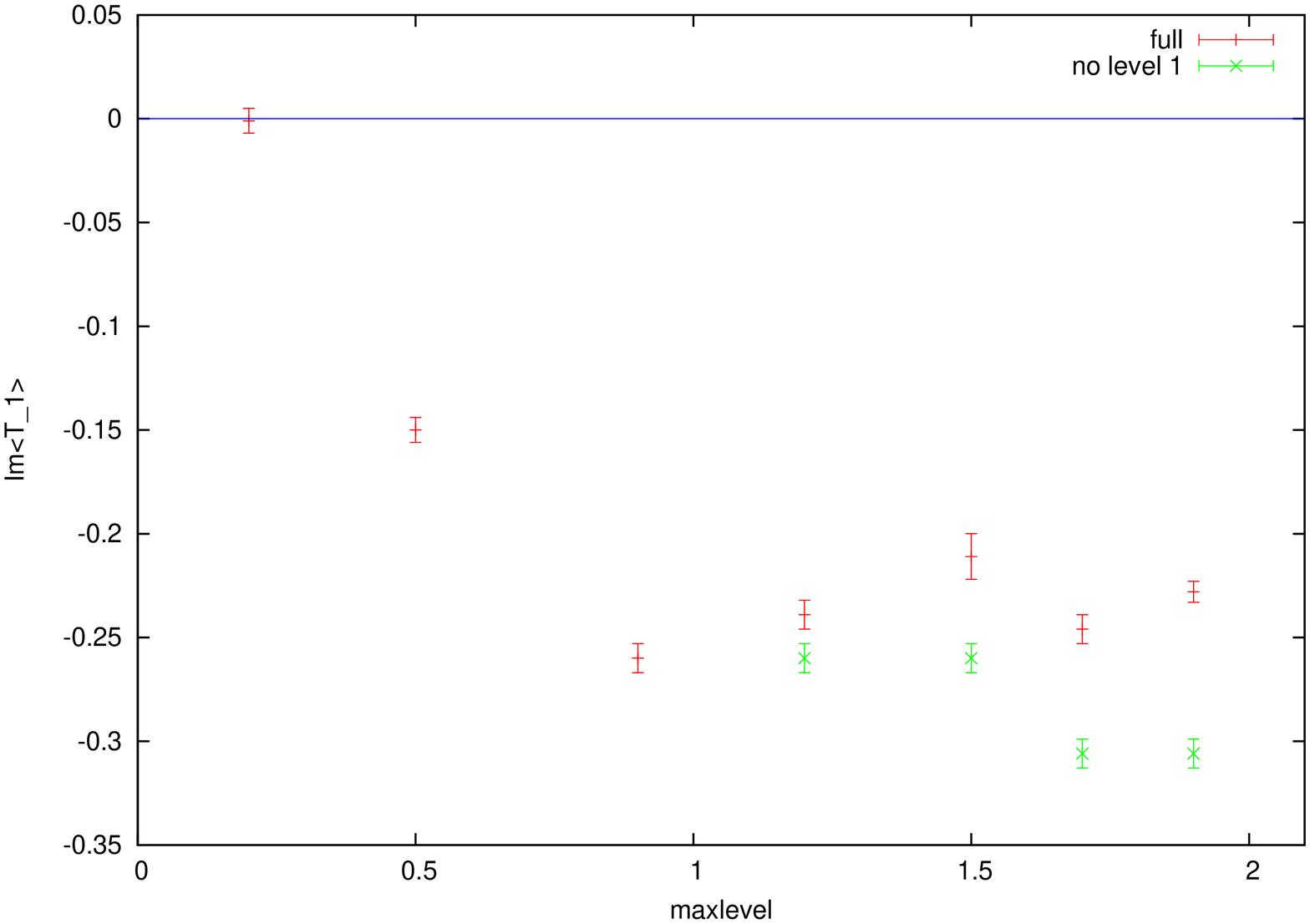,scale=0.31,angle=-90,clip}  
                \caption{$x_{min}=-10$}
        \end{subfigure}
        \hspace{4 ex}
        \begin{subfigure}{0.45\textwidth}
                \centering
                \epsfig{file=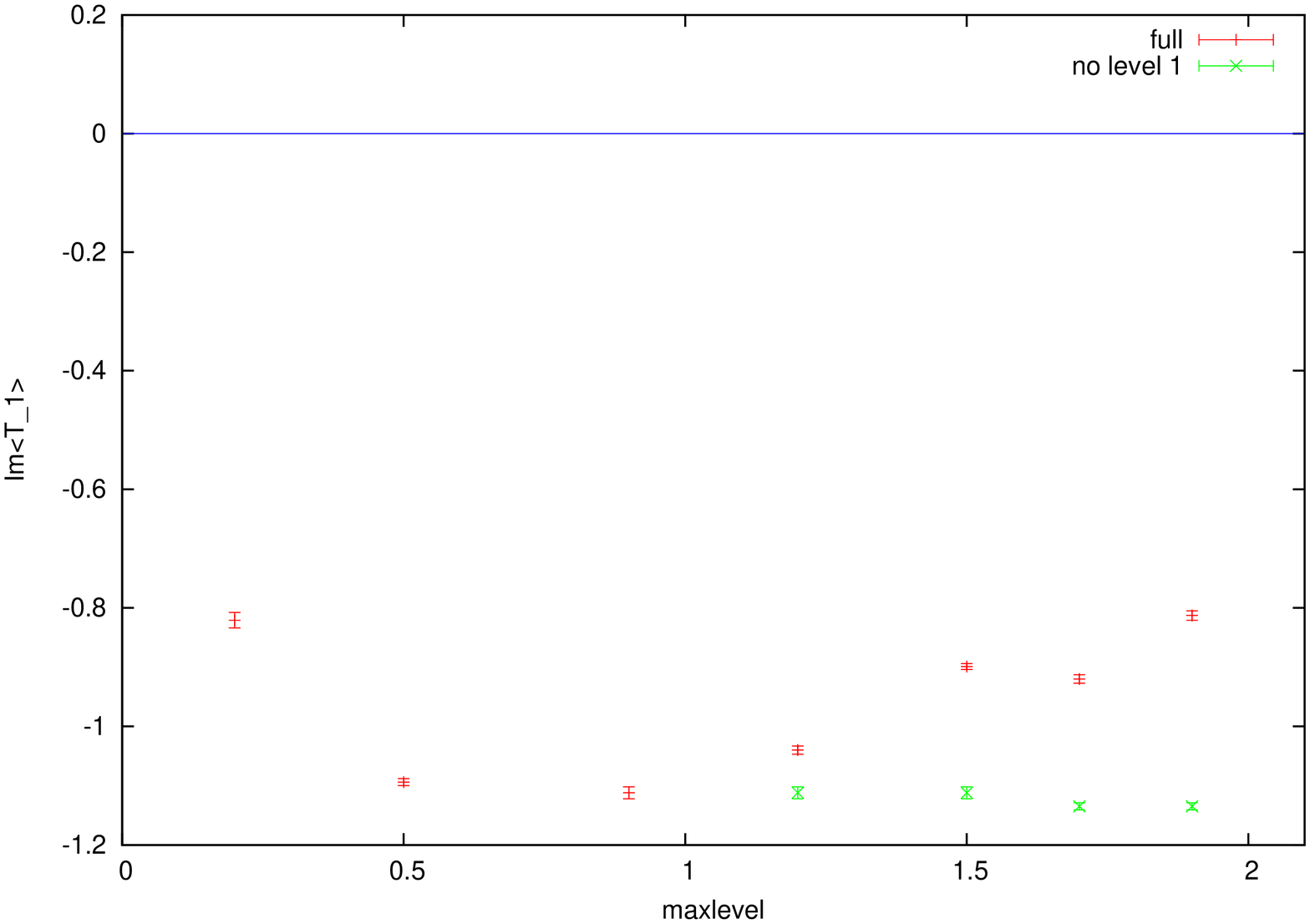,scale=0.31,angle=-90,clip}  
                \caption{$x_{min}=-9$}
        \end{subfigure}\ \\
        \vspace{2 ex}
        \hspace{-10 ex}
        \begin{subfigure}{0.45\textwidth}
                \centering
                \epsfig{file=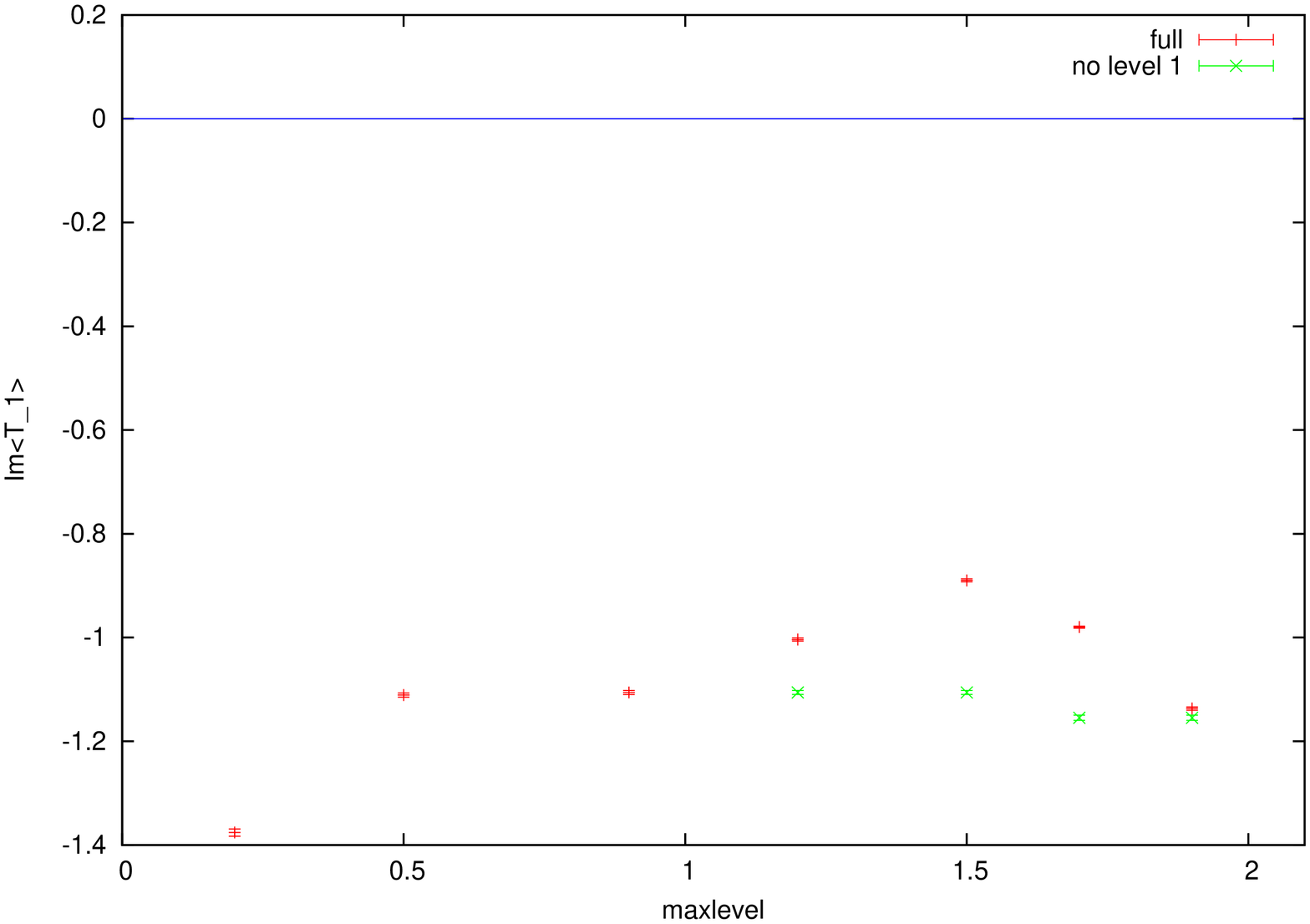,scale=0.31,angle=-90,clip}  
                \caption{$x_{min}=-8$}
        \end{subfigure}        
        \caption{$\Im\langle T_1 \rangle$ for $L=10$ for the full theory (red),
				         and without level-1 fields (green), for $x_{min}=-10,-9,-8$.}
\label{l10t1fig}
\end{figure}

The strong coupling region begins around $x_{min}=-10$, and we are able to
get results with reasonably small errors up to
$x_{min}=-8$.
We show results for the imaginary parts of the action and the $T_1$ mode in
Figs.~\ref{l10actfig} and~\ref{l10t1fig}. Roughly speaking, we would expect
results at a given $x_{min}$ for $L=10$ to match those at $x_{min}-10$ for $L=20$,
since both will extend the same distance into the strong coupling region, so we have
chosen the appropriate values of $x_{min}$ to allow this comparison to be made with
the results of section~\ref{sec:l=20}. In fact this turns out to be not exactly true,
and we see no instability at all at $x_{min}=-11$
(the imaginary parts are at most of order $10^{-5}$),
so we do not show any plots for this case.

We see that the results look rather similar to those for $L=20$. One
potentially interesting area in these plots is just above
$l_{max}=1$, since the density of level-1 modes is much higher here
for $L=20$ than for $L=10$.
However, comparing the plots, nothing interesting seems to happen in
this range. Overall, it appears there is not much difference
between $L=10$ and $L=20$, which at least suggests that the number
of modes we are looking at is not too small. 

\subsubsection{Including a large number of modes at $L=30$}
\label{l30sec}

This is a large interval, with the number of modes and their
density also becoming quite large - specifically we have $n_0=13$ and
$n_1=9$ for $l_{max}=2$. If there is any effect that
requires many modes, it would be surprising if it did not yet appear
here. However, the large number of modes also makes $L=30$ very
expensive\footnote{The computational cost goes very roughly as
  $n_0^3$.}. The largest $x_{min}$ we have been able to reach is
$x_{min}=-30$, corresponding to $x_{max}=0$ and so only at the
beginning of strong coupling.

\begin{figure}
  \centering
        \hspace{-9 ex}
        \begin{subfigure}{0.45\textwidth}
                \centering
                \epsfig{file=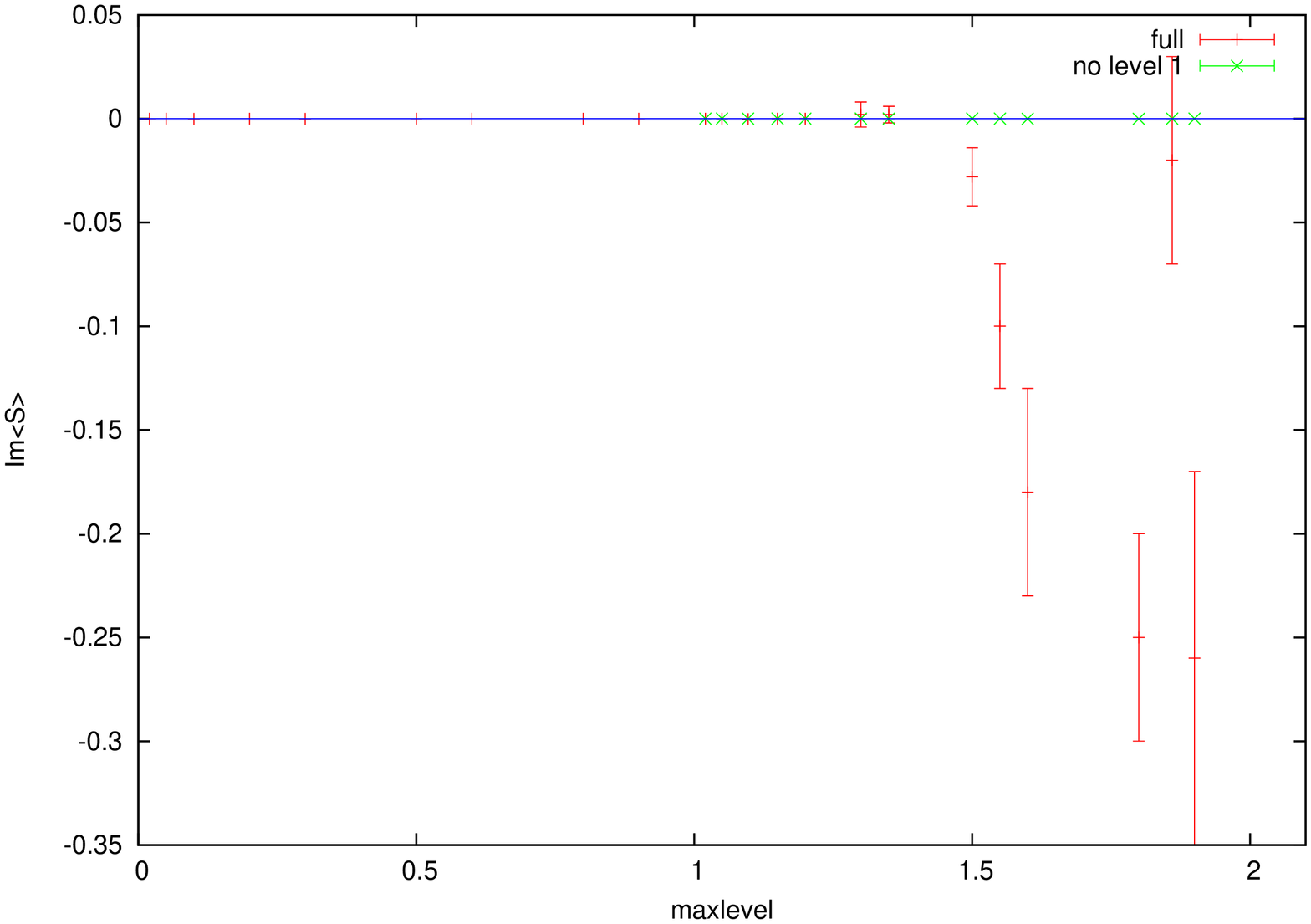,scale=0.31,angle=-90,clip}  
                \caption{$x_{min}=-31$}
        \end{subfigure}
        \hspace{4 ex}
        \begin{subfigure}{0.45\textwidth}
                \centering
                \epsfig{file=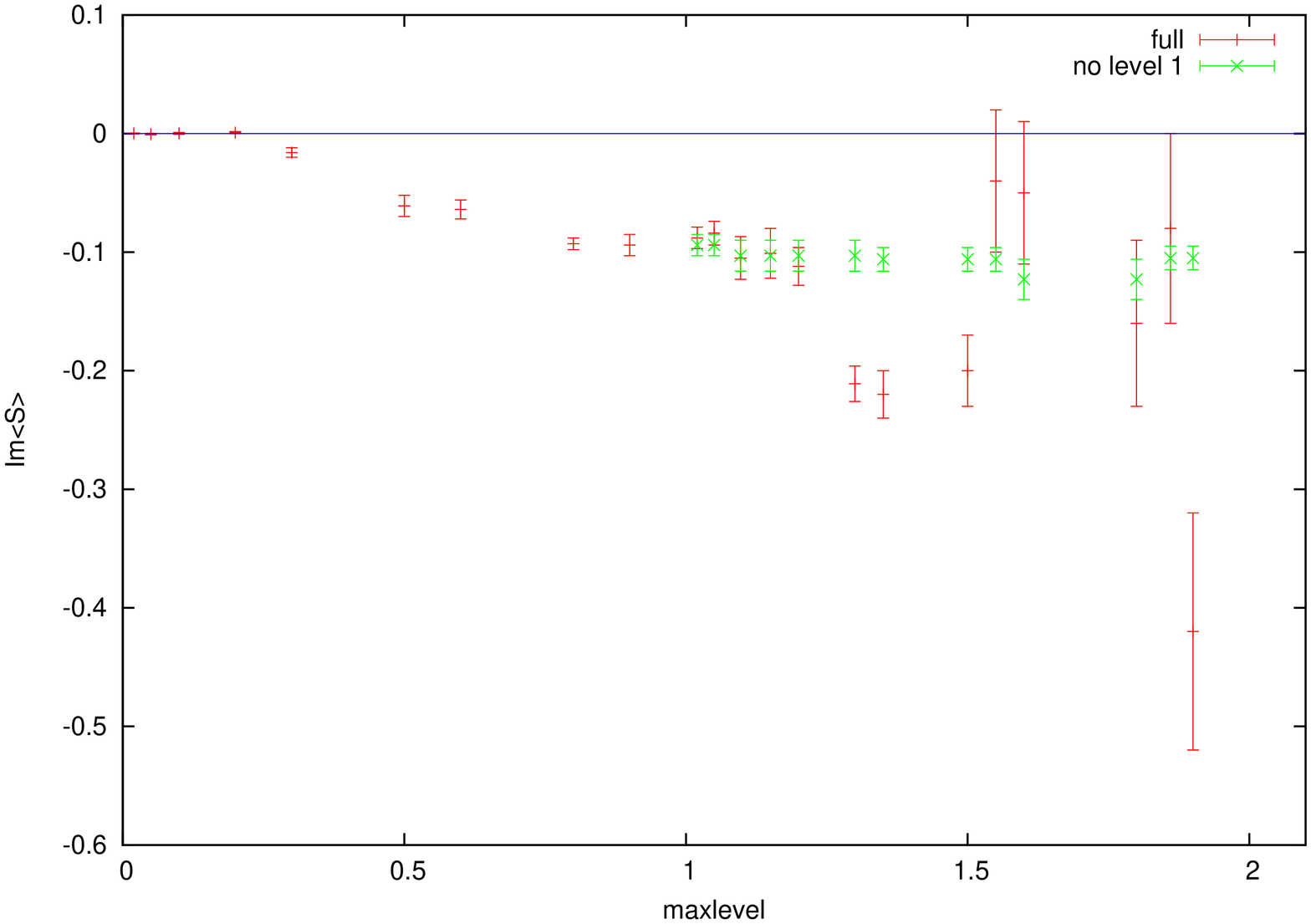,scale=0.31,angle=-90,clip}  
                \caption{$x_{min}=-30$}
        \end{subfigure}
        \caption{$\Im\langle S \rangle$ for $L=30$ for
the
full theory (red),
and without level-1 fields (green), for $x_{min}=-31$ and $-30$.}
\label{l30actfig}
\end{figure}

\begin{figure}
  \centering
        \hspace{-9 ex}
        \begin{subfigure}{0.45\textwidth}
                \centering
                \epsfig{file=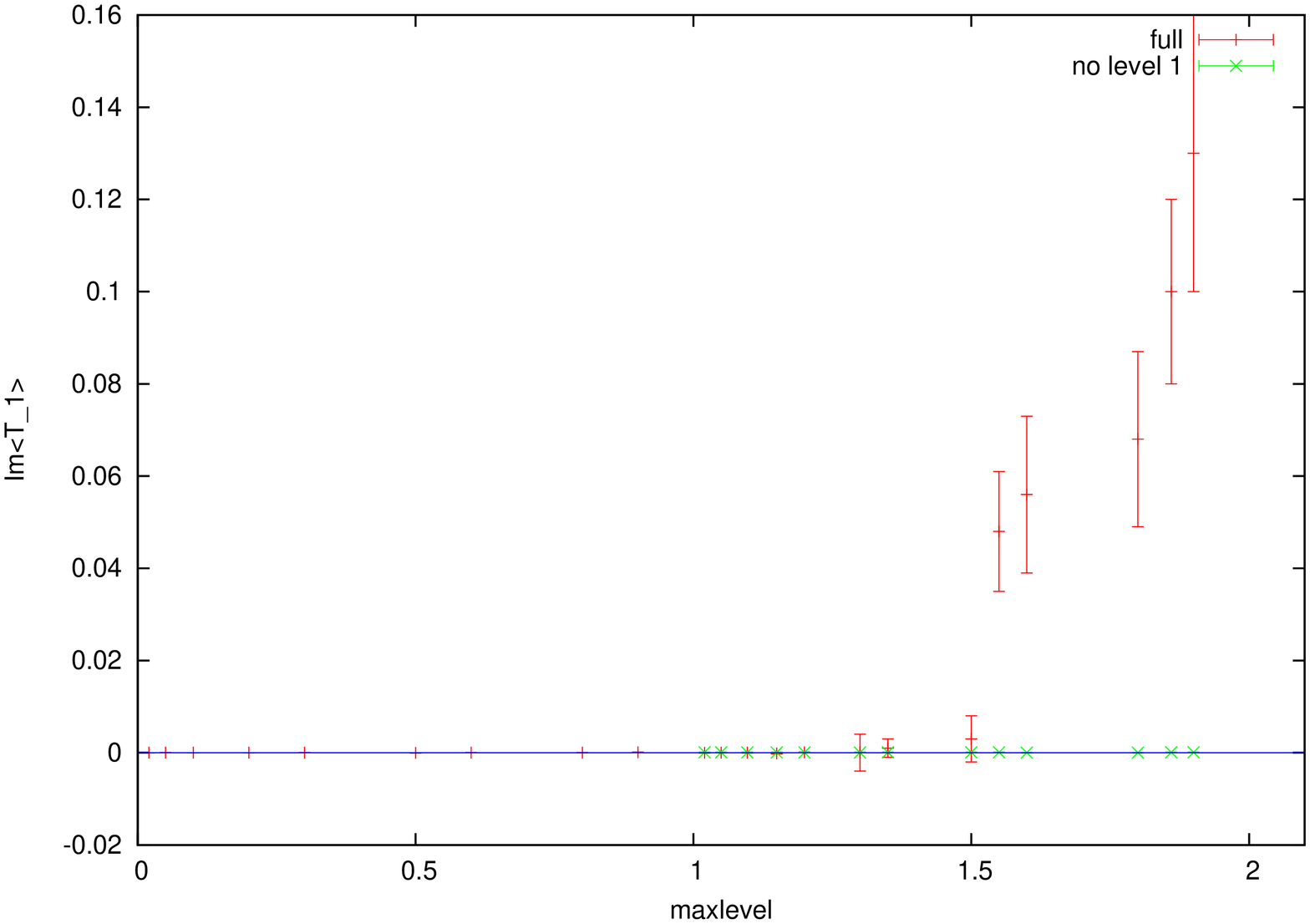,scale=0.31,angle=-90,clip}  
                \caption{$x_{min}=-31$}
        \end{subfigure}
        \hspace{4 ex}
        \begin{subfigure}{0.45\textwidth}
                \centering
                \epsfig{file=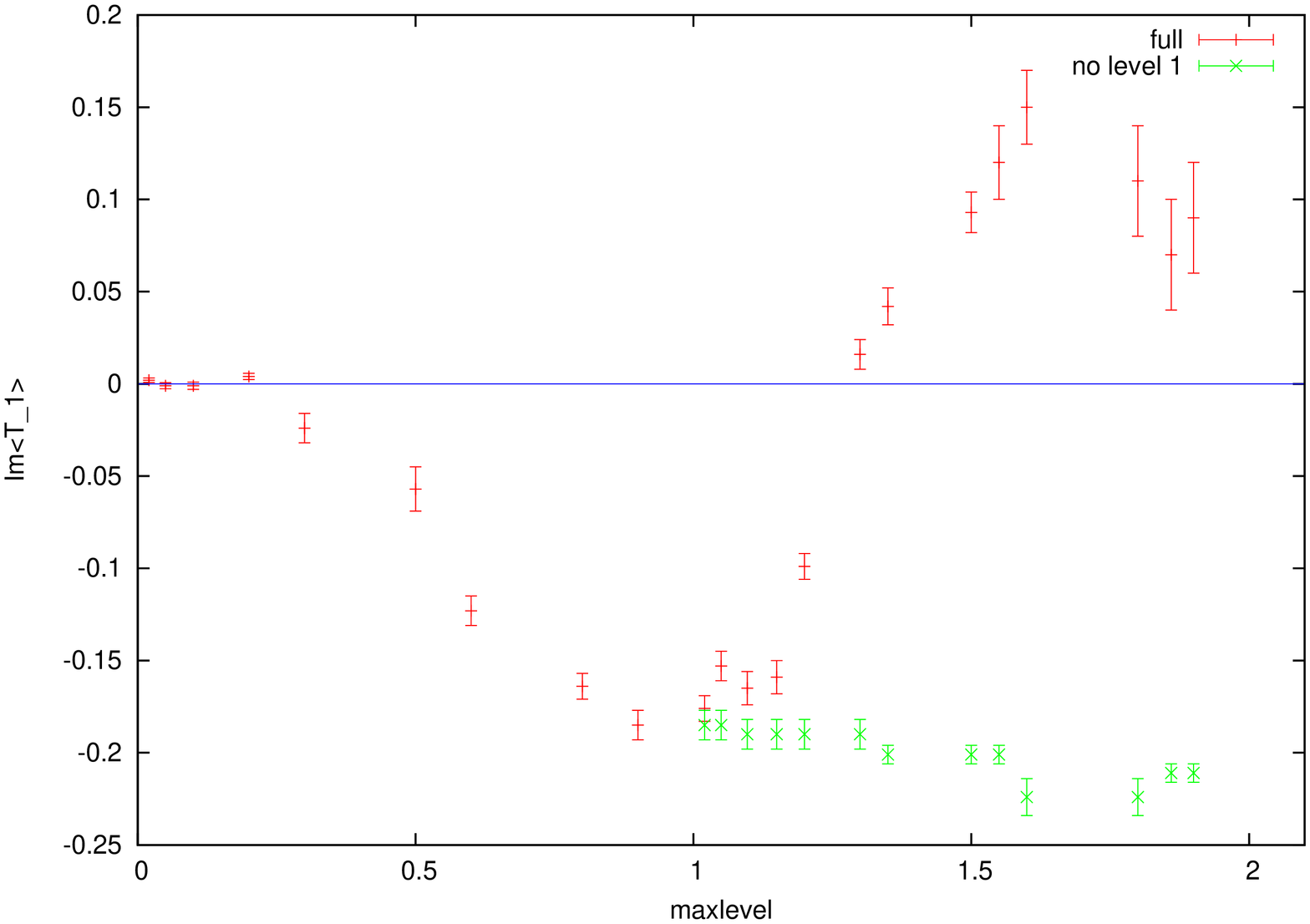,scale=0.31,angle=-90,clip}  
                \caption{$x_{min}=-30$}
        \end{subfigure}
        \caption{$\Im\langle T_1 \rangle$ for  $L=30$ for
the
full theory (red),
and without level-1 fields (green), for $x_{min}=-31$ and $-30$.}
\label{l30t1fig}
\end{figure}

Again, we have results for every $l_{max}$ between 0 and 2 which gives a
different number of modes. We show results for the imaginary parts of the
action and the $T_1$ mode in Figs.~\ref{l30actfig} and~\ref{l30t1fig}.
The
overall results look rather similar, though in detail they are different;
for example, $\Im\langle T_1 \rangle$ is now positive for some values of
$l_{max}$, which was not the case for smaller $L$.


Unlike for the case of $L=20$, $x_{min}=-21$, and $L=10$,
$x_{min}=-11$
we see that at
$x_{min}=-31$
there are
rather large imaginary parts for the larger values of $l_{max}$.
Therefore it looks like increasing $L$ at constant $x_{max}$ is
destabilising. This seems a bit counter-intuitive when we think about it another way:
going from $L=20$, $x_{min}=-21$ to $L=30$, $x_{min}=-31$ is a change
in $x_{min}$ while keeping $x_{max}$ fixed, so we are only adding a
region of extremely weak coupling --- how can this make things less
stable? Probably the answer is connected to the fact that increasing $L$ means we
have more modes, and thus a
shift in one boundary affects the physics throughout the interval.
Moreover, the non-locality of the action might also source some
influence of one boundary on the rest of the interval.
These effects lead to a worsening of the numerical sign problems and therefore
to larger errors.

\subsection{Changing $x_{min}$}
\label{xminsec}

$x_{min}$ is a very
important parameter as it controls the strength of the cubic terms and
hence the instabilities. We expect that for small $x_{min}$ there will
be no instabilities and so the imaginary parts of observables will be
zero. They will then increase, and finally, if they behave as in the
case of a single mode (see section~\ref{singlemodesec}) go to zero
(except for the action which should go to a finite value).

Since we have established above that the observables have a rather
smooth dependence on $l_{max}$, we have mostly concentrated on a few
values of $l_{max}$. We do not present results for $L=30$ in this
section since, as discussed above, we have been unable to reach very
strong couplings in this case.

\subsubsection{Extensive study at $L=20$}
\label{l20changexminsec}

We begin again with our results for $L=20$, where we have the most data.
For most of this scan in
$x_{min}$, we only use the values
$l_{max}=\{0.05,0.5,0.9,1.5,1.999\}$. We also include
$l_{max}=\{1.5,1.999\}$ with level-1 fields removed.
We have tried to reach the highest values of $x_{min}$ possible, in
order to try to reach the extreme strong-coupling region where all the
observables go to zero. In
practice this has only been possible for the lower values of
$l_{max}$.

\begin{figure}
  \centering
        \hspace{-9 ex}
        \begin{subfigure}{0.45\textwidth}
                \centering
                \epsfig{file=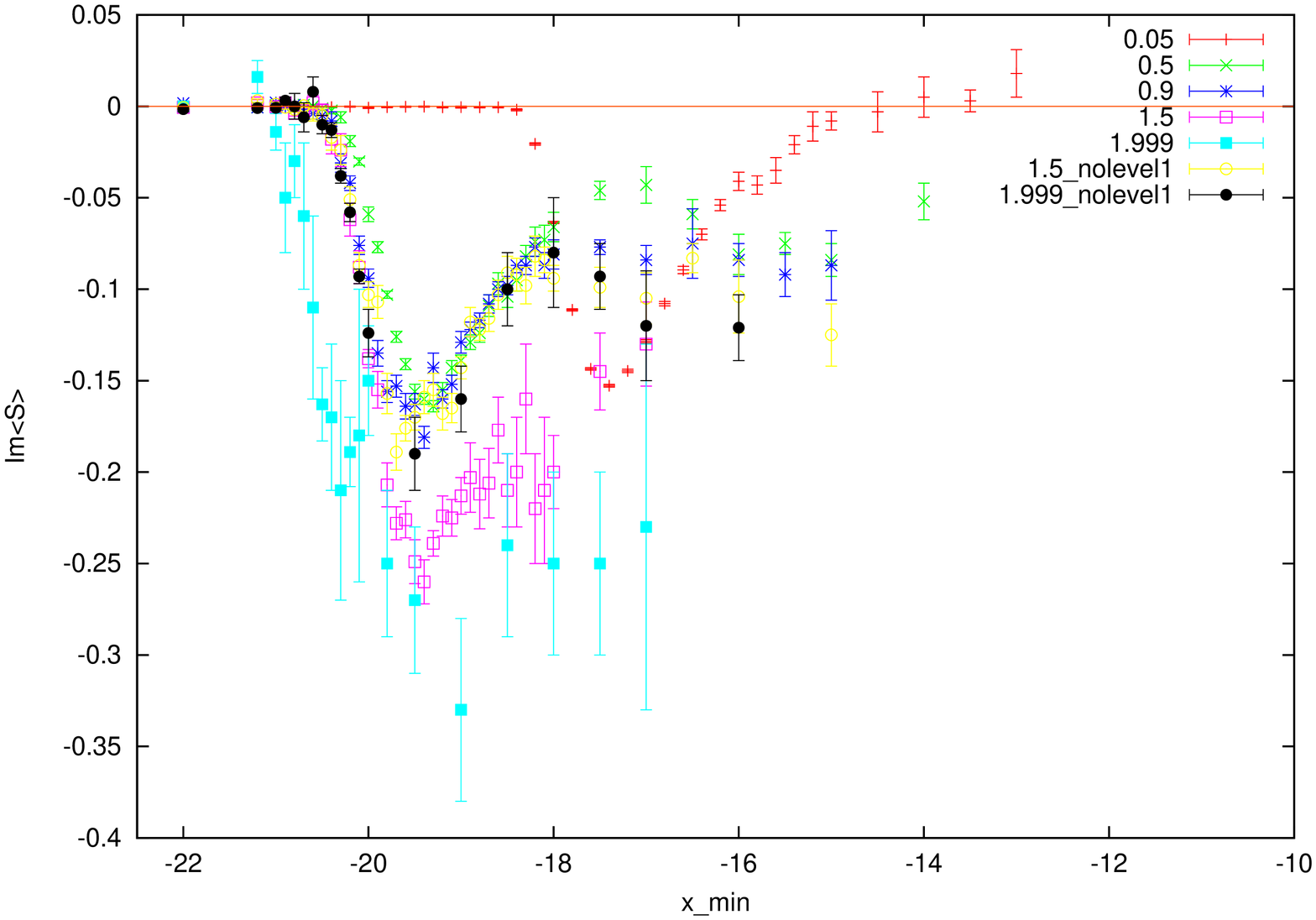,scale=0.31,angle=-90,clip}  
			\footnotesize (a) $\Im\langle S \rangle$\\ $\phantom{a}$\vspace{-.3mm}\\ $\phantom{X_{X_X}}$
        \end{subfigure}
        \hspace{4 ex}
        \begin{subfigure}{0.45\textwidth}
                \epsfig{file=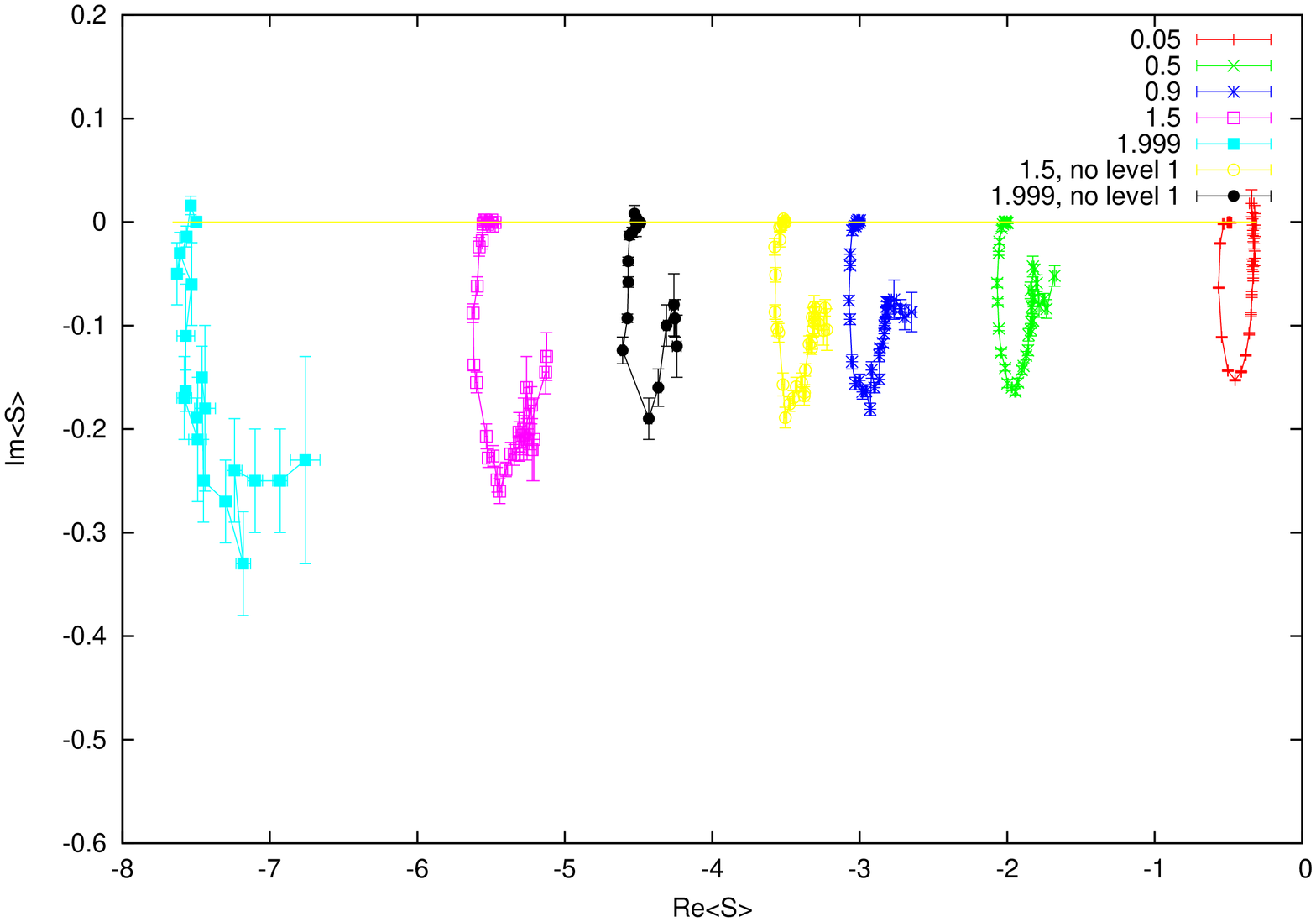,scale=0.31,angle=-90,clip}  
			\footnotesize $\phantom{XXX}\mbox{(b) Trajectory of $\langle S \rangle$ in the complex plane.}\\
			               \phantom{XXX}\mbox{In all cases increasing $x_{min}$ corresponds to}\\
			               \phantom{XXX}\mbox{anticlockwise movement along the trajectory.}$
        \end{subfigure}
        \caption{$\langle S \rangle$ for $L=20$ as a function of
    $x_{min}$ for various $l_{max}$.}
\label{l20varyxminactfig}
\end{figure}

First we show a plot of $\langle S \rangle$, in
Fig.~\ref{l20varyxminactfig}, showing both the behaviour of
the
imaginary part and the trajectory in the complex plane.
We see that there is a similar trend for all $l_{max}$.
This even includes the single-mode $l_{max}=0.05$,
except for a shift in $x_{min}$. Presumably the fact that there is a
single mode means that there is less instability. Actually, this is
the opposite of what we would expect from the hope that ``the
instabilities become of zero measure'' as the number of degrees of
freedom increases. 

In the complex plane, it appears that the
action executes a qualitatively similar 'loop' 
for all $l_{max}$. In all cases the
imaginary part only seems to become zero again when the trajectory reaches
its final point, although for the higher $l_{max}$ the data does not go
far enough to be certain of this. Note that the large offsets between
the loops for the different $l_{max}$ are simply due to the $-1/2$ of
action per mode that we get without the cubic terms. Apart from this all the
$l_{max}$ are rather similar --- things appear to get neither better
nor worse as more modes are added.

In some cases there appear to be cusps in the trajectories in
Fig.~\ref{l20varyxminactfig}, although this is not quite
clear. If they are present they are probably related to the
loops and cusps seen at small $L$ (see section~\ref{smalllsec} below.)

\begin{figure}
  \centering
  \epsfig{file=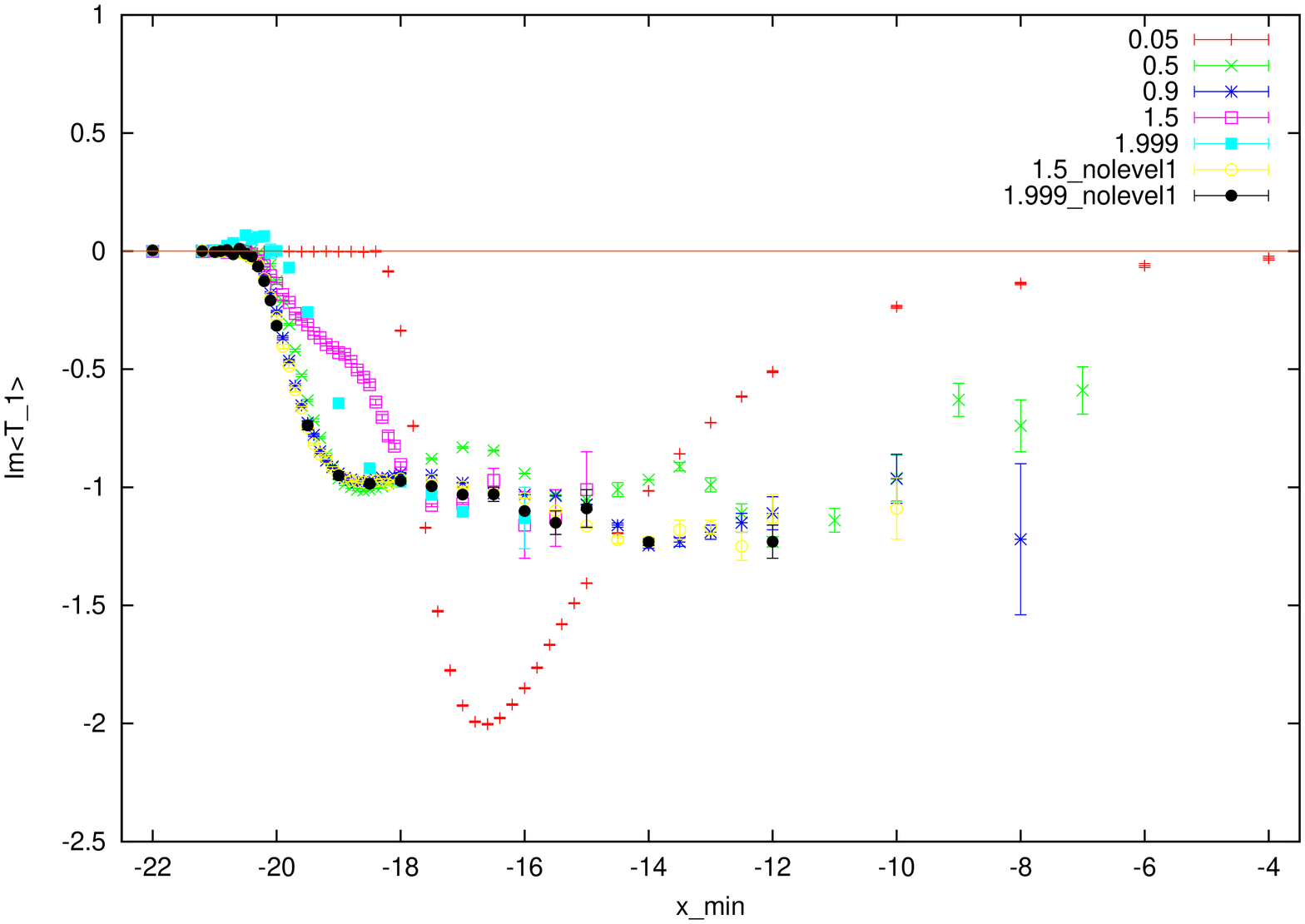,scale=0.43,angle=-90,clip}  
  \caption{$\Im\langle T_1 \rangle$ for $L=20$ as a function of
    $x_{min}$ for various $l_{max}$.}
  \label{l20varyxmint1fig}
\end{figure}

The imaginary part of $\langle T_1 \rangle$ has a similar behaviour,
plotted in Fig~\ref{l20varyxmint1fig}. In this case the return to zero
is slower, but this is partially compensated for by the fact that the errors are
smaller so we are able to go to higher $x_{min}$. Again there is not
much difference between the different values of $l_{max}$, except for
$l_{max}=0.05$, where again the instability begins at higher
$x_{min}$. There are clear oscillations in the data; these correspond
to cusps in the complex plane like those seen for the action. (These
can be seen more clearly at smaller $L$, e.g. in Fig.~\ref{l6varyxmincompt1fig}.)

\subsubsection{Results at $L=10$}

This is quite a small interval, but has the advantage that it is possible to
get to stronger coupling than in the $L=20$ case.
We use the values $l_{max}=\{0.2,0.9,1.9\}$, and also $l_{max}=1.9$ with level-1
fields removed.

\begin{figure}
  \centering
        \hspace{-9 ex}
        \begin{subfigure}{0.45\textwidth}
                \centering
                \epsfig{file=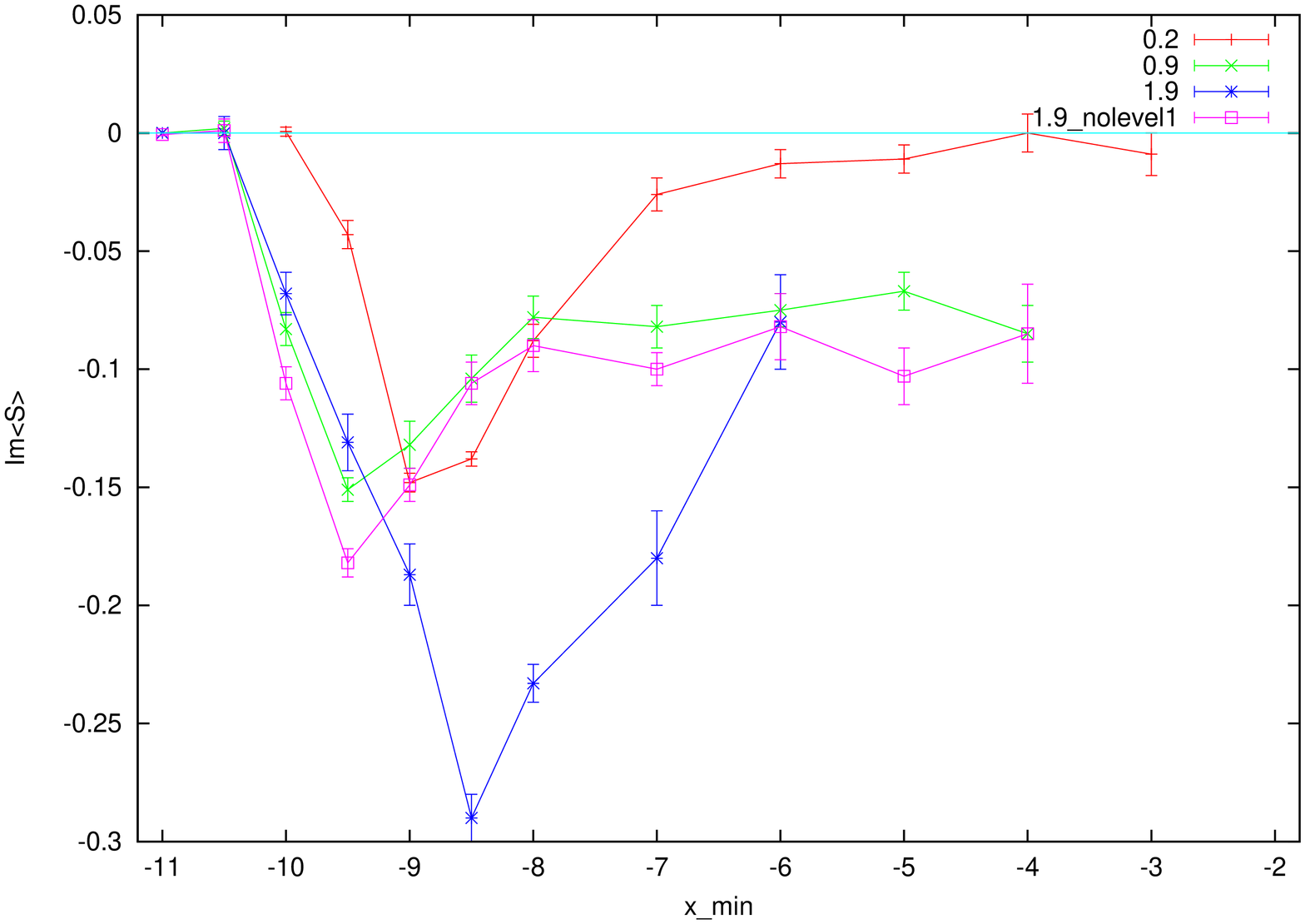,scale=0.31,angle=-90,clip}
							\footnotesize (a) $\Im\langle S \rangle$\\ $\phantom{a}$\vspace{-.3mm}\\ $\phantom{X_{X_X}}$
        \end{subfigure}
        \hspace{4 ex}
        \begin{subfigure}{0.45\textwidth}
                \epsfig{file=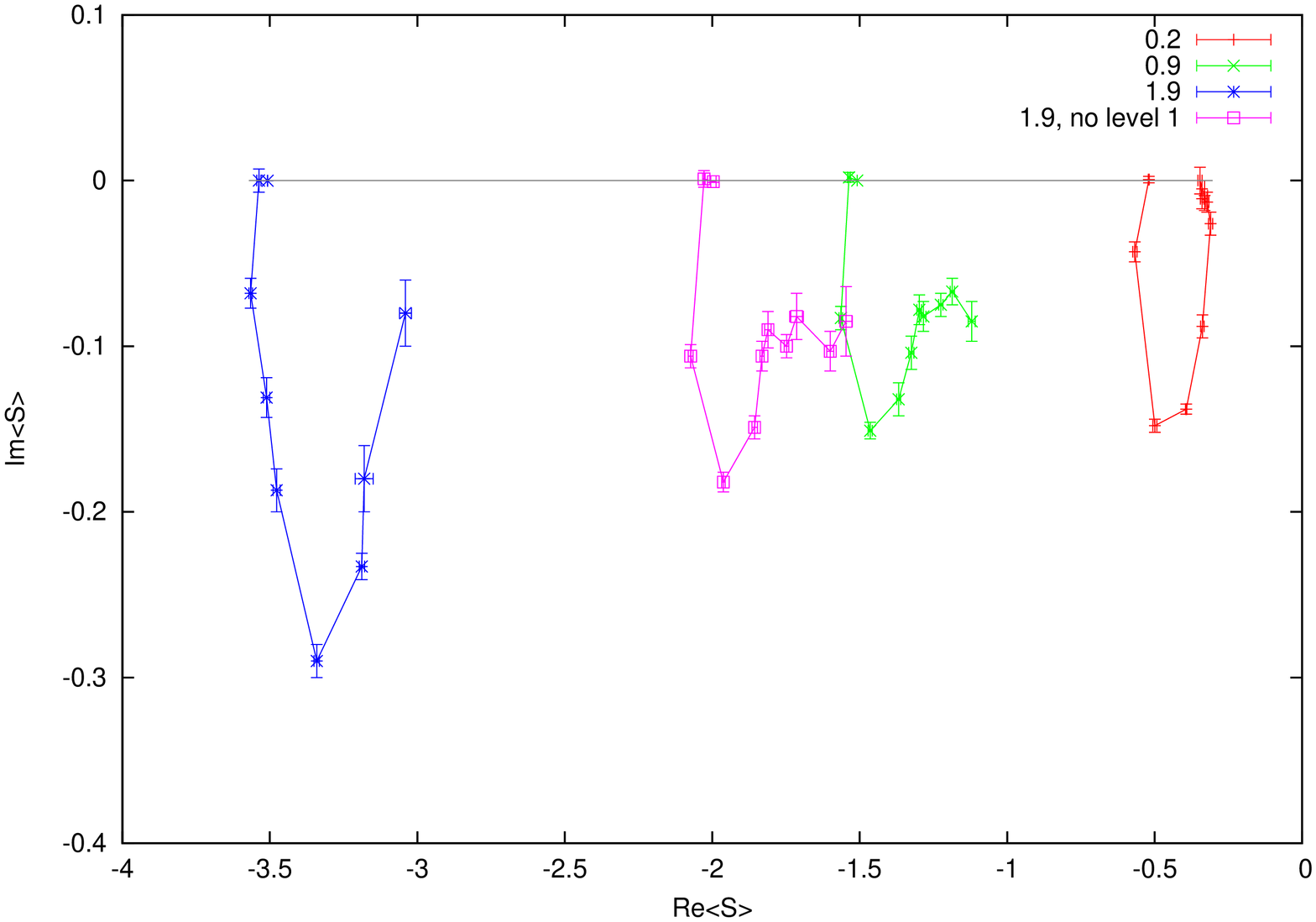,scale=0.31,angle=-90,clip}  
 			\footnotesize $\phantom{XXX}\mbox{(b) Trajectory of $\langle S \rangle$ in the complex plane.}\\
			               \phantom{XXX}\mbox{In all cases increasing $x_{min}$ corresponds to}\\
			               \phantom{XXX}\mbox{anticlockwise movement along the trajectory.}$
       \end{subfigure}
        \caption{$\langle S \rangle$ for $L=10$ as a function of
    $x_{min}$ for various $l_{max}$.}
\label{l10varyxminactfig}
\end{figure}

We begin with our results for $\langle S \rangle$, which are plotted in
Fig.~\ref{l10varyxminactfig}. The main features are the same as for $L=20$, but now we can follow
them to stronger coupling, giving us more confidence in what happens
there. It is clearer that $\langle S \rangle$ is returning to the real
axis for all $l_{max}$. The cusps in the complex plane are still
there, and the single-mode case (in this case $l_{max}=0.2$) again becomes
unstable at larger $x_{min}$ than the others, although the difference
seems smaller this time.

\begin{figure}
  \centering
  \epsfig{file=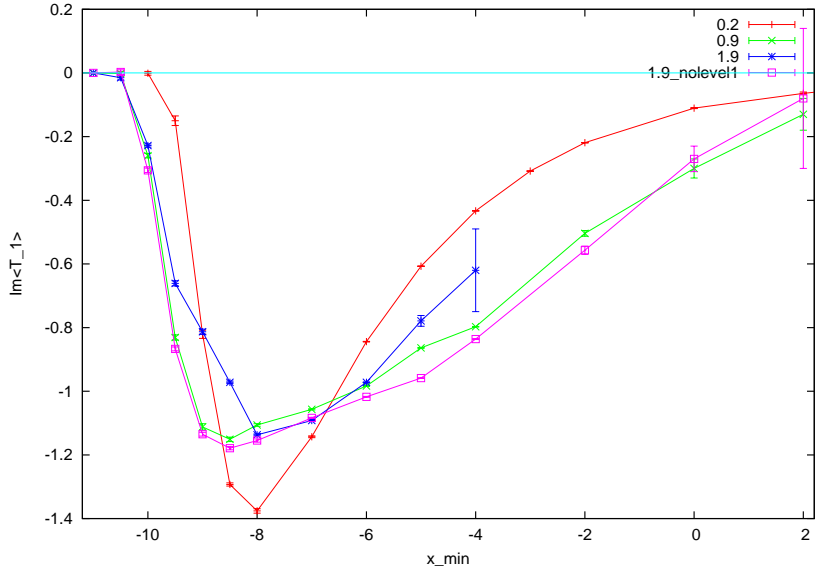,scale=0.43,angle=-90,clip}  
  \caption{$\Im\langle T_1 \rangle$ for $L=10$ as a function of
    $x_{min}$ for various $l_{max}$.}
  \label{l10varyxmint1fig}
\end{figure}

For $\Im\langle T_1 \rangle$, plotted in Fig.~\ref{l10varyxmint1fig},
we can follow the behaviour almost back to $\Im\langle T_1 \rangle=0$
for some $l_{max}$. In general, the behaviour is similar to that for
$L=20$. The oscillations are still present, for example
for $l_{max}=0.9$, though they are harder to see as we have not sampled
so densely in $x_{min}$. 

\subsection{Small interval length}
\label{smalllsec}

Some of the results in the sections above are rather complicated, due
to the presence of many modes. This is probably necessary to reach the
large-volume and continuum limits. However, it may also be useful to
look at a small number of modes to try to interpolate between the
well-understood single mode case (see section~\ref{singlemodesec}) and
the more complicated cases with many modes. Another advantage is that
by keeping the number of modes small we can go to stronger couplings.

In this section we keep the number of modes small by taking the
interval length $L$ small, specifically 6 or less.
Note that
the smallest value of $L$ we can take
while keeping at least one mode below level 2 is $L=\pi/\sqrt{2}\approx
2.22$, and that for $L < \pi$ we have only a single mode. 

Another reason to look at small $L$ is that it may be possible that
the continuum and large-volume limits are tied together, such that we
need to take $L$ and $l_{max}$ to infinity while keeping their ratio
$l_{max}/L$ fixed.
Such a requirement seems natural, for example, from the perspective of T-duality.
Since we are restricted to $l_{max} < 2$ we may
need to take $L$ small
too, to
keep this ratio at least moderately large.

\subsubsection{$L=6$}

\begin{figure}
  \centering
        \hspace{-9 ex}
        \begin{subfigure}{0.45\textwidth}
                \centering
                \epsfig{file=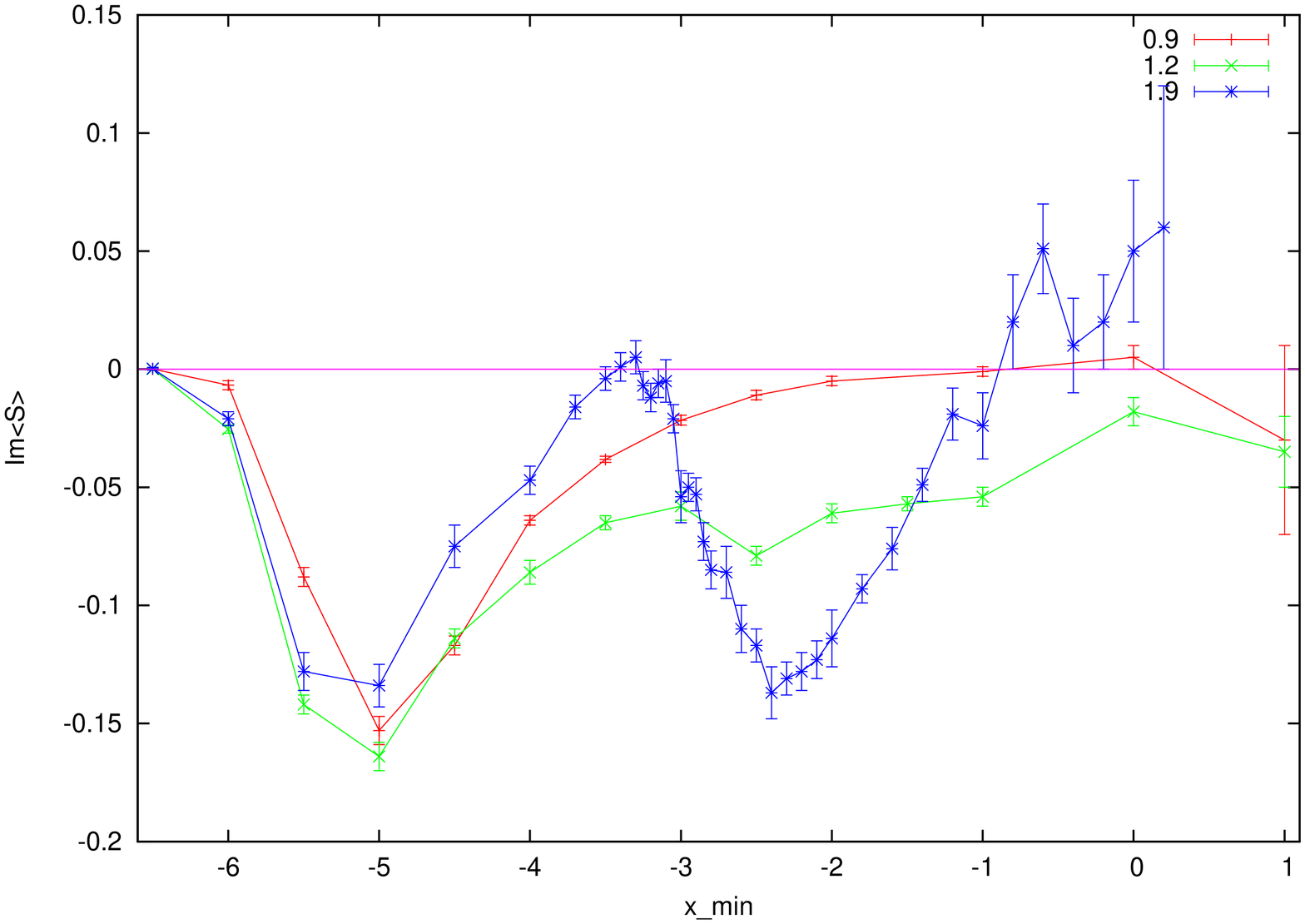,scale=0.31,angle=-90,clip}  
								\footnotesize (a) $\Im\langle S \rangle$\\ $\phantom{a}$\vspace{-.3mm}\\ $\phantom{X_{X_X}}$
        \end{subfigure}
        \hspace{4 ex}
        \begin{subfigure}{0.45\textwidth}
                \epsfig{file=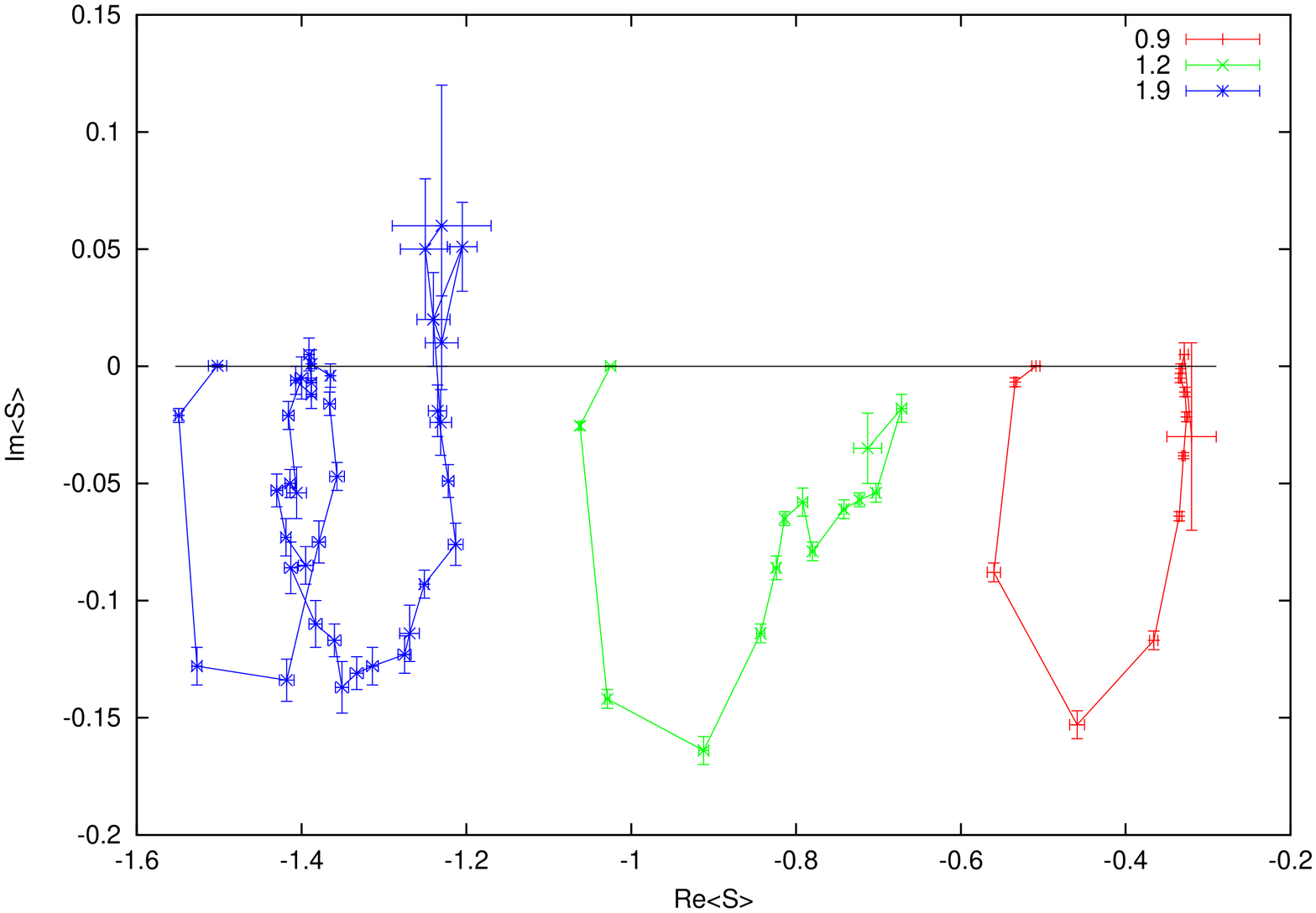,scale=0.31,angle=-90,clip}
			\footnotesize $\phantom{XXX}\mbox{(b) Trajectory of $\langle S \rangle$ in the complex plane.}\\
			               \phantom{XXX}\mbox{In all cases increasing $x_{min}$ corresponds to}\\
			               \phantom{XXX}\mbox{anticlockwise movement along the trajectory.}$
        \end{subfigure}
\caption{$\langle S \rangle$ for $L=6$ as a function of $x_{min}$ for various $l_{max}$.}
\label{l6varyxminactfig}
\end{figure}

\begin{figure}
  \centering
        \hspace{-9 ex}
        \begin{subfigure}{0.45\textwidth}
                \centering
                \epsfig{file=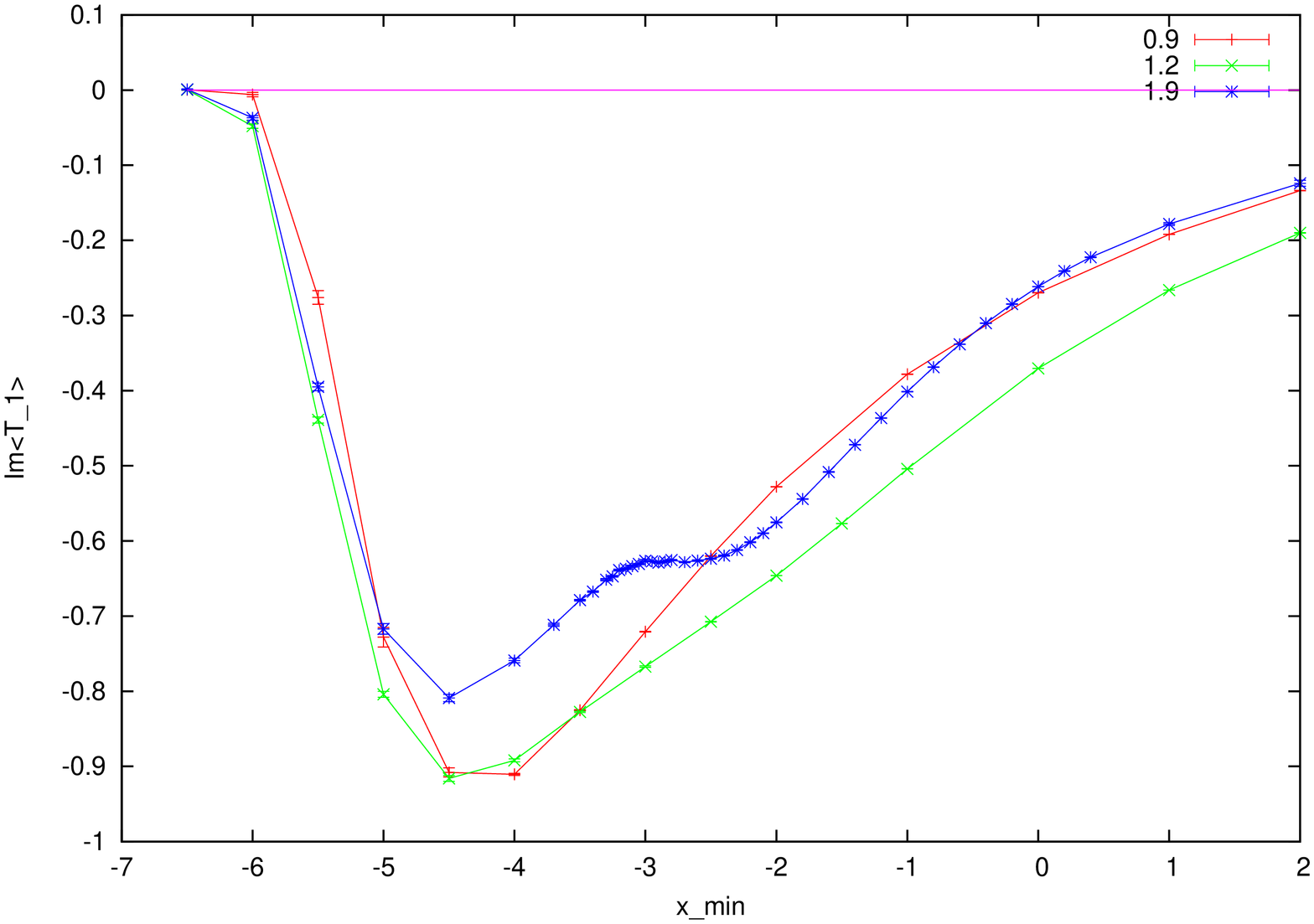,scale=0.31,angle=-90,clip}  
			\footnotesize (a) $\Im\langle T_1 \rangle$\\ $\phantom{a}$\vspace{-.3mm}\\ $\phantom{X_{X_X}}$
        \end{subfigure}
        \hspace{4 ex}
        \begin{subfigure}{0.45\textwidth}
                \epsfig{file=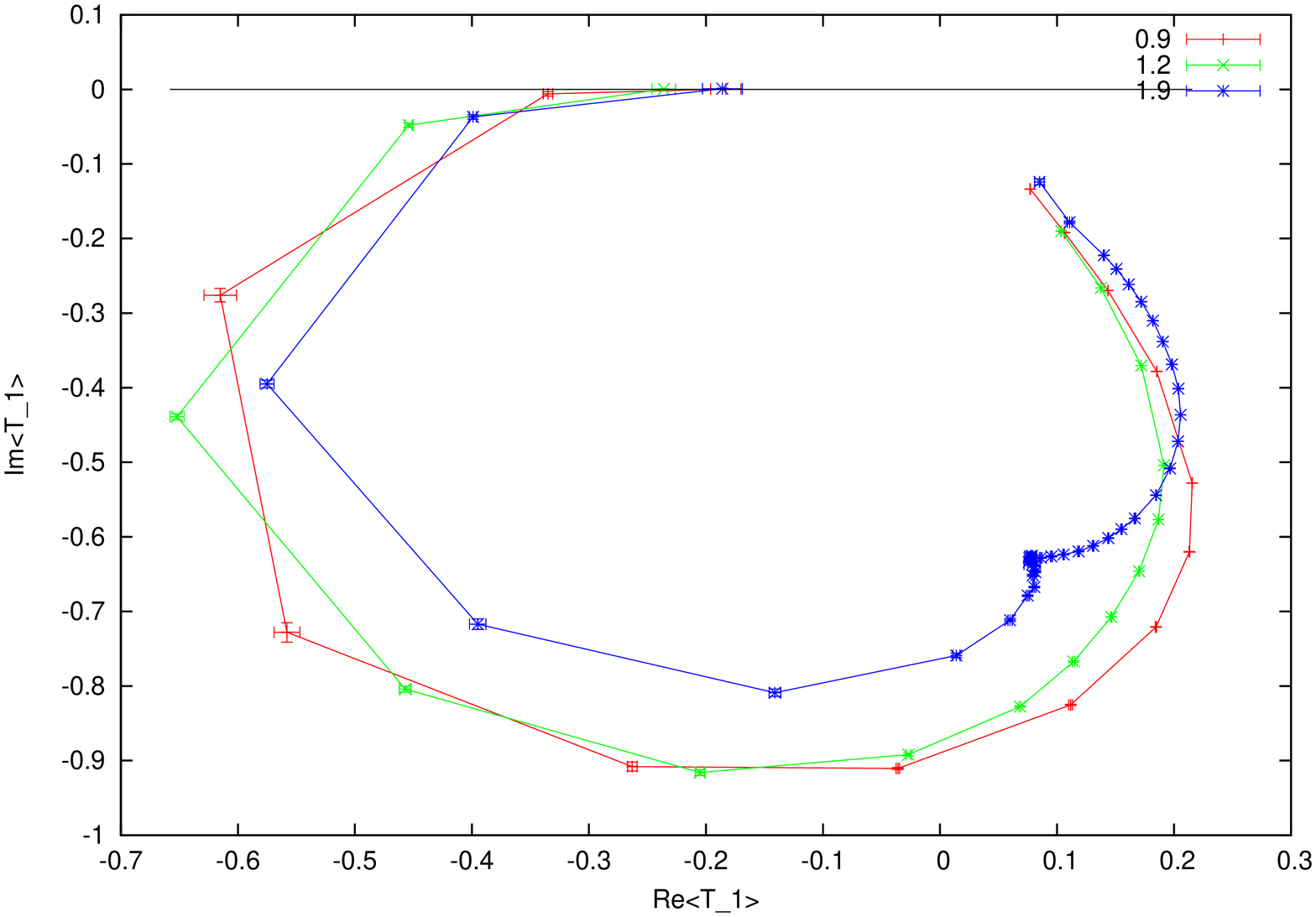,scale=0.31,angle=-90,clip}  
			\footnotesize $\phantom{XXX}\mbox{(b) Trajectory of $\langle T_1 \rangle$ in the complex plane.}\\
			               \phantom{XXX}\mbox{In all cases increasing $x_{min}$ corresponds to}\\
			               \phantom{XXX}\mbox{anticlockwise movement along the trajectory.}$
        \end{subfigure}
        \caption{$\langle T_1 \rangle$ for $L=6$ as a function of
    $x_{min}$ for various $l_{max}$.}
\label{l6varyxmincompt1fig}
\end{figure}

For $L=6$ the maximum number of modes for $l_{max}$ below 2 is
two for the $T$ field and one for the $A$ field. Since the number of modes
is small, we can look at all of them without the plots becoming excessively
complicated. We first plot $\langle S \rangle$ and $\langle T_1 \rangle$ 
in Figs.~\ref{l6varyxminactfig} and~\ref{l6varyxmincompt1fig} respectively.


The case $l_{max}=0.9$ is the single-mode case, and we see this behaves as
usual (see section~\ref{singlemodesec}). However the behaviour for
$l_{max}=1.2$ and in particular $l_{max}=1.9$ is very
different from
the behaviour seen at larger $L$. There are very large oscillations in
$\Im\langle S \rangle$, and an inflection point in $\Im\langle T_1
\rangle$.
From the right-hand panels of these figures we see that the oscillation
in $\Im\langle S \rangle$ is actually a loop in the complex plane, and
the inflection point in $\Im\langle T_1
\rangle$ is actually a cusp.

The remaining observables\footnote{There is also $A_1$, but this is always
consistent with zero.
We could have studied also expectation values of higher powers of the fields,
but didn't do that for simplicity.} are $T_2$, $T_1^2$, $T_2^2$, and
$A_1^2$ (the superscripts are powers).
These give a somewhat mixed picture: the trajectories in the complex plane
look smooth for $T_2$ and $A_1^2$, but have cusps for $T_1^2$ and
$T_2^2$. In general, they become cuspier as $l_{max}$ increases.

We do not understand what the cause of these loops and cusps is.
One odd feature is that the cusps do not appear in all variables at
the same $x_{min}$. For example, for $l_{max}=1.9$, the cusp in $T_1$
is around $x_{min}=-2.9$, and $T_2^2$ is smooth in this region, and 
the cusp in $T_2^2$ is around $x_{min}=-4.0$, where $T_1$ is
smooth. This possibly seems to indicate that they do not have a single cause.

Another issue is whether these features are present at larger $L$.
It is in fact quite likely that they are present, but that they
are more difficult to see. This is for two reasons: firstly the large
errors will
obscure
small features in the complex plane, and secondly
it is not possible to go to as strong couplings, where the features
seem to show up. In fact there are some indications for features like
this at larger $L$ in some of the plots, e.g. in
Fig~\ref{l10varyxminactfig}, and the oscillations in
Fig.~\ref{l20varyxmint1fig}.

For large $x_{min}$ nearly all the observables approach the
origin. The exception is the action, where we already saw for the case
of a single mode that the action went to a finite value despite the
fact that $T^2$ and $T^3$ go to zero.

However, it is still interesting to ask how the observables approach
the origin. In particular, do they approach along the real axis? If so
this would be a good sign, since it would mean that as the origin is approached
the ratio of imaginary part and the real part tends to zero. This is not necessary,
but could be an indication that we really approach a good limit,
since for expectation values that approach a non-zero real limit this
ratio tends to zero regardless of the angle of approach.
However, this does not seem to happen
for any of the quantities we measure; they all approach the origin from a
complex direction. There is also no trend of this getting better as
$l_{max}$ increases.

\subsubsection{$L=4$}

In this case the maximum number of modes is two; in comparison to
$L=6$ we have lost the $T_2$ ``tachyon'' mode. Any significant
differences between $L=4$ and $L=6$ could thus be due to having more
than one mode of the $T$ field.

\begin{figure}
  \centering
        \hspace{-9 ex}
        \begin{subfigure}{0.45\textwidth}
                \centering
                \epsfig{file=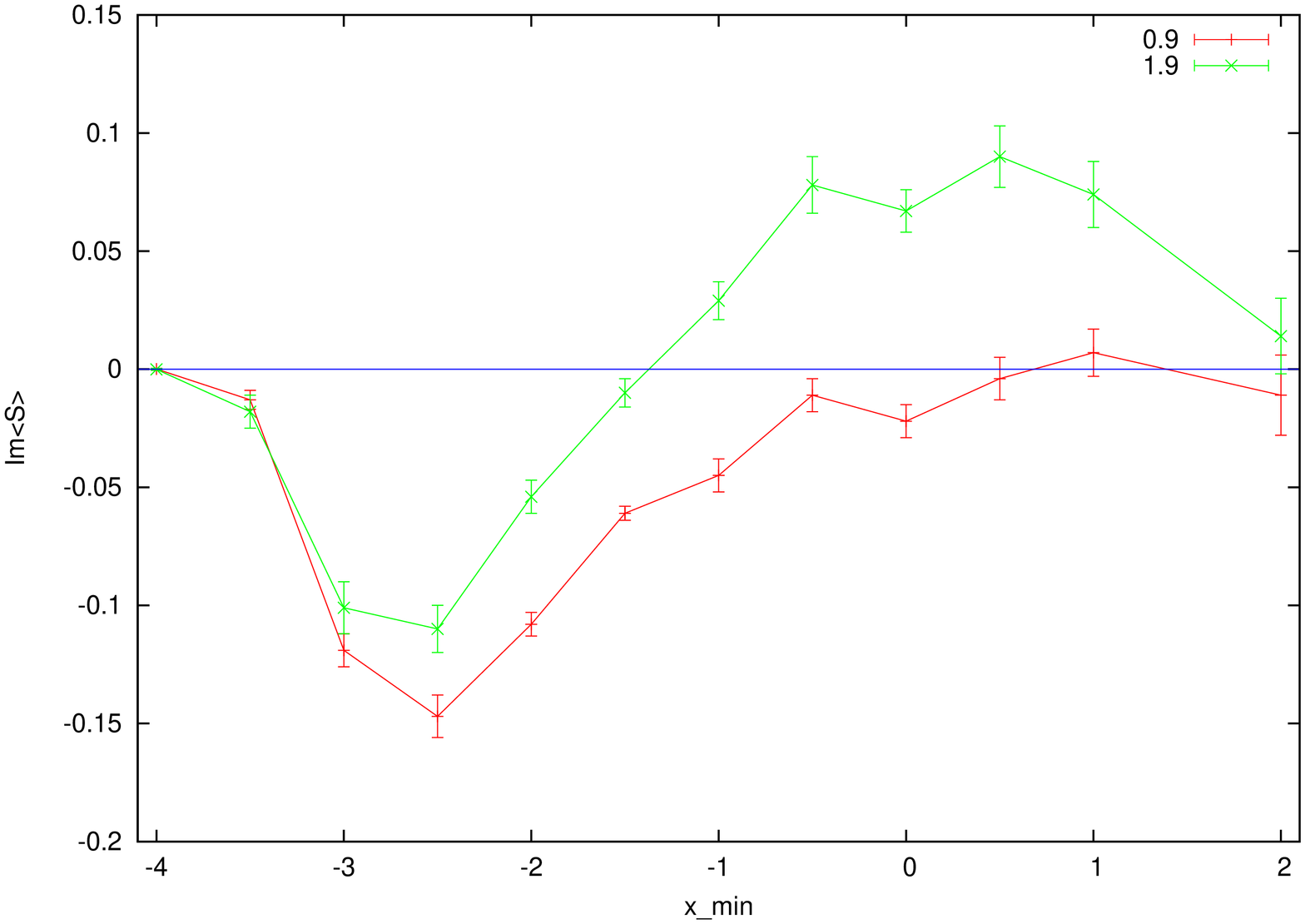,scale=0.31,angle=-90,clip}  
								\footnotesize (a) $\Im\langle S \rangle$\\ $\phantom{a}$\vspace{-.3mm}\\ $\phantom{X_{X_X}}$
        \end{subfigure}
        \hspace{4 ex}
        \begin{subfigure}{0.45\textwidth}
                \epsfig{file=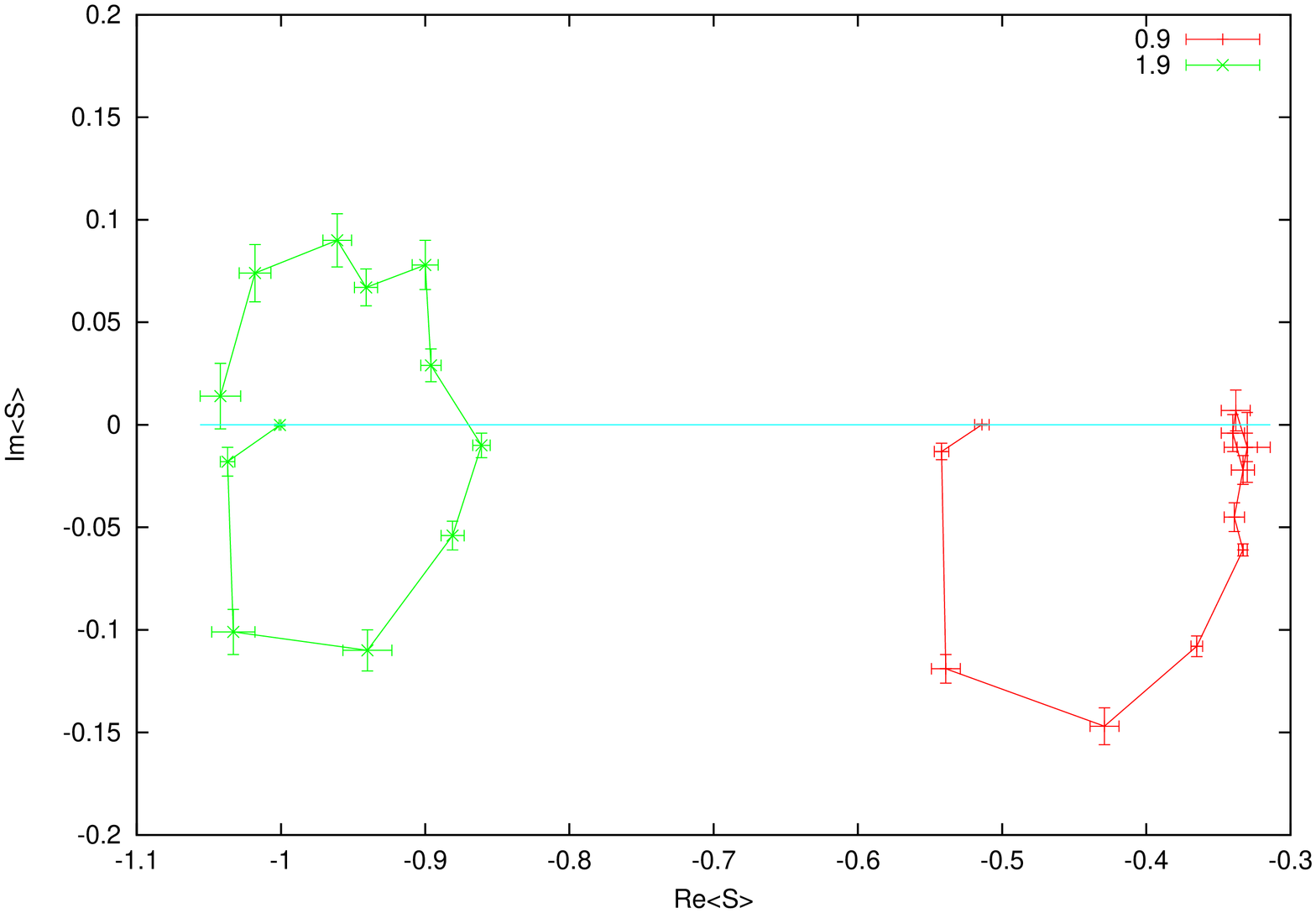,scale=0.31,angle=-90,clip}  
			\footnotesize $\phantom{XXX}\mbox{(b) Trajectory of $\langle S \rangle$ in the complex plane.}\\
			               \phantom{XXX}\mbox{In all cases increasing $x_{min}$ corresponds to}\\
			               \phantom{XXX}\mbox{anticlockwise movement along the trajectory.}$
        \end{subfigure}
        \caption{$\langle S \rangle$ for $L=4$ as a function of
    $x_{min}$ for various $l_{max}$.}
\label{l4varyxminactfig}
\end{figure}

\begin{figure}
  \centering
        \hspace{-9 ex}
        \begin{subfigure}{0.45\textwidth}
                \centering
                \epsfig{file=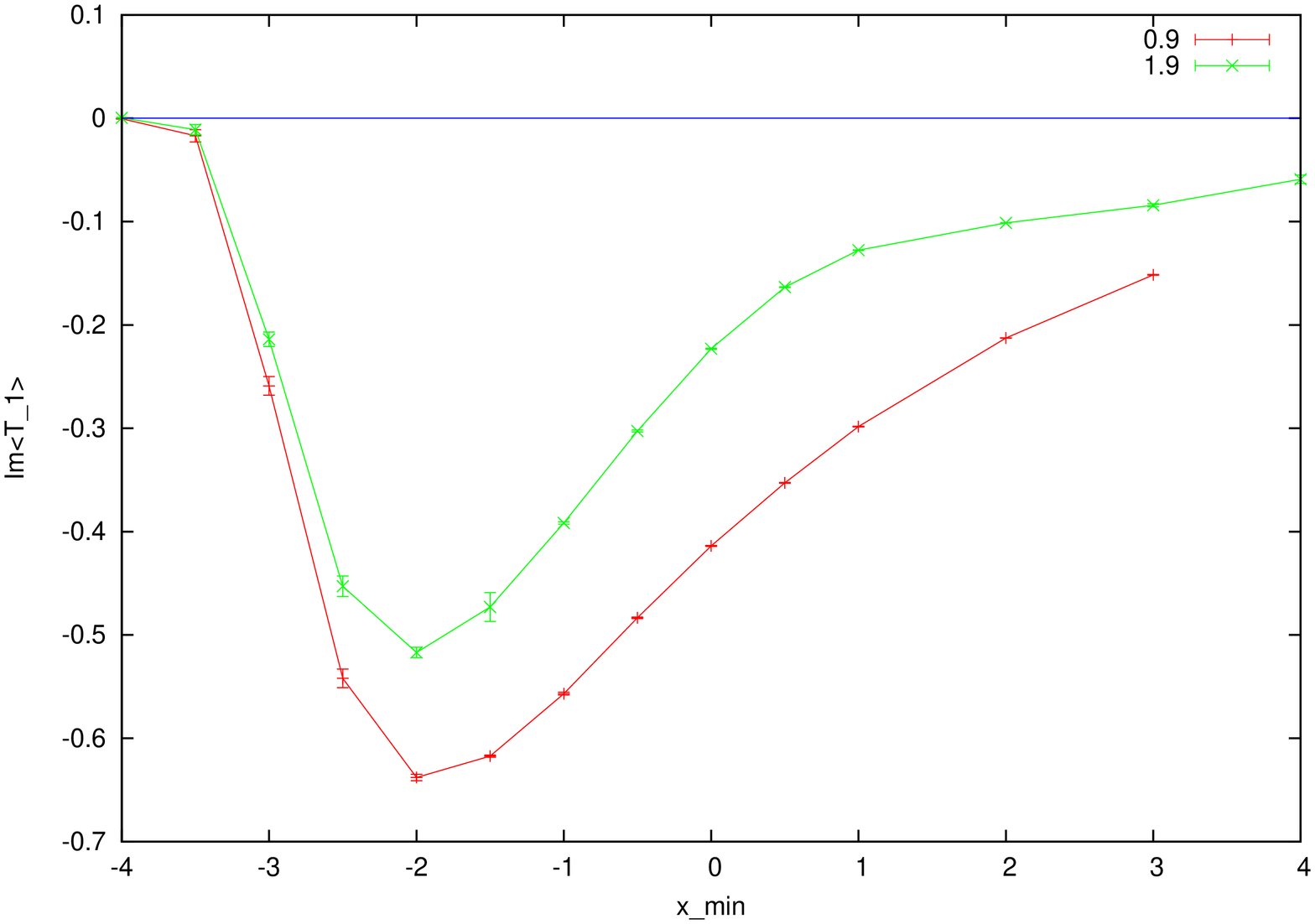,scale=0.31,angle=-90,clip}  
								\footnotesize (a) $\Im\langle T_1 \rangle$\\ $\phantom{a}$\vspace{-.3mm}\\ $\phantom{X_{X_X}}$
        \end{subfigure}
        \hspace{4 ex}
        \begin{subfigure}{0.45\textwidth}
                \epsfig{file=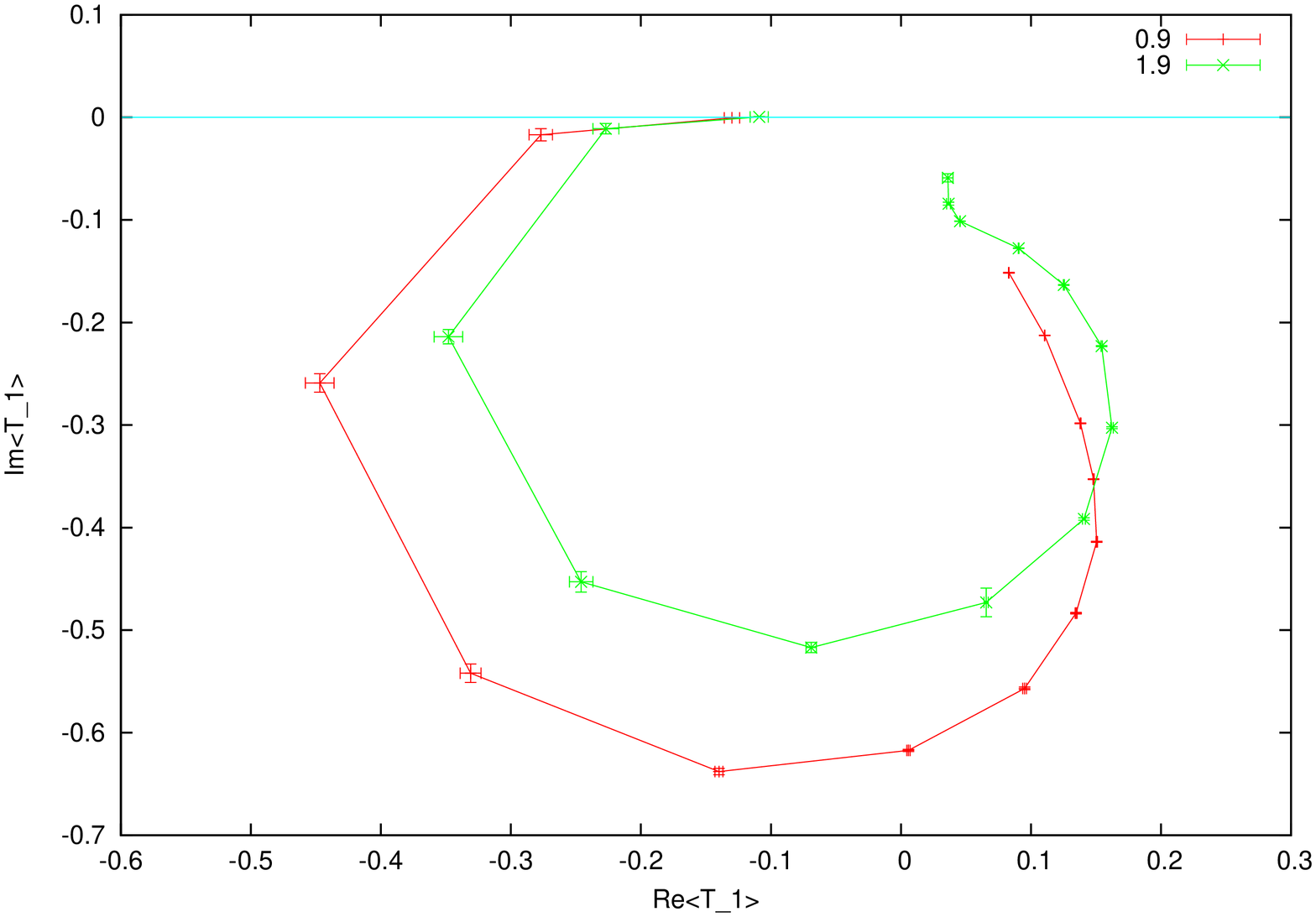,scale=0.31,angle=-90,clip}  
  		\footnotesize $\phantom{XXX}\mbox{(b) Trajectory of $\langle T_1 \rangle$ in the complex plane.}\\
			               \phantom{XXX}\mbox{In all cases increasing $x_{min}$ corresponds to}\\
			               \phantom{XXX}\mbox{anticlockwise movement along the trajectory.}$
       \end{subfigure}
        \caption{$\langle T_1 \rangle$ for $L=4$ as a function of
    $x_{min}$ for various $l_{max}$.}
\label{l4varyxmint1fig}
\end{figure}

We show $\langle S \rangle$ and $\langle T_1 \rangle$,
in Figs.~\ref{l4varyxminactfig} and~\ref{l4varyxmint1fig} respectively.
We see a change in behaviour: $\Im\langle S \rangle$ becomes positive for $l_{max}=1.9$
and large $x_{min}$, corresponding to a large loop
in the complex plane. Also $\langle T_1 \rangle$ has a rather sharp
curve, although not a cusp, for $l_{max}=1.9$. Note that $\langle S
\rangle$ has only a single loop, whereas for $L=6$ it had two; it is
tempting to speculate that this is connected to the number of modes of the
$T$ field in each case.

The remaining observables, that is $T_1^2$ and $A_1^2$,
behave similarly to $L=6$.
Taking all the observables together, the
differences from
$L=6$ are not
very great. In particular, the sizes of the imaginary parts are
similar, and there is no change in the tendency to approach the origin
from a complex direction (as opposed to along the real axis). The
biggest difference is probably that the cusps, loops etc. are more
pronounced and more complicated for $L=6$.

\subsubsection{$L=3.15$}
Finally,
we report results for $L=3.15$. This is almost the smallest
interval we can have while keeping more than one mode --- the second
mode would go above level 2 at $L=\pi$. The modes
present are in fact the same as at $L=4$; it is just their masses and
the cubic terms that are different. For example, evaluating the masses explicitly,
we find for $L=4$
\begin{eqnarray}
m_{T_1} &=&0.812\ldots \\
m_{A_1} &=&1.288\ldots,
\end{eqnarray}
and for $L=3.15$
\begin{eqnarray}
m_{T_1} &=&1.018\ldots \\
m_{A_1} &=&1.427\ldots.
\end{eqnarray}
So as we decrease $L$ the masses increase, while coming closer together.

\begin{figure}
  \centering
  \epsfig{file=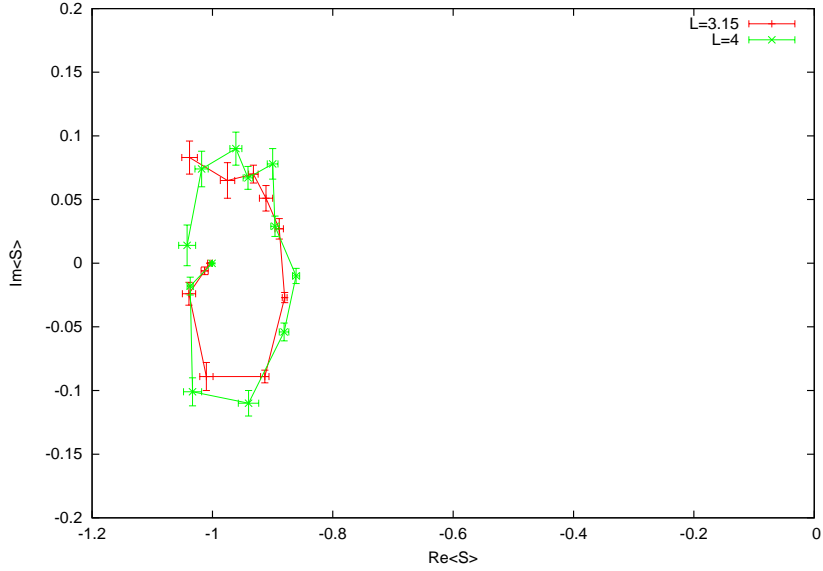,scale=0.43,angle=-90,clip}  
  \caption{Trajectory of $\langle S \rangle$
	in the complex plane for $L=4$ and
    $L=3.15$, both for
    $l_{max}=1.999$. Increasing
    $x_{min}$ corresponds to anticlockwise movement along the trajectory.}
  \label{compare4vs3.15fig}
\end{figure}

We find that in general the results are rather similar to those for
$L=4$. As an example we show the action in the complex plane in
Fig.~\ref{compare4vs3.15fig}, where clearly the differences are rather
small. We find similar results for the other observables.
This suggests that the changes in the number of modes, rather than the
changes in the parameters in the action, are more important. Also,
this is the largest $l_{max}/L$ we are able to reach, and nothing
very helpful seems to happen.

\subsection{Comparing different interval lengths}

In this section we do not present any results not already mentioned
above. Instead we show some plots comparing results at different
$L$, mostly showing that there are no strong trends in $L$.
In particular, there is not much sign of instabilities reducing as $L$
increases. This would be seen by either the imaginary parts of
observables getting smaller, or by the approach to the origin at large
$x_{max}$ happening at a smaller angle to the real axis (or both).

When comparing different $L$, it is not always clear what are the
equivalent things to compare. For example, if $L$ is doubled, should
the mode $T_1$ be now compared to $T_2$, which has the same
wavelength? Or should it still be compared to $T_1$, which is the
lowest mode? We will mention issues of this sort as they become relevant below.

\begin{figure}
  \centering
        \hspace{-9 ex}
        \begin{subfigure}{0.45\textwidth}
                \centering
                \epsfig{file=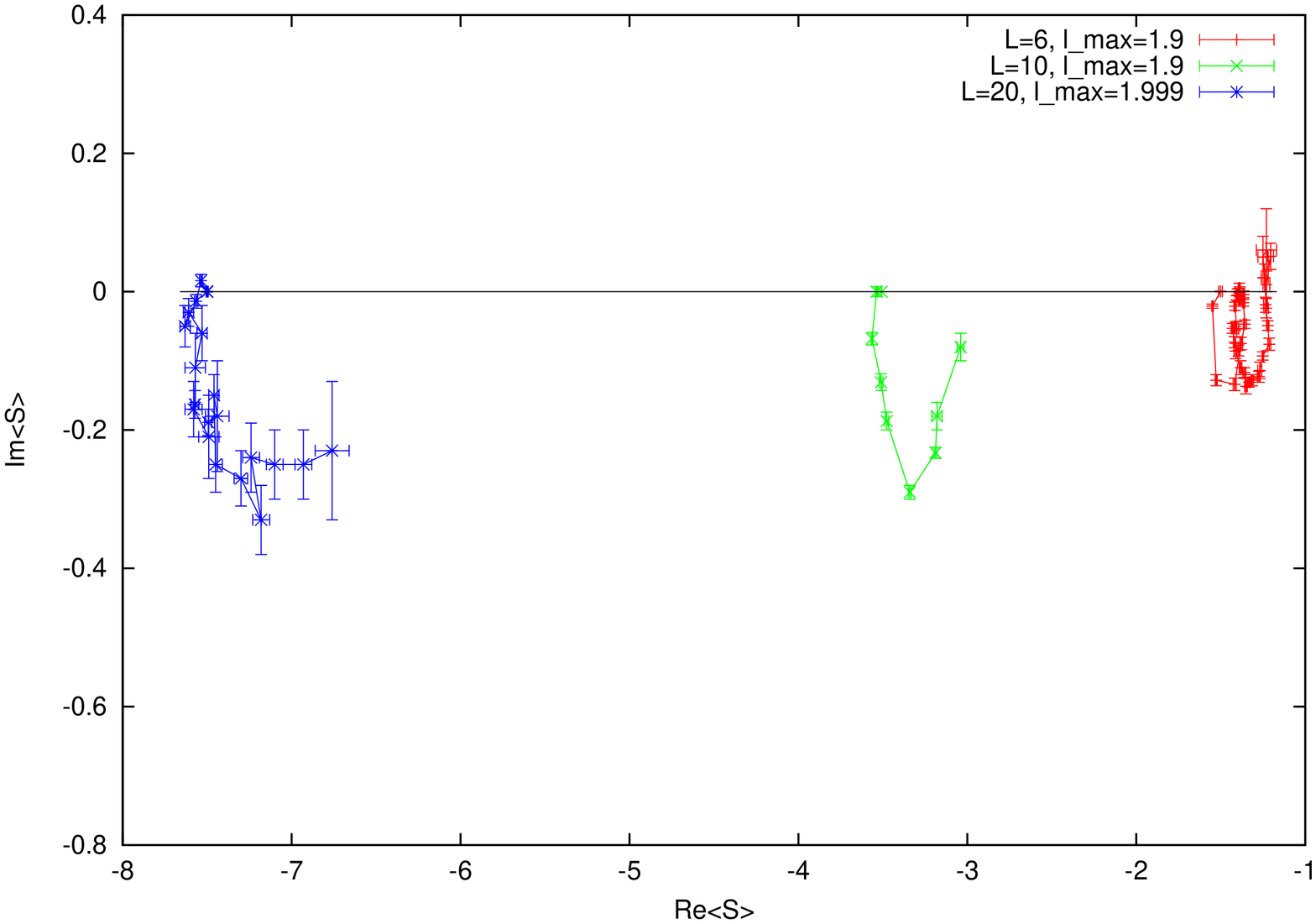,scale=0.31,angle=-90,clip}  
                \caption{$l_{max}=1.999$}
        \end{subfigure}
        \hspace{4 ex}
        \begin{subfigure}{0.45\textwidth}
                \centering
                \epsfig{file=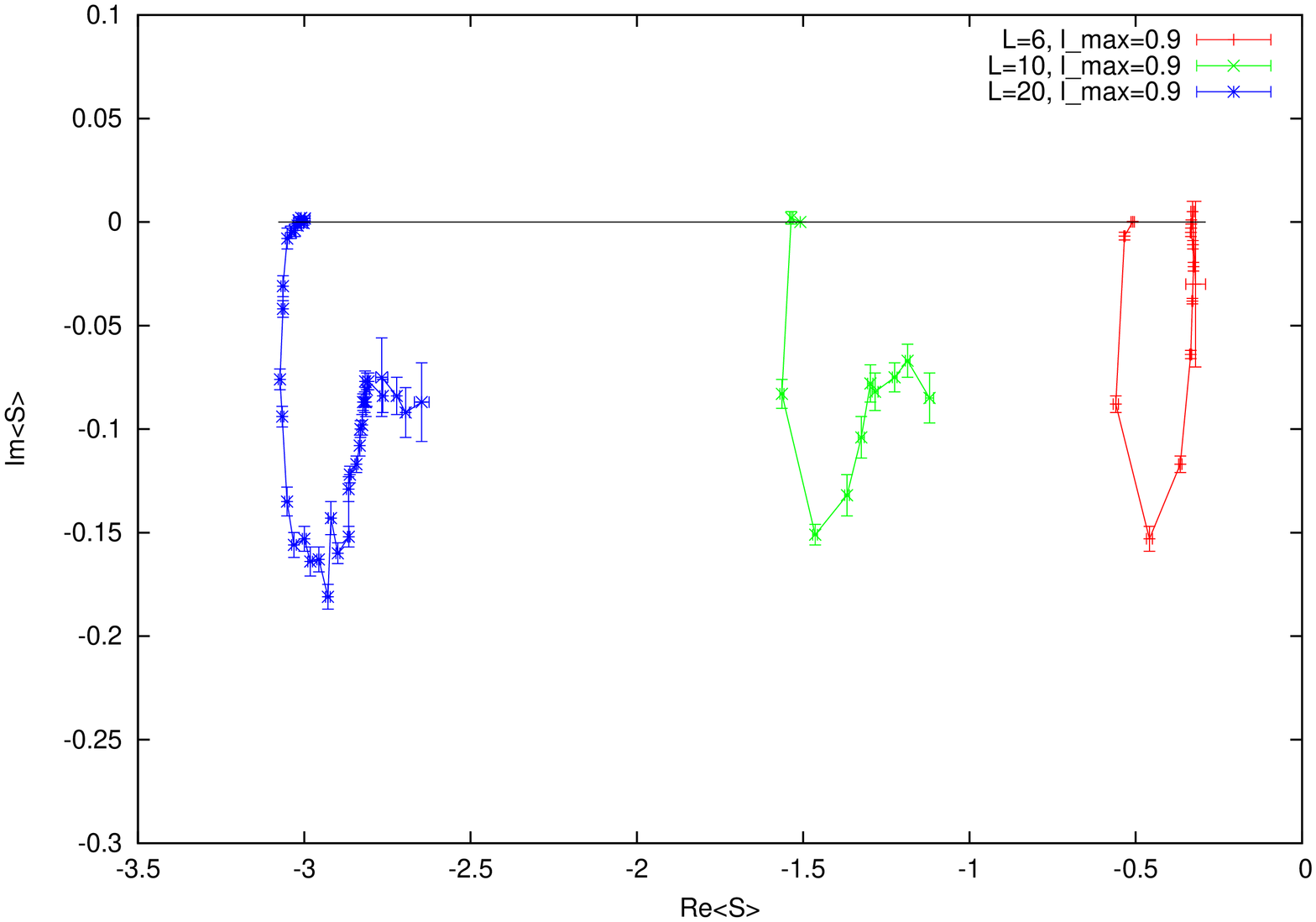,scale=0.31,angle=-90,clip}  
                \caption{$l_{max}=0.9$}
        \end{subfigure}
        \caption{Trajectory of $\langle S \rangle$
	in the complex plane for $L=6$ (red),
    $L=10$ (green), and $L=20$ (blue). Increasing
    $x_{min}$ corresponds to anticlockwise movement along the trajectory.}
\label{compareact3fig}
\end{figure}

First we compare $\langle S \rangle$ at two different values of $l_{max}$ for
$L=6$, $10$ and $20$, in Fig.~\ref{compareact3fig}. We see that in both cases there is a
very large shift of $\langle S \rangle$ for the different
$L$. However, most of this is just due to the fact that the number of
modes is increasing and the free
action is $-1/2$ per mode. Apart from this we see a small increase in
the size of $\Im \langle S \rangle$. There is also perhaps a tendency for
the trajectories to become more complicated for higher $L$.

\begin{figure}
  \centering
  \epsfig{file=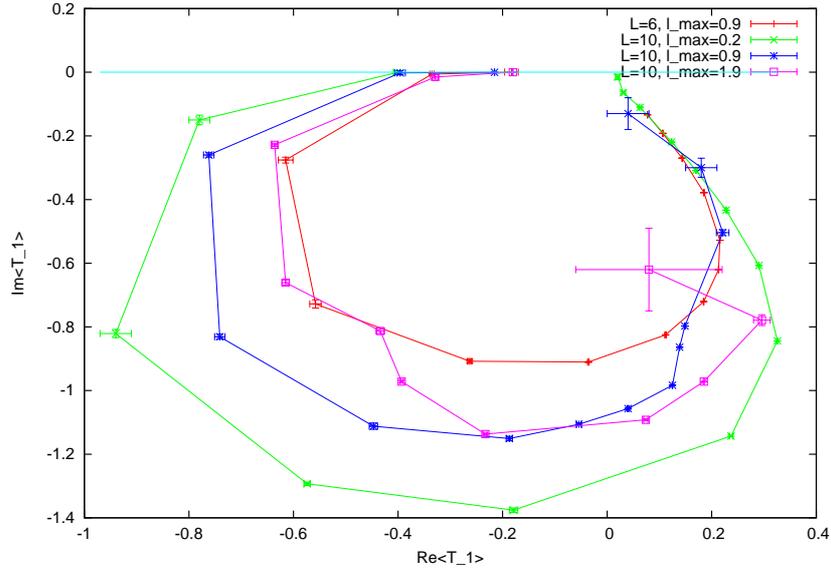,scale=0.43,angle=-90,clip}  
  \caption{Trajectory of $\langle T_1 \rangle$
in the complex plane for $L=6$,
    $l_{max}=0.9$ (red), and for $L=10$ with $l_{max}=0.2$ (green),
    $l_{max}=0.9$ (blue), and $l_{max}=1.9$ (magenta). Increasing
    $x_{min}$ corresponds to anticlockwise movement along the trajectory.}
  \label{comparet1fig}
\end{figure}

We now turn to comparisons of the field $T$. First we consider the
lowest mode $T_1$, for which we show a plot in Fig.~\ref{comparet1fig}.
The cases $L=6$, $l_{max}=0.9$ and $L=10$, $l_{max}=0.2$ are both for
the single-mode case: the reason they differ is that the mass is
smaller for the latter case. The other trajectories show what happens
as $l_{max}$ is then increased for $L=10$. We see that both the real
and imaginary parts of $T_1$ decrease, and the trajectory also becomes
more complicated.

\begin{figure}
  \centering
  \epsfig{file=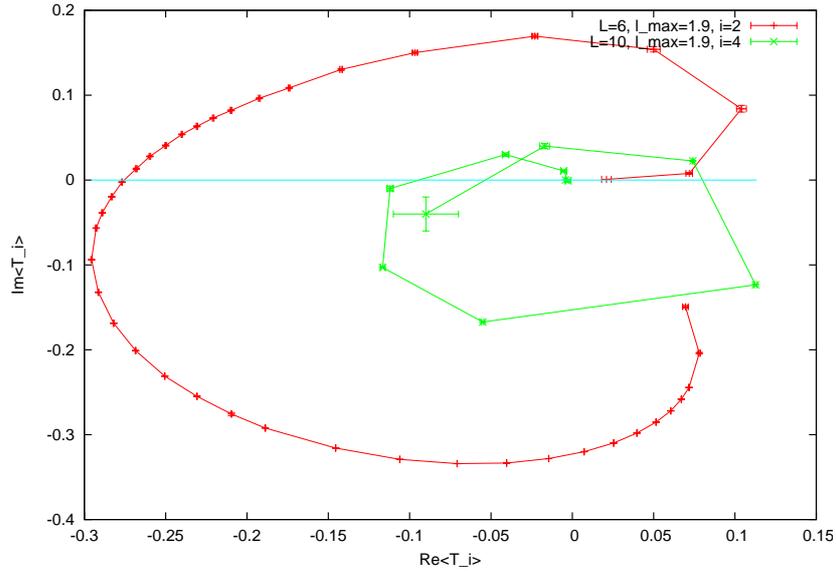,scale=0.43,angle=-90,clip}  
  \caption{Trajectory of $\langle T_2 \rangle$
	  in the complex plane for $L=6$ (red),
    and of $\langle T_4 \rangle$
	  for $L=10$ (green), both for
    $l_{max}=1.9$. Increasing
    $x_{min}$ corresponds to anticlockwise movement along the trajectory.}
  \label{comparet2fig}
\end{figure}

Next we compare two modes with roughly the same wavelength, namely
$T_2$ at $L=6$ and $T_4$ at $L=10$, both for $l_{max}=1.9$. Their
trajectories in the complex plane are plotted in Fig.~\ref{comparet2fig}.
We see in this case that there is a large change in scale.
Apart from this we see that whereas the
$L=6$ trajectory has one loop, the trajectory for $L=10$ has at least
the start of a second.

\begin{figure}
  \centering
        \hspace{-9 ex}
        \begin{subfigure}{0.45\textwidth}
                \centering
                \epsfig{file=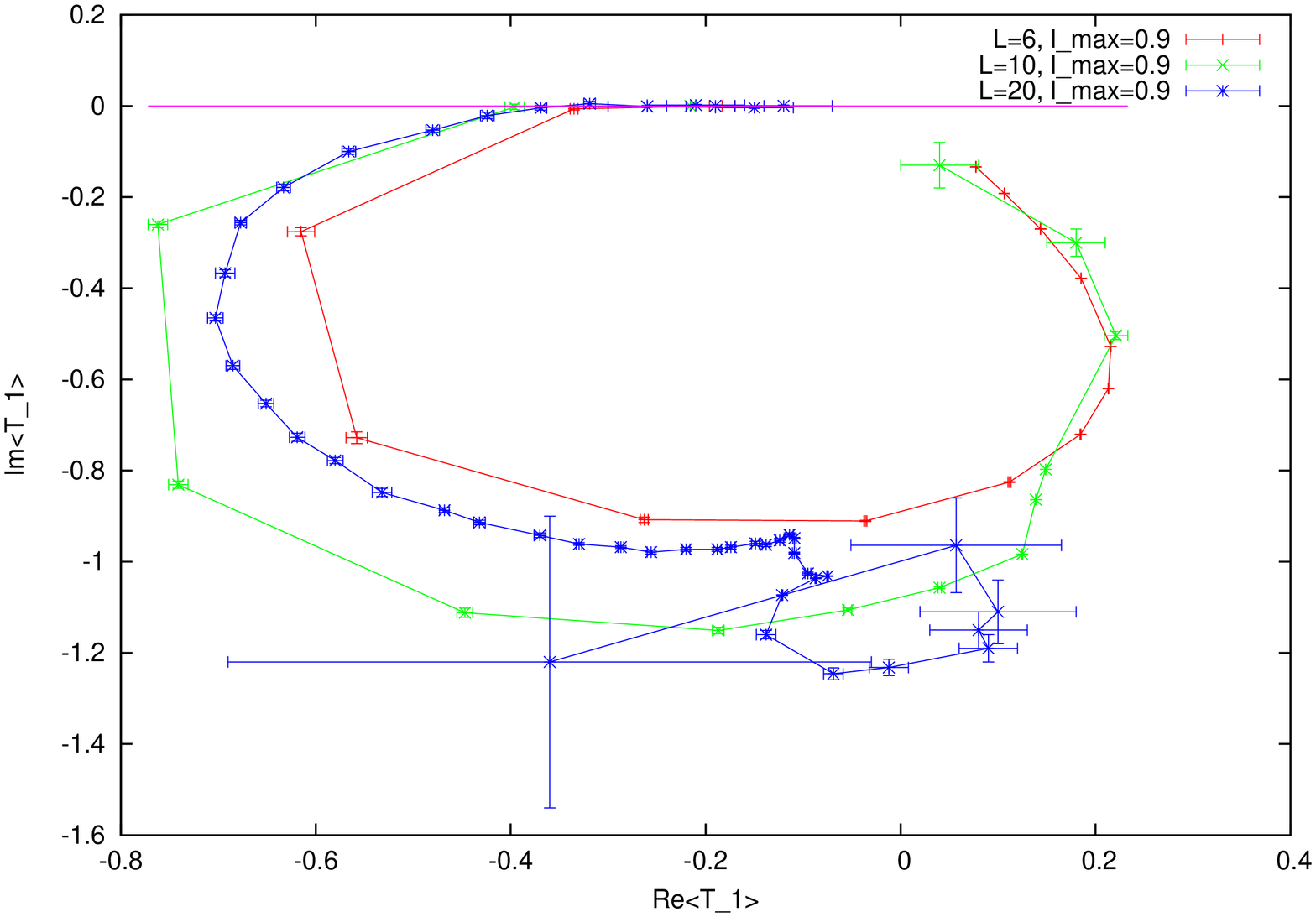,scale=0.31,angle=-90,clip}  
                \caption{$l_{max}=0.9$}
        \end{subfigure}
        \hspace{4 ex}
        \begin{subfigure}{0.45\textwidth}
                \centering
                \epsfig{file=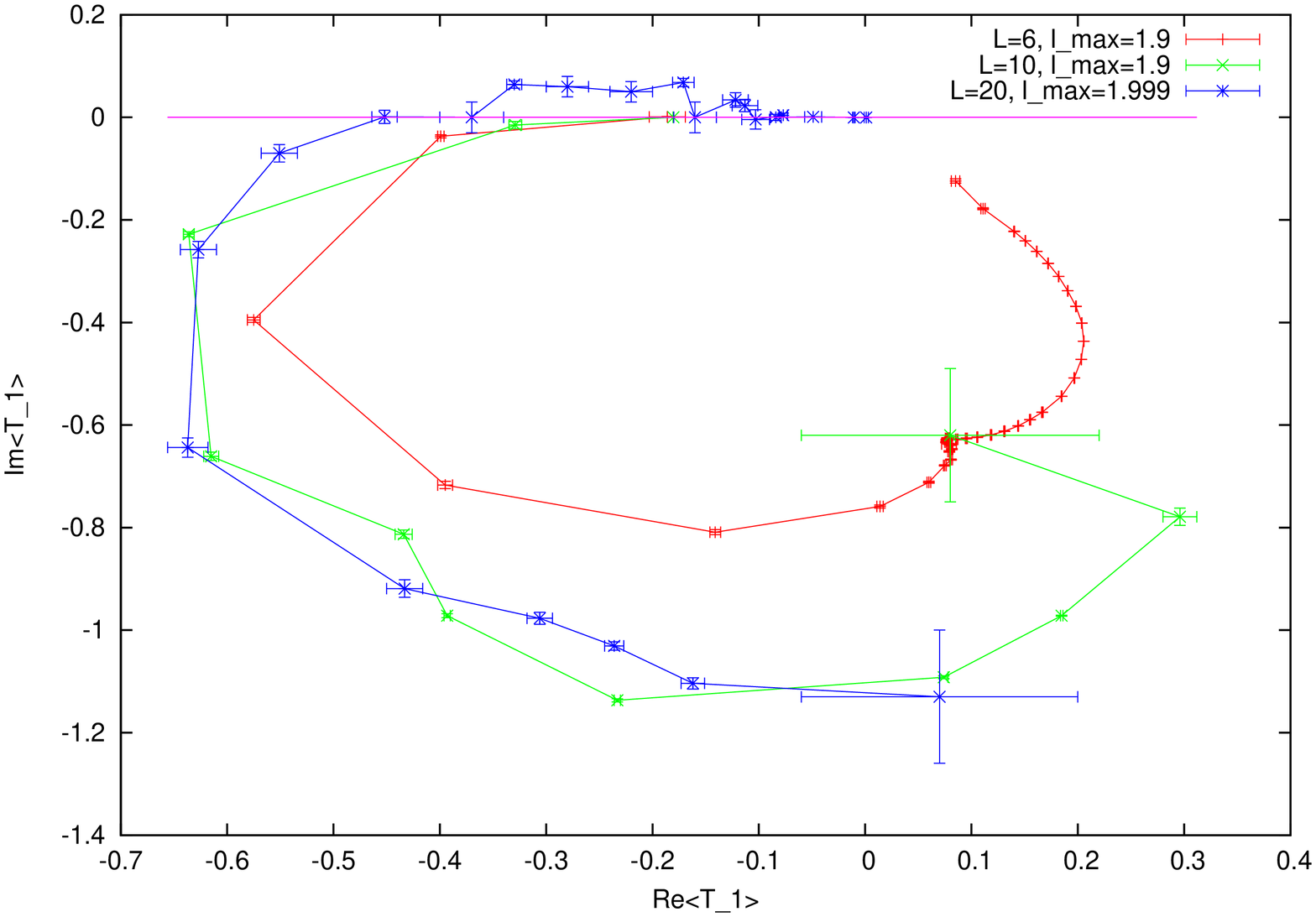,scale=0.31,angle=-90,clip}  
                \caption{$l_{max}=1.999$}
        \end{subfigure}
        \caption{Trajectory of $\langle T_1 \rangle$	
	in the complex plane for $L=6$ (red),
    $L=10$ (green), and $L=20$ (blue). Increasing
    $x_{min}$ corresponds to anticlockwise movement along the trajectory.}
\label{comparet3fig}
\end{figure}

Keeping instead the mode number fixed, we plot $\langle T_1 \rangle$ in
Fig.~\ref{comparet3fig} for $L=6$, $10$, and $20$. We see that now
there is quite good matching in the magnitudes of $\langle T_1
\rangle$. (The fact that the trajectory looks smoother for some cases is
simply because we have made measurements closer together in $x_{min}$.)
There is no sign of the imaginary parts decreasing, or of the approach
to the origin
occurring
at a smaller angle to the real axis. However,
we once again see that trajectory becomes more complicated as $L$
increases, with e.g. two cusps and a loop being present for $L=20$ with $l_{max}=0.9$.
It also appears that the behaviour for the two values of $l_{max}$
is similar, though unfortunately it is not possible to go to high enough
$x_{min}$ with $l_{max}=1.999$
to see well the approach to the origin.
For the same reason,
we cannot really see if the trajectories are becoming more complicated.

\begin{figure}
  \centering
  \epsfig{file=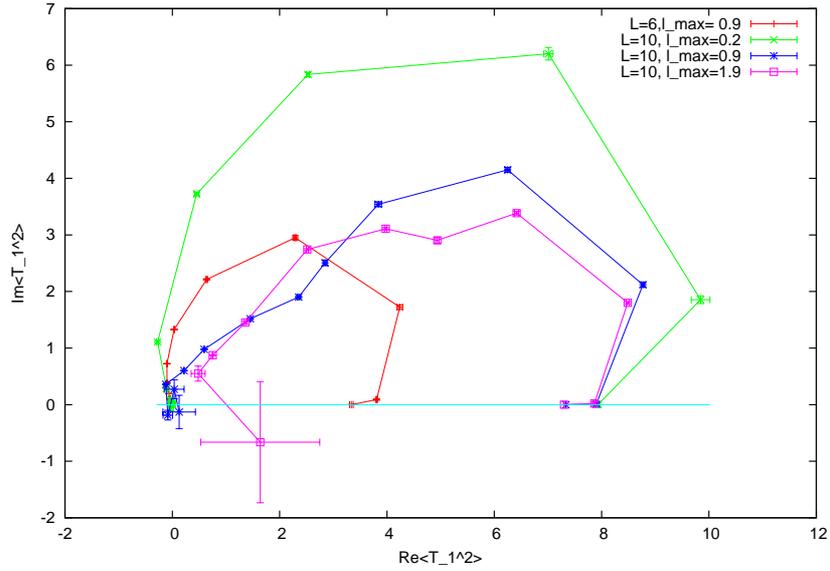,scale=0.43,angle=-90,clip}  
  \caption{As	Fig.~\ref{comparet1fig}, but for $\langle T_1^2 \rangle$
  rather than $\langle T_1 \rangle$.}
  \label{comparetsq1fig}
\end{figure}

We now turn to comparisons of the $\langle T_n^2 \rangle$. As an example we plot
the lowest mode $\langle T_1^2 \rangle$ in
Fig.~\ref{comparetsq1fig}. The cases $L=6$, $l_{max}=0.9$ and $L=10$,
$l_{max}=0.2$ in this plot both have only a single mode; the large
increase in $\langle T_1^2 \rangle$
is due to the mass decrease.
The other two trajectories
then
show what happens when $L$ is kept fixed
and $l_{max}$ is increased. We see that $\langle T_1^2 \rangle$
decreases, which roughly means that the fluctuations of $\langle T_1 \rangle$
decrease. Also the trajectories become more complicated,
although they do not seem to be as complicated as the trajectories of 
the $\langle T_n \rangle$ we saw above.

We find similar behaviour for the other modes
$\langle T_n^2 \rangle$ and also for $\langle A_n^2 \rangle$:
generally there are some complicated trajectories in the complex plane,
with in some cases cusps or loops appearing. There is no trend for
imaginary parts to get smaller or the approach to the origin to occur
at a smaller angle to the real axis as we increase $L$ or $l_{max}$.

\subsection{Relations among different expectation modes $\langle T_n \rangle$}

\begin{figure}
  \centering
  \psfig{file=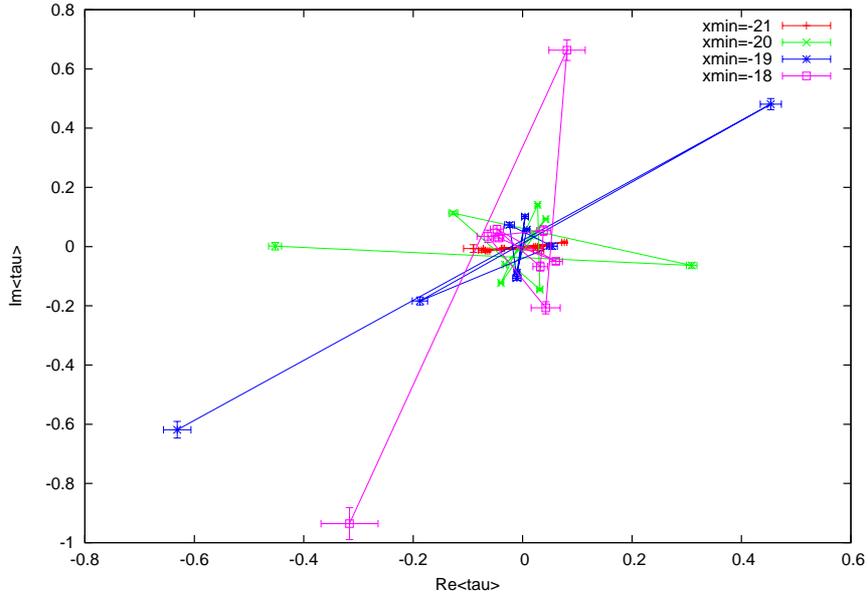,scale=0.45,angle=-90,clip}  
  \caption{The $\langle T_n \rangle$ in the complex plane for $L=20$,
    $l_{max}=1.999$, and various $x_{min}$. In each case $\langle T_1
    \rangle$ is the point furthest to the left/bottom.}
  \label{l20complextsfig}
\end{figure}

Apart from looking at how an individual mode, say $\langle T_1 \rangle$
moves in the
complex plane as $x_{min}$ increases, we can get additional information
by examining how all the modes move together. We show an example for
$L=20$ for a range of $x_{min}$ from $-21$ to $-18$, i.e. over the
transition from weak to strong coupling, with $l_{max}=1.999$,
in Fig.~\ref{l20complextsfig}. The behaviour appears to be quite
complicated, with the $\langle T_n \rangle$
all moving off the real axis as $x_{min}$
is increased, and then forming a complicated star-shaped pattern.
We see similar behaviour for other values of the parameters.

We do not understand the full behaviour here, but some features can be
understood. Firstly there is a rougly oscillatory behaviour, with
$\langle T_{n+1} \rangle$ generally being on the opposite side of the
origin to $\langle T_n \rangle$. This is presumably because the cubic
terms change sign when $n$ increases by one. This in turn is because
the cubic terms are proportional to integrals whose values are
dominated by contributions near $x_{max}$, and the sine-wave modes will
change sign in this region when $n$ increases by one.
Secondly there is a general decrease in magnitude as $n$
increases. This is presumably because the masses of the modes
increase, constraining them to be closer to the origin.

\subsubsection{$T(x)$ in position space}
\label{sec:positionSpace}

Another way to look at the behaviour of the $T_n$ is to recall that
they are the Fourier modes of the field $T$. Thus we can translate
back to position space to obtain the expectation value of $T(x)$:
\begin{equation}
\langle T(x) \rangle = \sqrt{\frac{2}{L}} \sum_{j=1}^{n_0} \langle T_j
\rangle \mathrm{sin} \left( \frac{j \pi (x-x_{min})}{L} \right)
\end{equation}
This will in general be complex of course.

\begin{figure}
  \centering
  \epsfig{file=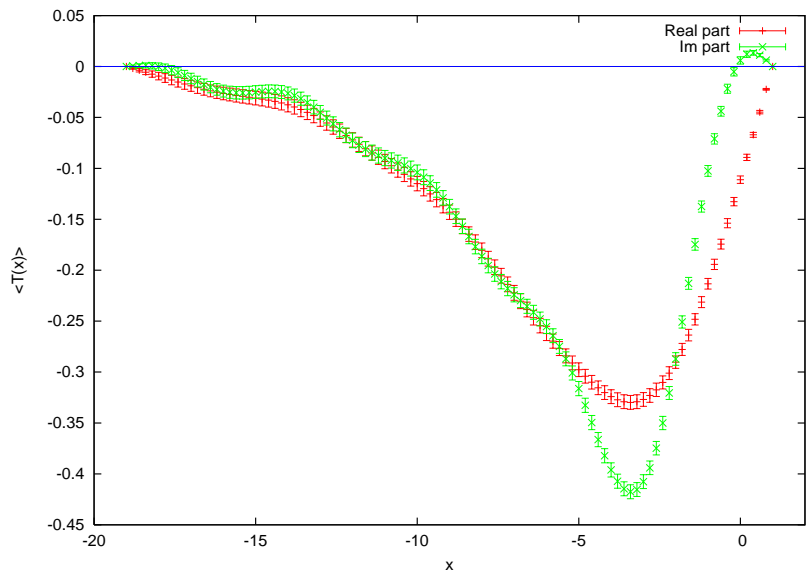,scale=0.43,angle=-90,clip}  
  \caption{The real (red) and imaginary (green) parts of $\langle T(x)
    \rangle$ for $L=20$, $l_{max}=1.999$, $x_{min}=-19$.}
  \label{l20xmin-19trealfig}
\end{figure}

The advantage of this is that the complicated behaviour of the $T_n$
in the complex plane can be packaged into a pair of
functions,
the behaviour of which may be easier to understand. We show an example in
Fig.~\ref{l20xmin-19trealfig}, where we plot $\langle T(x) \rangle$ for $L=20$,
$l_{max}=1.999$ and $x_{min}=-19$,
which is
one of the cases plotted in Fig.~\ref{l20complextsfig}.

The behaviour indeed appears simpler when plotted in this way. We see
that $T(x)=0$ as both ends of the interval, as it must with our
boundary conditions. Away from the boundaries it is very assymetric,
with a maximum near the large-$x$ end. This is not surprising, since
the cubic terms
that
pull $T(x)$ away from zero are largest
there. Note that in this case $T$ has 9 modes, so the shortest
lengthscale that can appear is roughly $L/9 \approx 2.2$, which is
comparable to the distance from the peak of $T$ to the boundary; the
peak is as far to the right as it can be. Indeed, we have found that this overall shape is typical. 

This in fact raises a general issue with our simulations. The
characteristic lengthscale over which we expect the fields to vary is something like
$1/V \approx 0.5$. But the finest scale we can probe is, say, half a
wavelength of the highest mode. We have $n_0=L \sqrt{l_{max}}/\pi$
modes on an interval of length $L$, so the highest mode has
half-wavelength $\pi/\sqrt{l_{max}}$. We can only go up to
$l_{max}=2$, giving a minimum lengthscale of $\pi/\sqrt{2} \approx
2.2$. To reach a lengthscale of 0.5, we would need to go up to
$l_{max}=39$, probably an impossible task.

\begin{figure}
  \centering
        \hspace{-9 ex}
        \begin{subfigure}{0.45\textwidth}
                \centering
                \epsfig{file=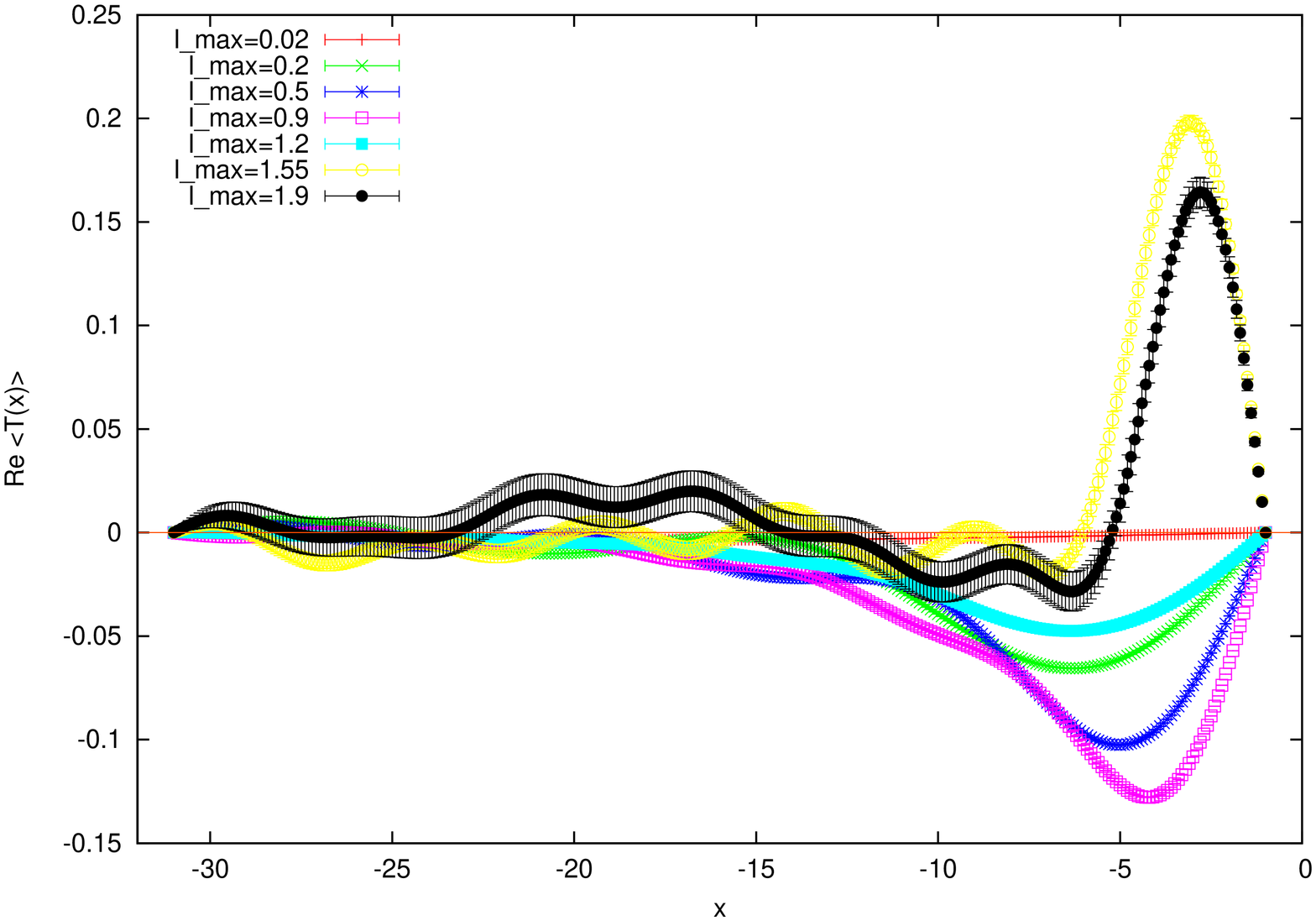,scale=0.31,angle=-90,clip}  
                \caption{Real parts}
        \end{subfigure}
        \hspace{4 ex}
        \begin{subfigure}{0.45\textwidth}
                \centering
                \epsfig{file=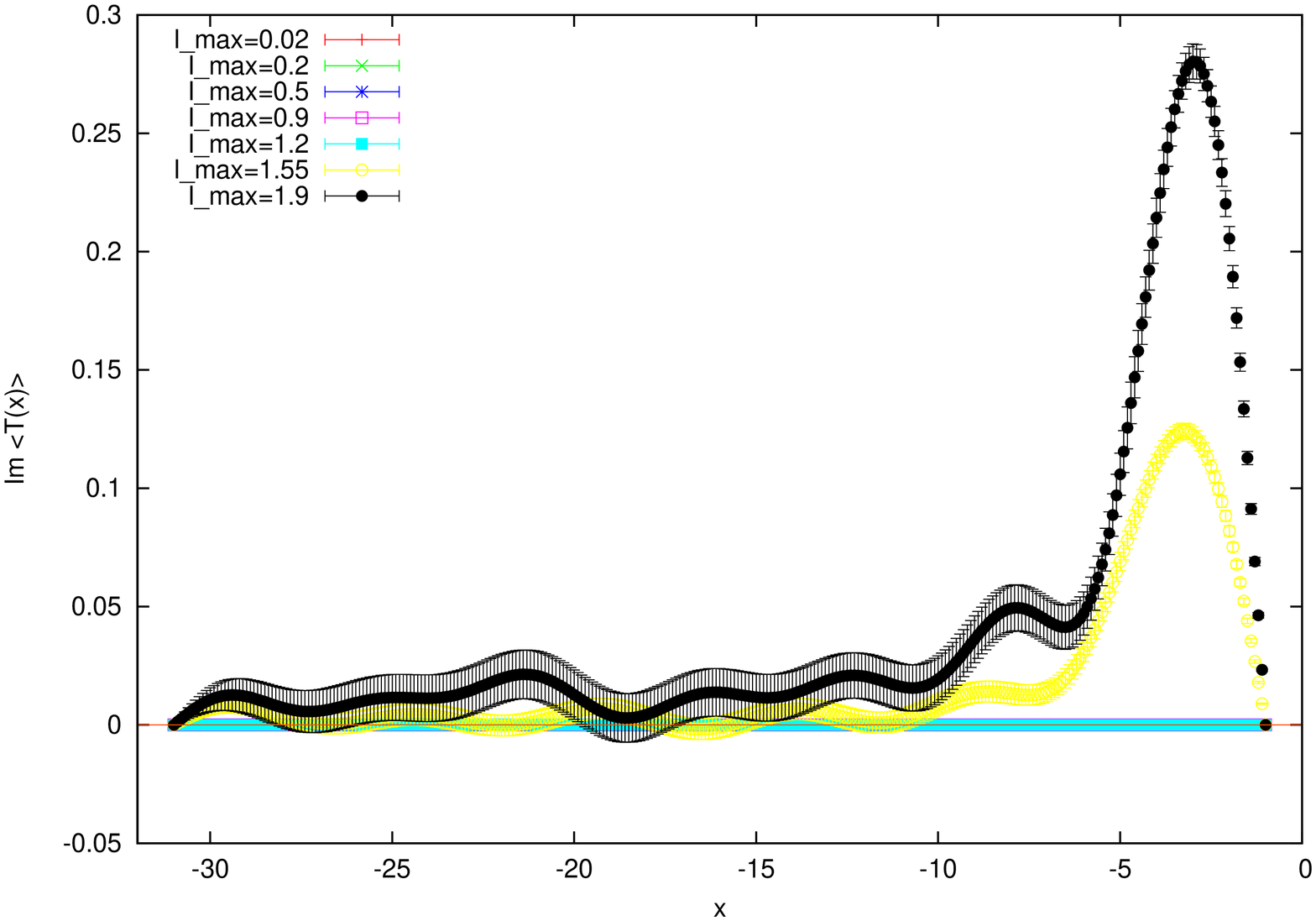,scale=0.31,angle=-90,clip}  
                \caption{Imaginary parts}
        \end{subfigure}
        \caption{$\langle T(x)
    \rangle$ for $L=30$, $x_{min}=-31$, and various values of $l_{max}$.}
\label{l30xmin-31treal}
\end{figure}

One way to look at this issue is to fix everything
except $l_{max}$ and see how things change as $l_{max}$ increases,
which corresponds to being able to look at smaller and smaller lengthscales.
We have done this for the case of
$L=30$,
$x_{min}=-31$ in
Fig~\ref{l30xmin-31treal}.

We see that as $l_{max}$ increases features on smaller length-scales appear.
However, there does not appear to be a smooth limit: $\langle T(x)
\rangle$ changes significantly every time $l_{max}$ is increased. In
fact this should not be surprising as we know from
section~\ref{l_maxsec} that the observables depend strongly on $l_{max}$.
This suggests that we may indeed not be resolving small enough lengthscales.

\subsubsection{Correlations}

It is also possible to look at correlations among the $T_n$, or
equivalently between $T(x)$ at different points. These are related by
\begin{equation}
\langle T(x_1) T(x_2) \rangle = \frac{2}{L}\sum_{n,k=1}^{n_0} \langle
T_n T_k \rangle \mathrm{sin} \left( \frac{n \pi (x-x_{min})}{L}
\right) \mathrm{sin} \left( \frac{k \pi (x-x_{min})}{L} \right).
\end{equation}

In principle, correlators could be used to extract the masses of the
states in the theory. However, since everything is space-dependent,
this will be complicated; the correlators will not simply decrease as
{\scalebox{.8}{$\displaystyle \sum_k e^{-m_k x}$}}.
We have not attempted to extract masses but have simply
looked at a few correlators as a first step in this direction.

\begin{figure}
  \centering
  \epsfig{file=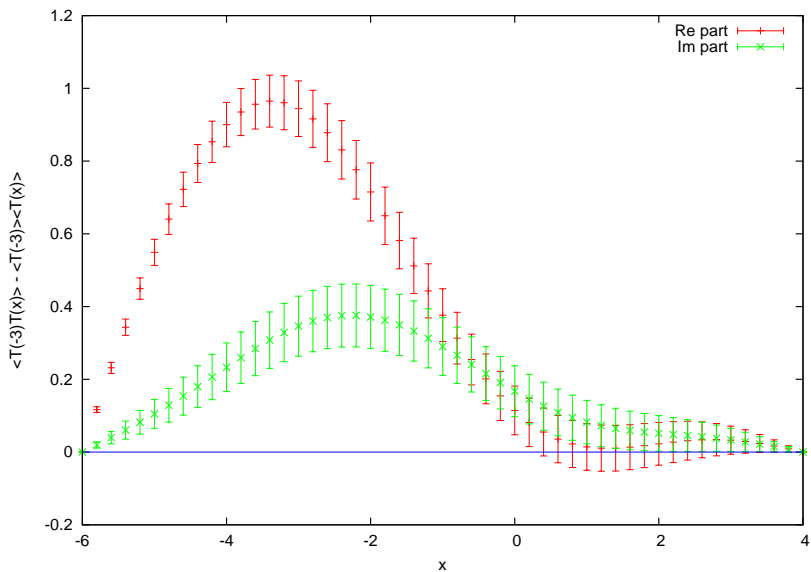,scale=0.43,angle=-90,clip}  
  \caption{Correlator $\langle T(x_1) T(x_2) \rangle - \langle T(x_1)
    \rangle \langle T(x_2) \rangle$
for fixed $x_1=-3$, with $x_{min}=-6$, $L=10$, and $l_{max}=-1.9$.}
  \label{tcorrfig}
\end{figure}

We show an example in Fig.~\ref{tcorrfig}. Here we have fixed $x_1$ and
let $x_2$ vary, and subtracted $\langle T(x_1) \rangle \langle T(x_2) \rangle$
to show just the fluctuations. The results look reasonable:
the largest correlations are at small separations (around $x_2=-3$).
They then fall off with distance, although rather slowly.
One thing to note is that the correlators are not symmetrical in
space --- as expected since the action is $x$--dependent.

\subsection{Varying the dilaton gradient}

In the full theory, the dilaton
gradient
$V$ is fixed to
\begin{equation}
V=-\sqrt\frac{25}{6\alpr}.
\end{equation}
Let us define
\begin{equation}
v \equiv 6\alpr V^2\,.
\end{equation}
Thus, the correct value for the dilaton slope is obtained at
\begin{equation}
v=25\,.
\end{equation}
However, in our level-truncated model we can choose any value, and it
would be interesting to see whether anything special happens at $v=25$.
In particular, we might hope that the
instability will be minimised at this value.

Another important value is $v=24$.
Here the mass-squared of the $T$ field, given by
\begin{equation}
\label{TmassV}
\alpr m_0^2=\frac{v}{24}-1
\end{equation}
becomes zero, and it becomes negative when $v$ is further decreased.
Hence for $v<24$ we should expect a tachyonic instability, different from
the cubic instability we have been concerned with before\footnote{Actually
for finite $L$ the lightest $T$-mode has mass-squared
$\alpr m_0^2=\frac{v}{24}-1+\alpr(\frac{\pi}{L})^2$
so the instability starts slightly below $v=24$.}. This quadratic instability
cannot be controlled by our analytic continuation with $\gamma=\pi/6$,
and we find the simulations indeed become unstable.
We have carried out several scans in $v$ which we describe below.

\subsubsection{$L=20$}

\begin{figure}
  \centering
        \hspace{-9 ex}
        \begin{subfigure}{0.45\textwidth}
                \centering
                \psfig{file=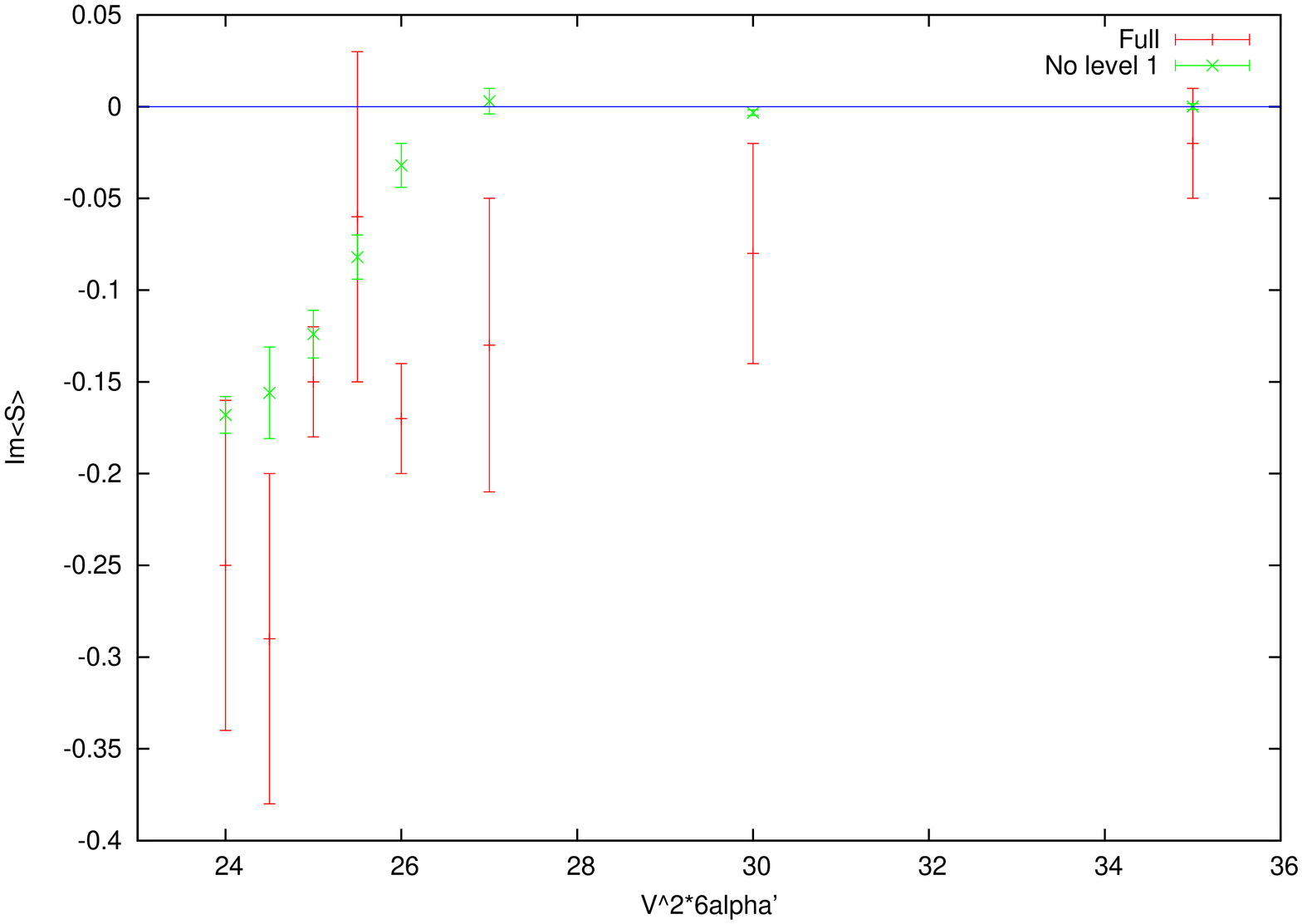,scale=0.31,angle=-90,clip}  
                \caption{$x_{min}=-20$}
        \end{subfigure}
        \hspace{4 ex}
        \begin{subfigure}{0.45\textwidth}
                \centering
                \epsfig{file=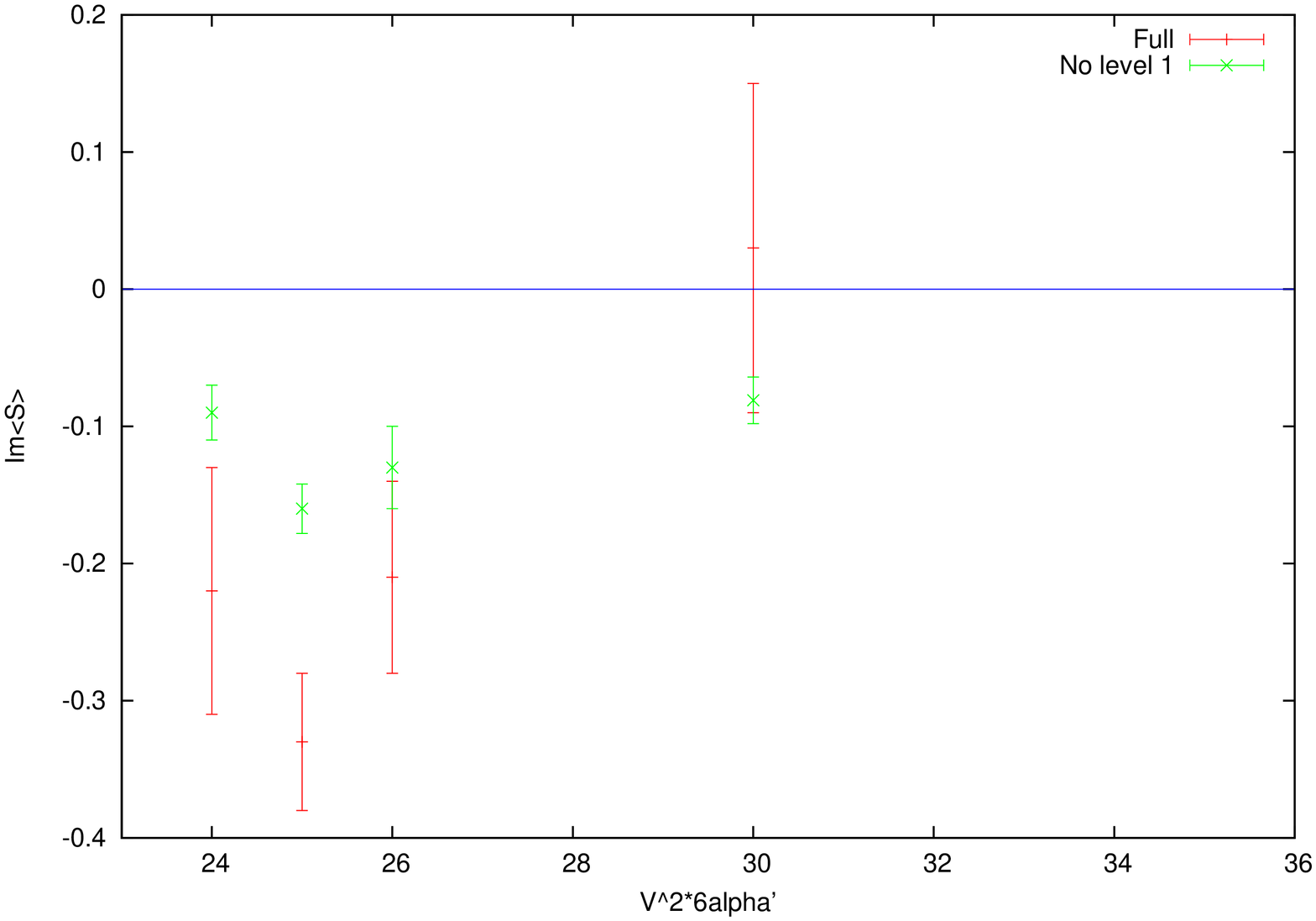,scale=0.31,angle=-90,clip}  
                \caption{$x_{min}=-19$}
        \end{subfigure}\ \\
       \vspace{2 ex}
        \hspace{-10 ex}
        \begin{subfigure}{0.45\textwidth}
                \centering
                \epsfig{file=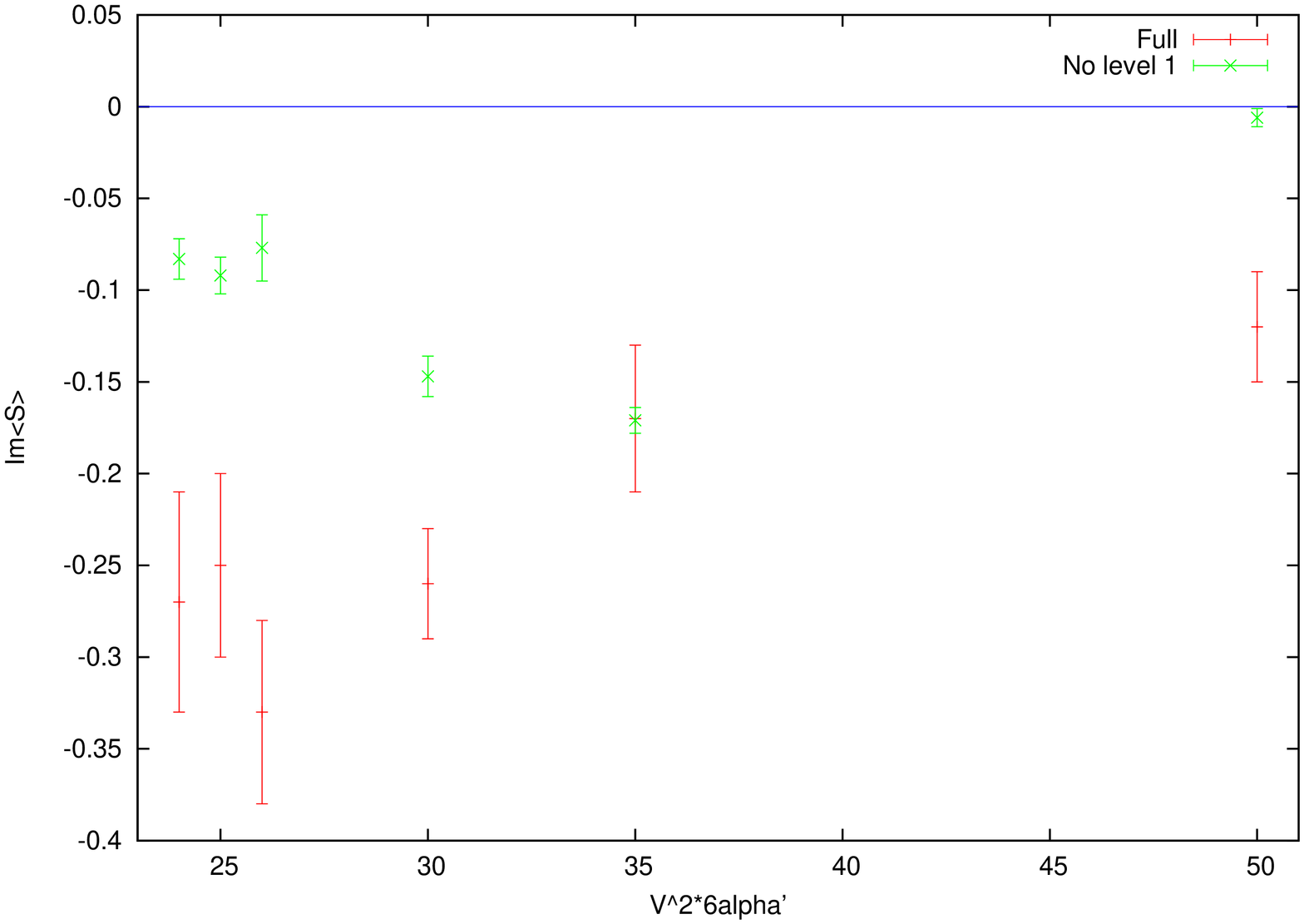,scale=0.31,angle=-90,clip}  
                \caption{$x_{min}=-18$}
        \end{subfigure}        
        \caption{$\Im\langle S \rangle$ as a function of $v$ for $L=20$,
		             $x_{min}=-20,-19,-18$, and $l_{max}=1.999$.}
\label{Vact20fig}
\end{figure}

\begin{figure}
  \centering
        \hspace{-9 ex}
        \begin{subfigure}{0.45\textwidth}
                \centering
                \psfig{file=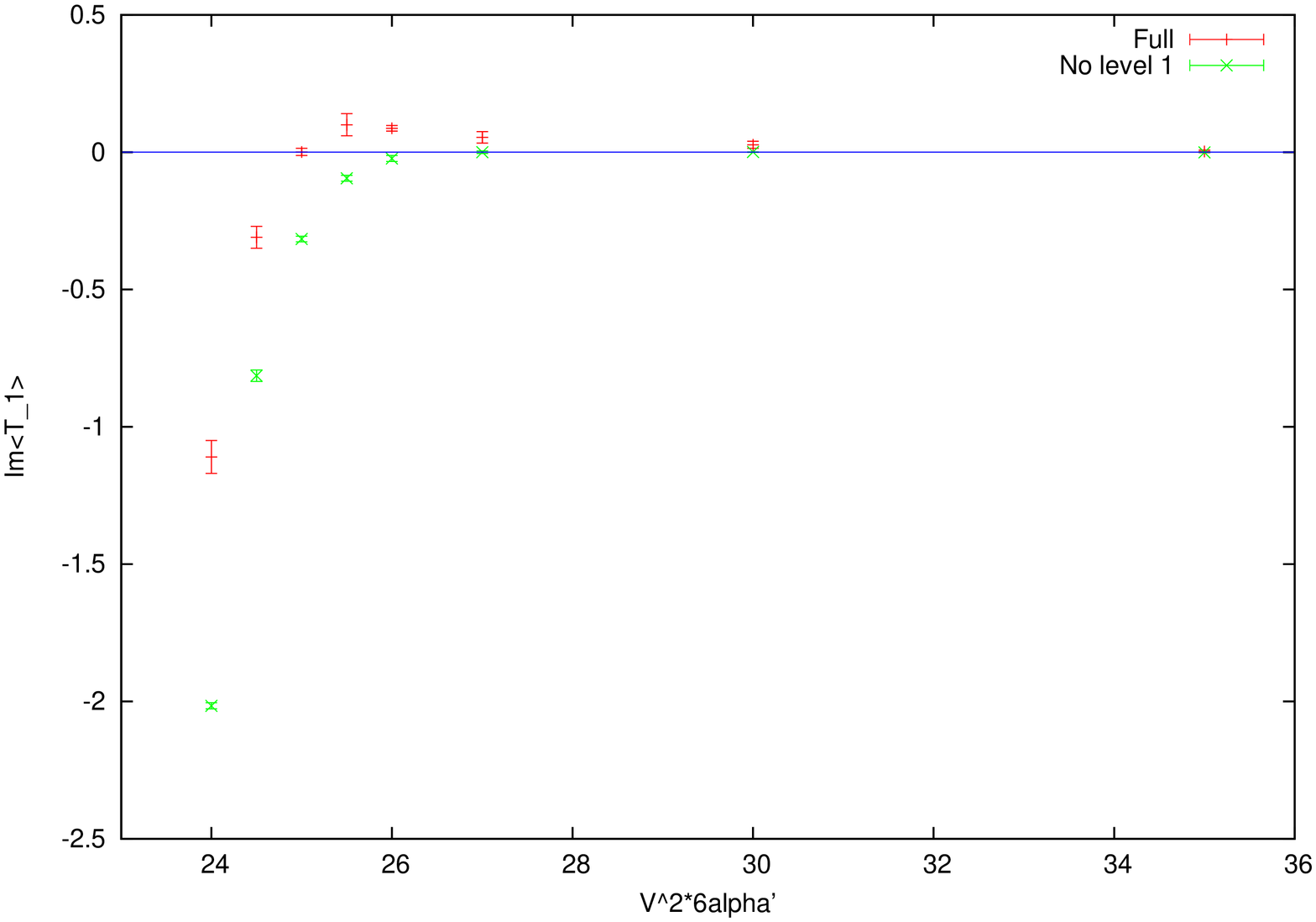,scale=0.31,angle=-90,clip}  
                \caption{$x_{min}=-20$}
        \end{subfigure}
        \hspace{4 ex}
        \begin{subfigure}{0.45\textwidth}
                \centering
                \epsfig{file=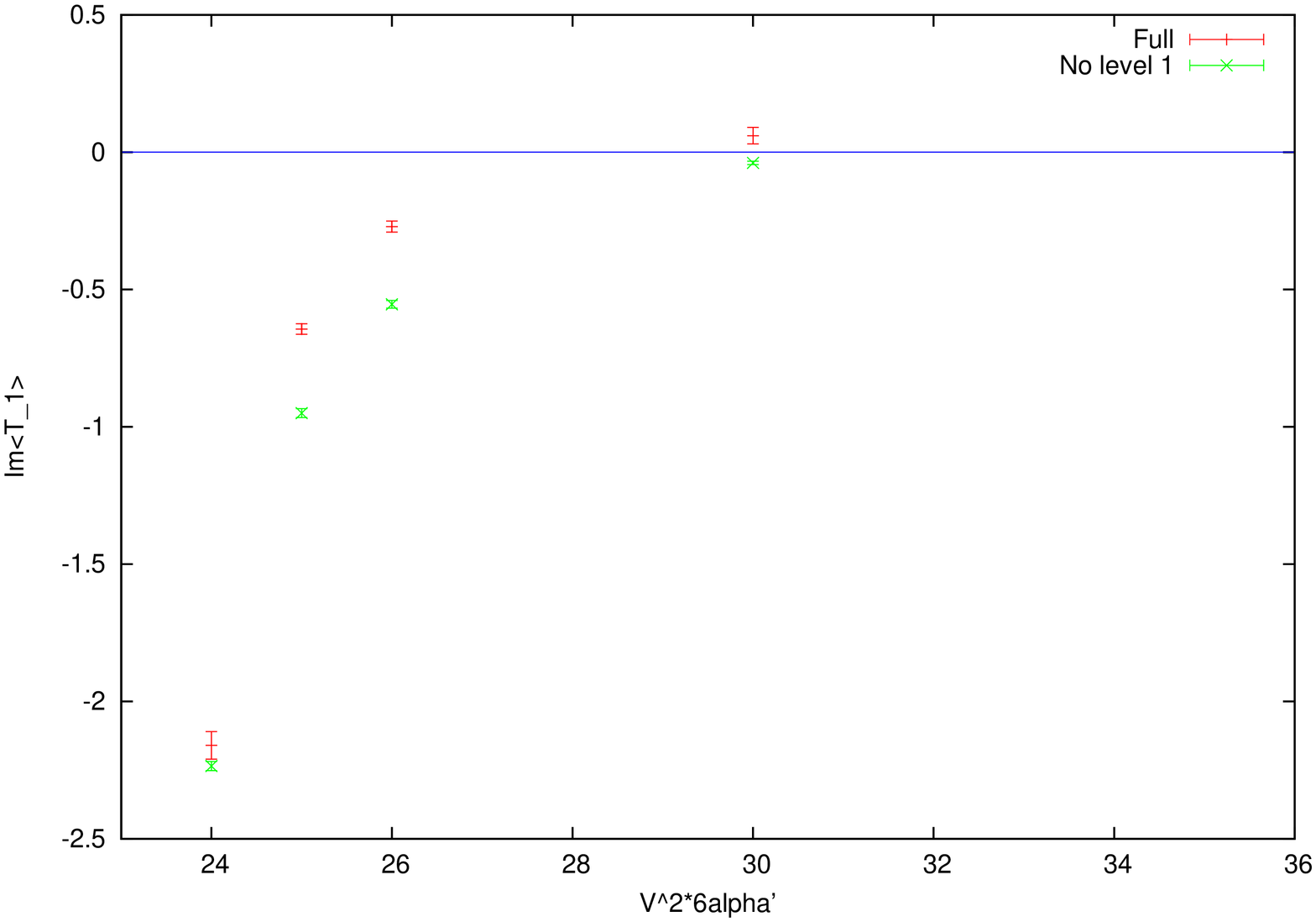,scale=0.31,angle=-90,clip}  
                \caption{$x_{min}=-19$}
        \end{subfigure}\ \\
        \vspace{2 ex}
        \hspace{-10 ex}
        \begin{subfigure}{0.45\textwidth}
                \centering
                \epsfig{file=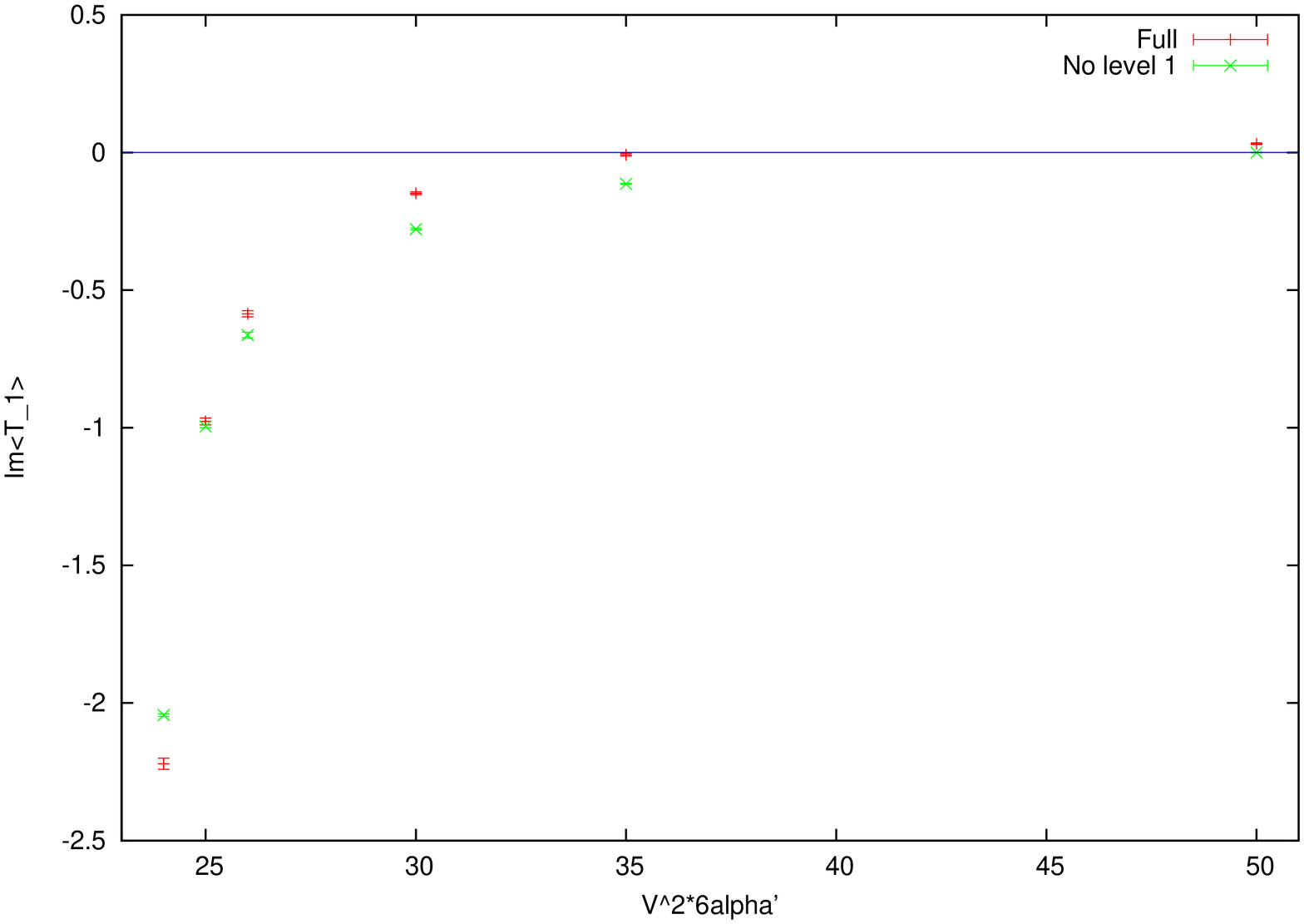,scale=0.31,angle=-90,clip}  
                \caption{$x_{min}=-18$}
        \end{subfigure}        
        \caption{$\Im\langle T_1 \rangle$ as a function of $v$
				for $L=20$,	$x_{min}=-20,-19,-18$, and $l_{max}=1.999$.}
\label{Vt120fig}
\end{figure}

Here we have results for $l_{max}=1.999$, and also for $l_{max}=1.999$
without level-1 fields. The imaginary parts of $\langle S \rangle$ and $\langle T_1 \rangle$
are shown in Figs.~\ref{Vact20fig} and~\ref{Vt120fig} respectively. For
both observables it appears that the instability decreases
monotonically as $v$ increases. It is also clear that $v=25$
is no better than neighbouring values of $v$.

\subsubsection{$L=10$}

\begin{figure}
  \centering
        \hspace{-9 ex}
        \begin{subfigure}{0.45\textwidth}
                \centering
                \epsfig{file=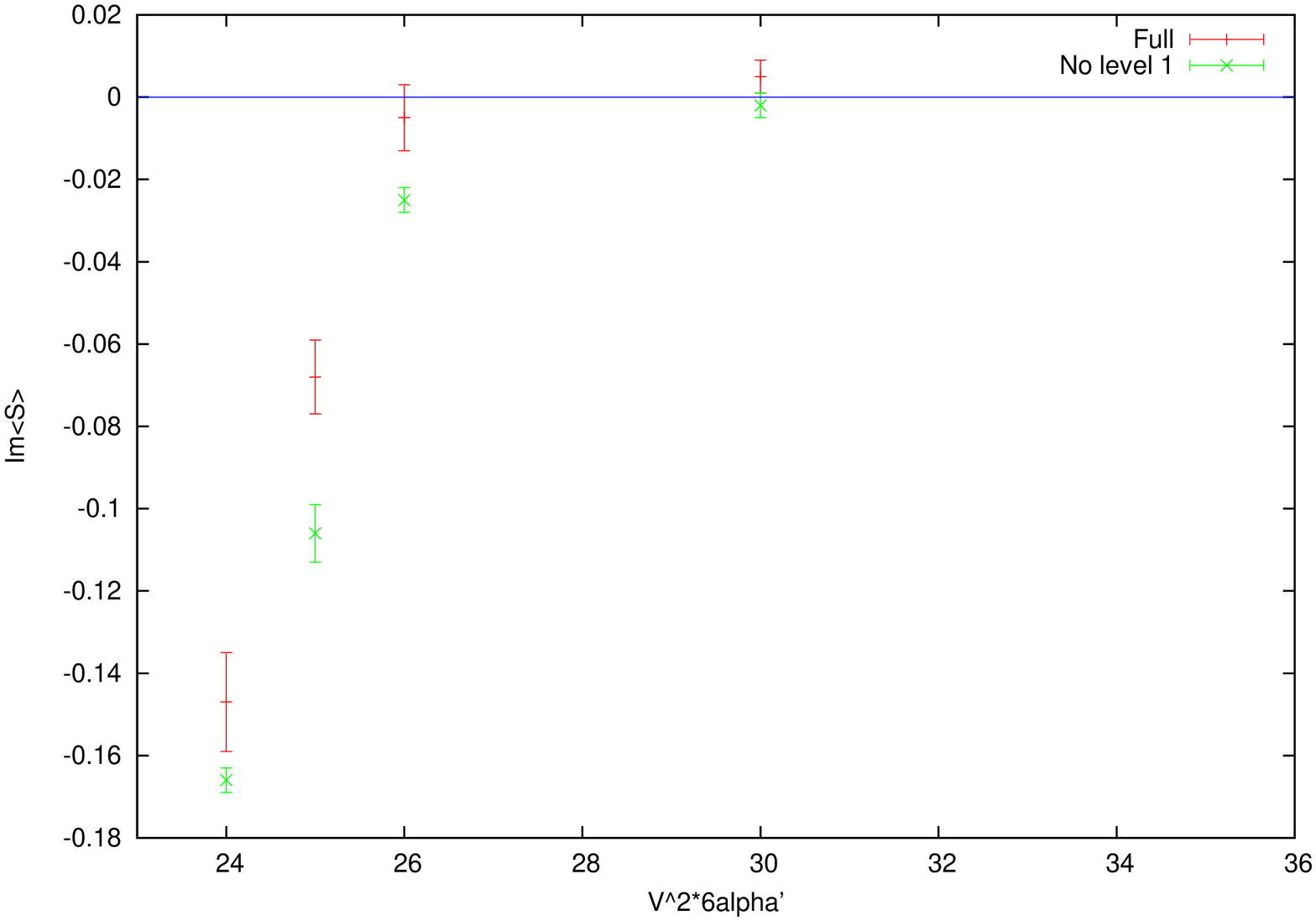,scale=0.31,angle=-90,clip}  
                \caption{$x_{min}=-10$}
        \end{subfigure}
        \hspace{4 ex}
        \begin{subfigure}{0.45\textwidth}
                \epsfig{file=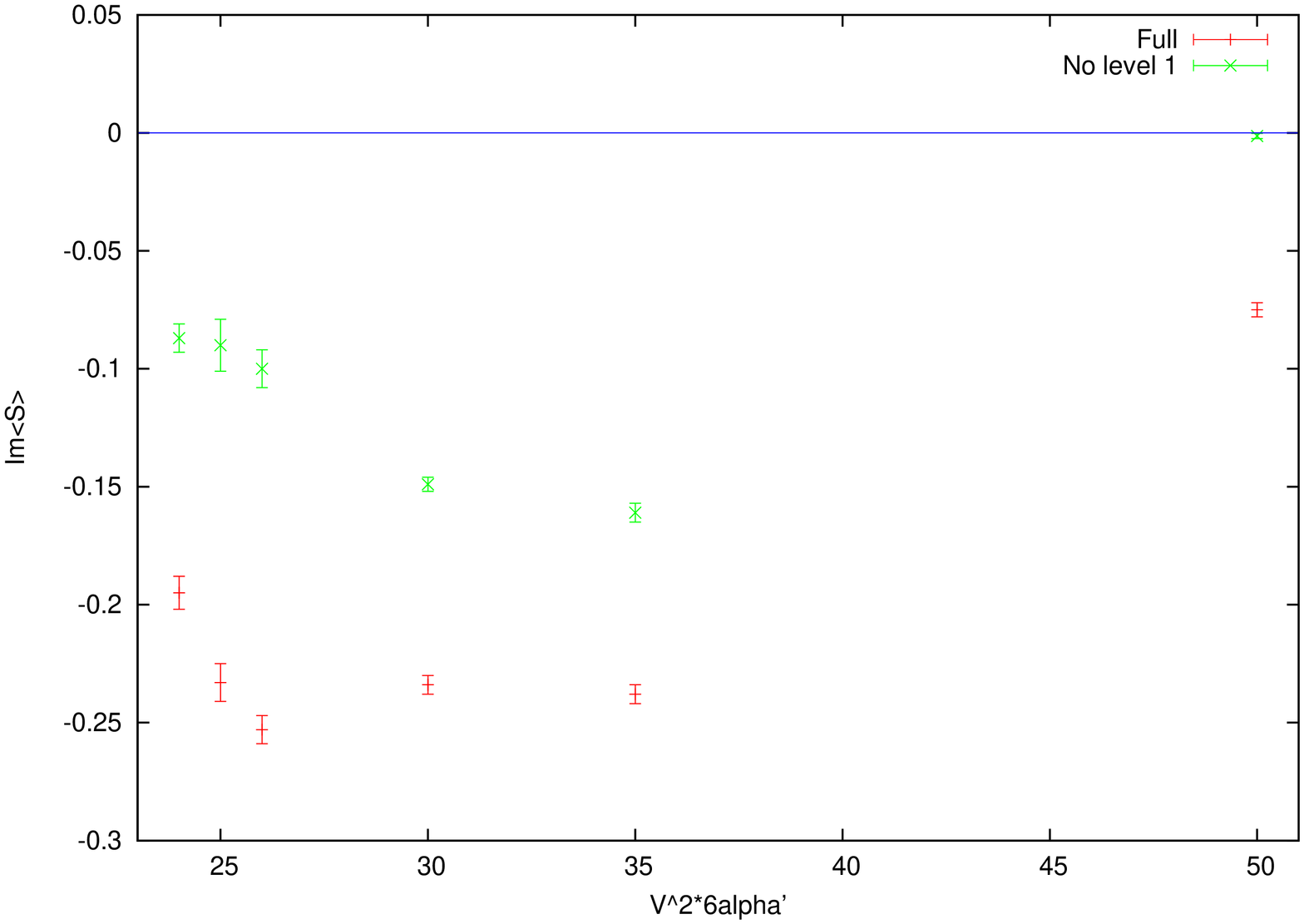,scale=0.31,angle=-90,clip}  
                \caption{$x_{min}=-8$}
        \end{subfigure}\ \\
        \vspace{2 ex}
   \centering
       \hspace{-10 ex}
        \begin{subfigure}{0.45\textwidth}
                \centering
                \epsfig{file=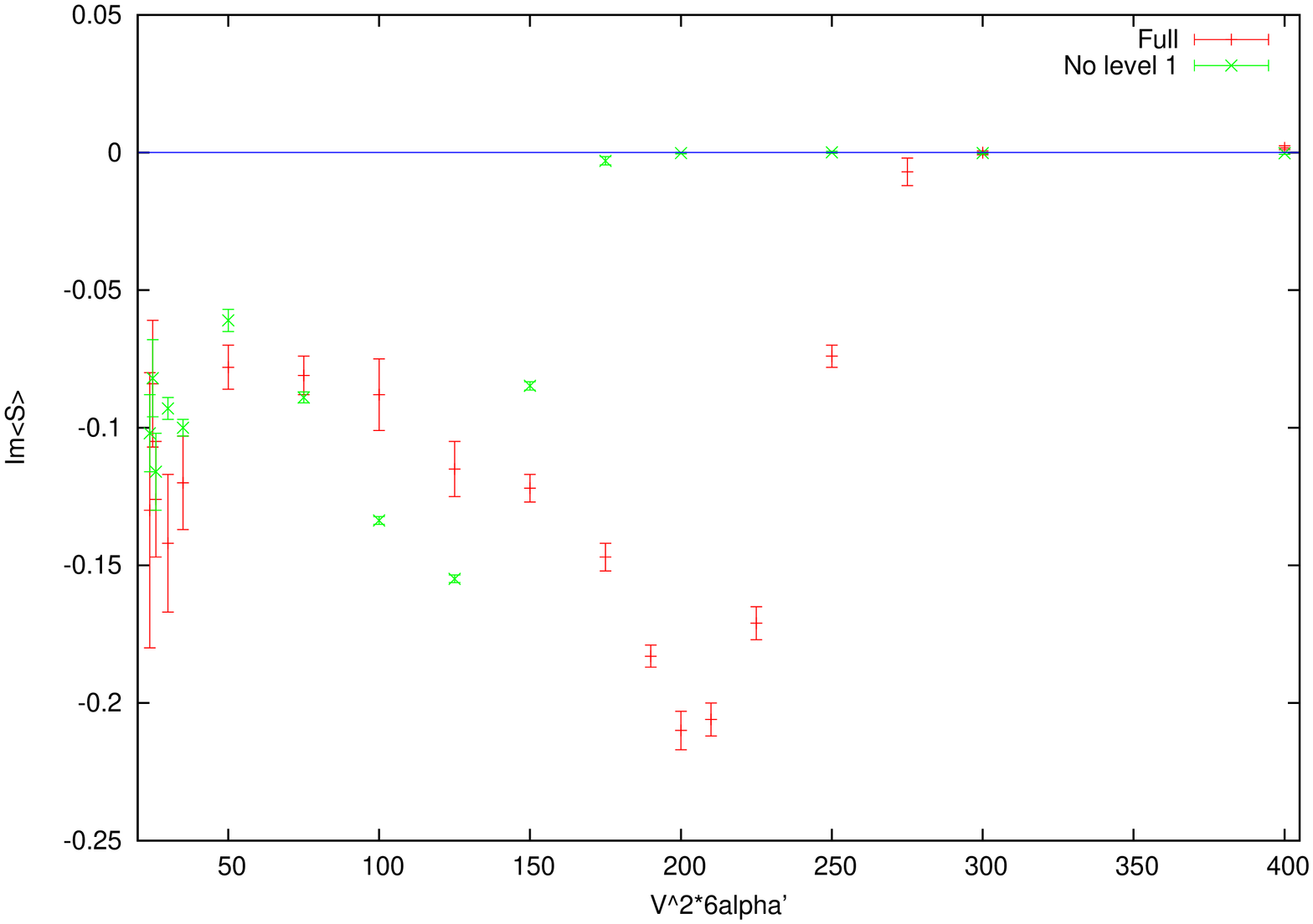,scale=0.31,angle=-90,clip}  
                \caption{$x_{min}=-6$}
        \end{subfigure}        
        \caption{$\Im\langle S \rangle$ as a function of $v$
				for $L=10$,	$x_{min}=-10,-8,-6$, and $l_{max}=1.9$.}
\label{Vact10fig}
\end{figure}

\begin{figure}
  \centering
        \hspace{-9 ex}
        \begin{subfigure}{0.45\textwidth}
                \centering
                \epsfig{file=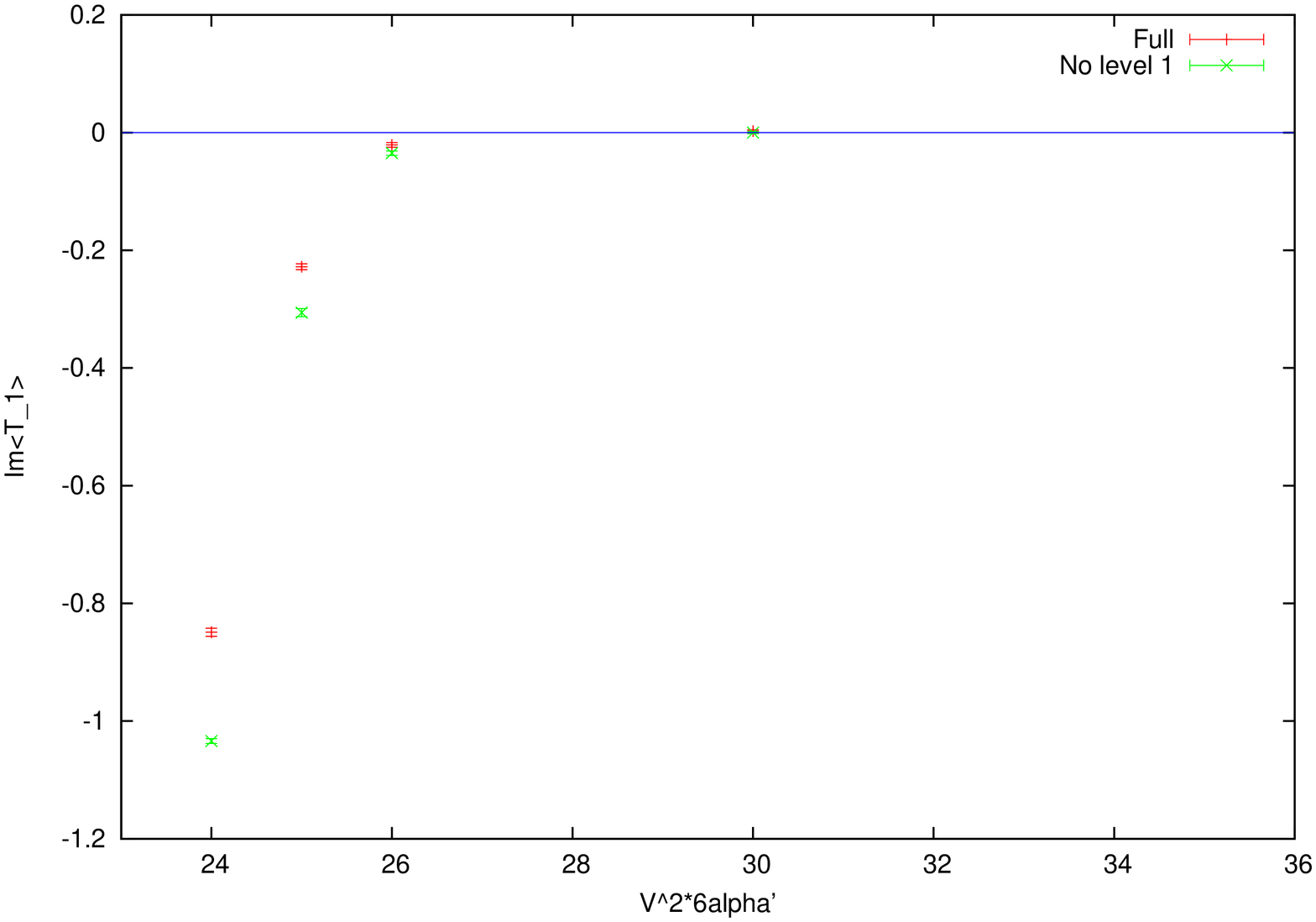,scale=0.31,angle=-90,clip}  
                \caption{$x_{min}=-10$}
        \end{subfigure}
        \hspace{4 ex}
        \begin{subfigure}{0.45\textwidth}
                \centering
                \epsfig{file=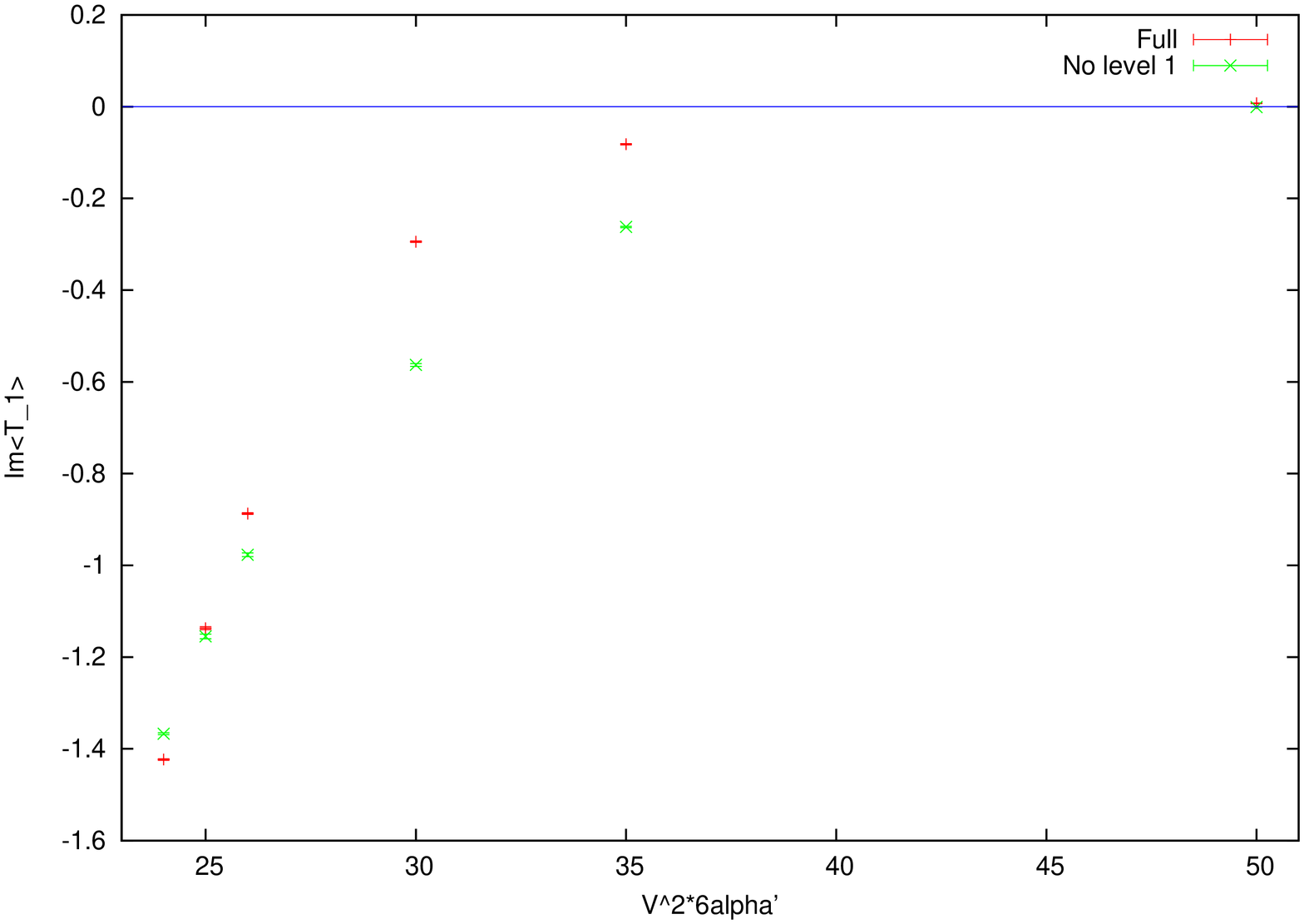,scale=0.31,angle=-90,clip}  
                \caption{$x_{min}=-8$}
        \end{subfigure}\ \\
        \vspace{2 ex}
        \hspace{-10 ex}
        \begin{subfigure}{0.45\textwidth}
                \centering
                \epsfig{file=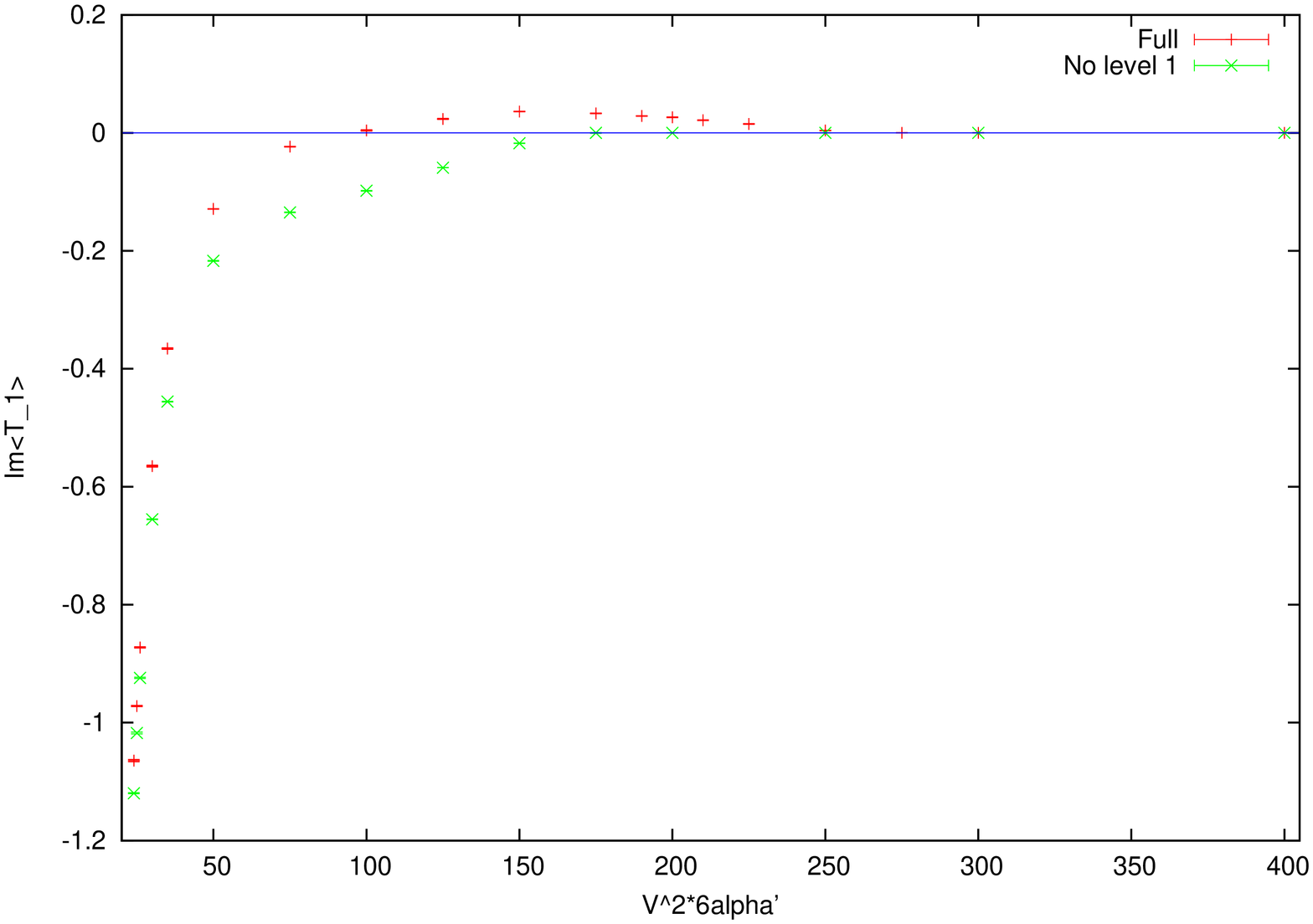,scale=0.31,angle=-90,clip}  
                \caption{$x_{min}=-6$}
        \end{subfigure}        
        \caption{$\Im\langle T_1 \rangle$ as a function of $v$
				for $L=10$,	$x_{min}=-10,-8,-6$, and $l_{max}=1.9$.}
\label{Vt110fig}
\end{figure}

We now turn to $L=10$ where the errors are smaller. Here we use
$l_{max}=1.9$ which is again the highest value of $l_{max}$ below 2. We begin with
$x_{min}=-10$ which is quite weak coupling. We plot $\Im \langle S \rangle$ in
Fig.~\ref{Vact10fig} and $\Im \langle T_1 \rangle$ in Fig.~\ref{Vt110fig}.

Again the overall trends are similar and nothing remarkable happens at $v=25$.
For the case $x_{min}=-6$,
which is extremely strong coupling, we have done a more extensive scan in
$v$, going up to $v=400$. 
(Note the $x$-axis is compressed in these plots compared to the
earlier plots in this section.)
Here we see the behaviour is more complicated: there are two minima in
$\Im \langle S \rangle$, and $\Im \langle T_1 \rangle$ crosses zero and
approaches zero from above for large $v$, although this does not
happen if level-1 fields are excluded. Disappointingly most of this
interesting behaviour is at large $v$, far from $v=25$.

\subsubsection{Interpretation}

Much of the rather complicated behaviour seen above can be understood, at least
qualitatively, by considering how the quadratic and cubic terms depend on $v$.

The behaviour of the quadratic (mass) terms is the more
straightforward. The mass of the $T$ field~(\ref{TmassV})
simply increases with $v$.

\begin{figure}
  \centering
        \hspace{-9 ex}
        \begin{subfigure}{0.45\textwidth}
                \centering
                \psfig{file=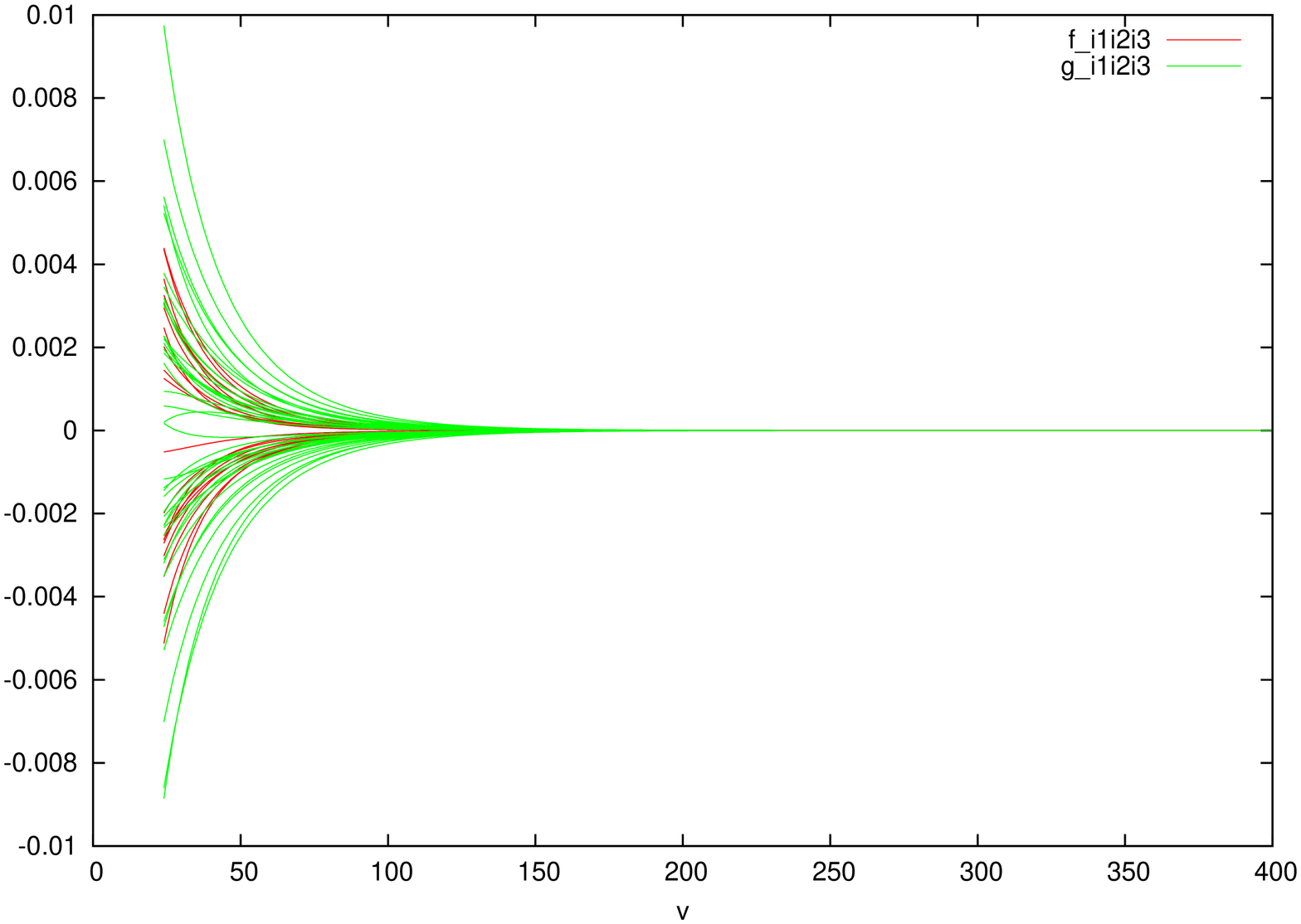,scale=0.31,angle=-90,clip}  
                \caption{$x_{min}=-10$}
        \end{subfigure}
        \hspace{4 ex}
        \begin{subfigure}{0.45\textwidth}
                \centering
                \psfig{file=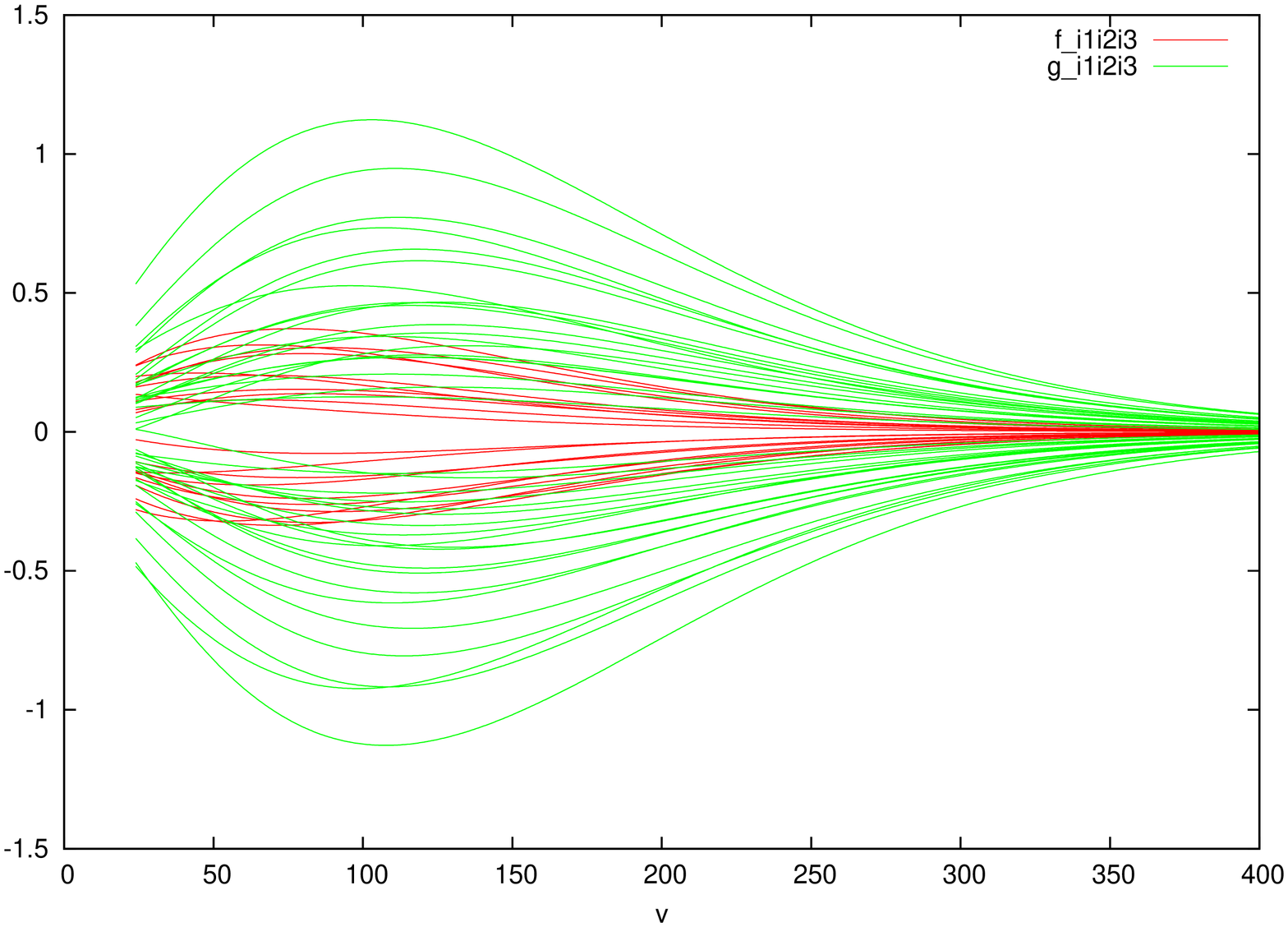,scale=0.31,angle=-90,clip}  
                \caption{$x_{min}=-6$}
        \end{subfigure}
        \caption{The $f_{i_1i_2i_3}$ and $g_{i_1i_2i_3}$
				as functions of	$v$	for $L=10$, $x_{min}=-10, -6$.}
\label{coeffsfig}
\end{figure}

The cubic terms are more complicated. There are many of these, given
by the coefficients $f_{i_1i_2i_3}$ and $g_{i_1i_2i_3}$. We have
plotted them in Fig.~\ref{coeffsfig} as functions
of $v$ for two of the sets of parameters above:
$L=10$, $x_{min}=-10$ and $L=10$, $x_{min}=-6$.
We see that for $x_{min}=-10$, the coefficients mostly decline
throughout the range of $v$, whereas for $x_{min}=-6$, they peak around
$v=100$ and only then begin to decline.

We can understand this analytically for the particular case of $f_{111}$. Taking the exact
expression and keeping only those terms which survive when $v$ is
large, one obtains
\begin{equation}
f_{111} \propto K^{-3v/24} e^{x_{max}\sqrt{\frac{v}{24}}}.
\label{f111eq}
\end{equation}
For sufficiently large $v$ the first term will dominate, so the
coefficient will decline. For somewhat smaller $v$ the second,
growing, term will matter as well, and there will be a peak at
$v=\frac{8}{3} x_{max}^2/\mathrm{ln}K \approx 10 x_{max}^2$. However
we will not see this peak if $x_{max}$ is small --- it would be at
small or negative $v$ and we are only interested in $v \ge 24$ where
the quadratic terms are stable\footnote{Also in this case we would
have to take into account terms ignored
in~(\ref{f111eq}).}.
Presumably
something similar happens for the other $f_{i_1i_2i_3}$ and the
$g_{i_1i_2i_3}$, and this accounts for the behaviour seen in
Fig.~\ref{coeffsfig}.

Putting all this together we can understand the
observed pattern of stability.
For small $x_{min}$ the quadratic terms increase and the cubic
terms decrease monotonically as $v$ increases. Both these trends
increase stability, so we see the imaginary parts of observables decrease.
However, for larger $x_{min}$ at first both terms increase. Near
$v=24$ the increase in the quadratic term should matter more, since
below this value it is negative. Thus at first the instability
decreases. Then the increase in the cubic terms becomes more
important, and the instability increases. Finally the cubic terms peak
and start decreasing, and the instability decreases again. This
matches the behaviour seen for $x_{min}=-10$. It does not fully
explain the behaviour seen for $x_{min}=-6$, but this argument
is rather rough and could be refined --- for example we have not
considered the fermion determinant at all.

To summarise, both our numerical results and our analysis of the behaviour
of the coefficients show that nothing special happens at $v=25$.
It is of course possible that this
will change if we worked at higher $l_{max}$
or $L$, especially in light of the observation made in~\ref{sec:positionSpace},
according to which we are far from being able to sample space in a
high enough resolution as compared to the scale set by the dilaton slope.

\subsection{Adding trivial parts to the action}
\label{sec:Adding_triv_terms}

As discussed in section~\ref{sec:trivial_terms},
we can add terms proportional to $(p_1+p_2+p_3 - iV/2)$,
which vanish using the delta function, to the action. These should not affect the results. 

Clearly there are many terms of this form we could test. We have chosen the specific example
\begin{equation}
S_{\mathrm{trivial}}=Z \sum_{n_{1,2,3}=1}^N (p_1+p_2+p_3 - iV/2) p_1^4 p_2^4 p_3^4 T(p_1) T(p_2) T(p_3).
\label{eq:trivial term}
\end{equation}
To keep this term simple, we have only considered it when only level-0 fields are
present, i.e. for $l_{max}<1$. Note that the extra cubic coefficients due to this
term are similar in magnitude to the existing ones when $Z$ is of order unity.

\begin{figure}
  \centering
  \psfig{file=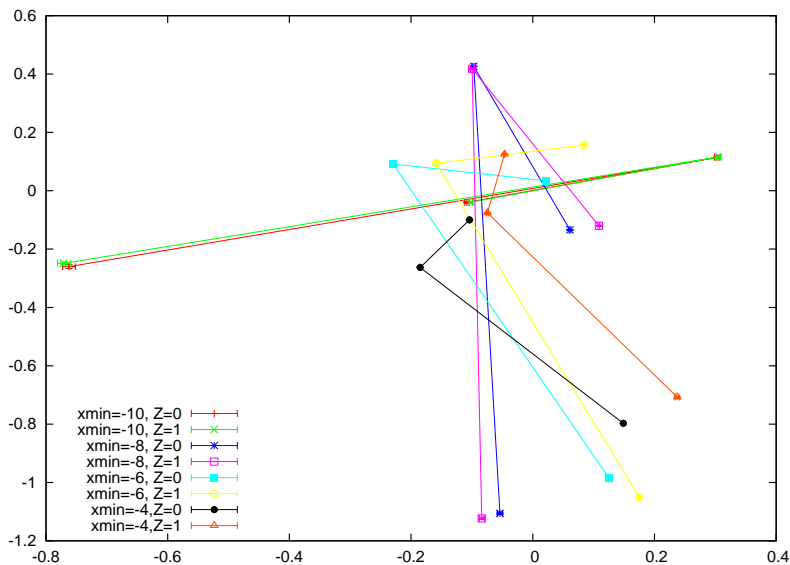,scale=0.43,angle=-90,clip}  
  \caption{The $\langle T_n \rangle$ in the complex plane for $L=10$ and $l_{max}=0.9$,
	for various $x_{min}$, both with and without the term in~(\ref{eq:trivial term}).}
  \label{l10trivialfig}
\end{figure}

We have found that adding this term can significantly affect the results.
We show an example of its effect in Fig.~\ref{l10trivialfig}.
Not surprisingly, the difference between simulations with and
without~(\ref{eq:trivial term}) increase with $Z$.
However, we are more interested in how this difference depends on the other
parameters, and in particular, whether it decreases in the continuum and/or large-volume limits.

To analyse this, we have compared results with $Z=0$ and $Z=1$
with several sets of parameters. In general, the full effect
of~(\ref{eq:trivial term}) is complicated, but there are some clear trends:
\begin{itemize}
  \item{The differences between $Z=0$ and $Z=1$ increase at larger $x_{min}$
	      (stronger coupling). This is presumably simply because the new term
				increases exponentially in $x$, just like the original cubic terms.}
  \item{The absolute differences are roughly similar for different mode numbers.
	      Since the $\langle T_n \rangle$ are usually smaller for higher
				modes this means the relative differences increase.}
  \item{The differences remain roughly constant as $L$ increases.}
\end{itemize}

This last point is presumably because the dependence on $L$ is mainly
in the terms $p_i^4 = \frac{(n_i \pi)^4}{L^4}$, and the numerator
and denominator cancel, since the highest mode number $n_i$
available is proportional to $L$ (for fixed $l_{max}$).

We conclude that we cannot really establish $Z$-independence
and can only continue with a rough principle of retaining the vertex
in a form that is ``as simple as possible''.
The reason for this awkward situation is the sine-expansion that we were forced
to use in light of the fact that the linear dilaton prevents us from using
periodic boundary conditions.
Moreover, at $l_0\geq 2$ the sine-expansion leads to non-real coefficient terms,
which after the analytical continuation lead to terms that are cubic, but not
purely imaginary. This situation prevents us from carrying out our analysis
in higher levels.

\subsection{Scheme 1}
\label{scheme1sec}

We now turn to the other `schemes', starting with scheme 1. This is the
same as scheme 4 except that the Grassmann-odd fields $B$ and $C$ are
not present, but the Grassmann-even field $\mathcal{B}$ is added instead.
The action in this scheme is given in~(\ref{Scheme2QuadAction})
and~(\ref{Scheme2CubicAction}).
Note that for $0 \le l_{max} < 1$ it is the same as scheme 4.

We have done only a few runs to compare with the results of scheme 4.
Specifically, we have carried out a scan in $x_{min}$ with the other parameters
fixed to $\alpr=1$, $V=-\sqrt\frac{25}{6\alpr}$, $L=20$, and with $l_{max}=1.999$, since this should
have the largest difference to scheme 4. As will be seen below  the
instability is greater in scheme 1 at the same $x_{min}$; in fact it
is so much greater that the simulations become prohibitively expensive
around $x_{min}=-22$, where the instability is still tiny in scheme
4. Hence the range of $x_{min}$ covered in scheme 1 is
$-23$ to $-22.2$,
which does not overlap with the range
$-22$ to $-18$
covered in scheme 4.

\begin{figure}
  \centering
  \epsfig{file=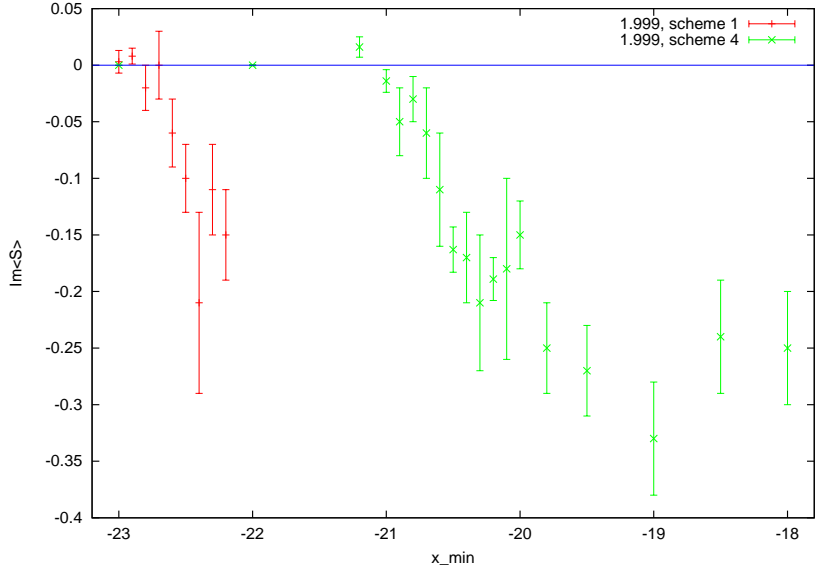,scale=0.43,angle=-90,clip}  
  \caption{$\Im\langle S \rangle$ as a function of $x_{min}$ for $L=20$,
    $l_{max}=1.999$ for scheme 1 (red) and scheme 4 (green).}
  \label{scheme1actfig}
\end{figure}

\begin{figure}
  \centering
  \epsfig{file=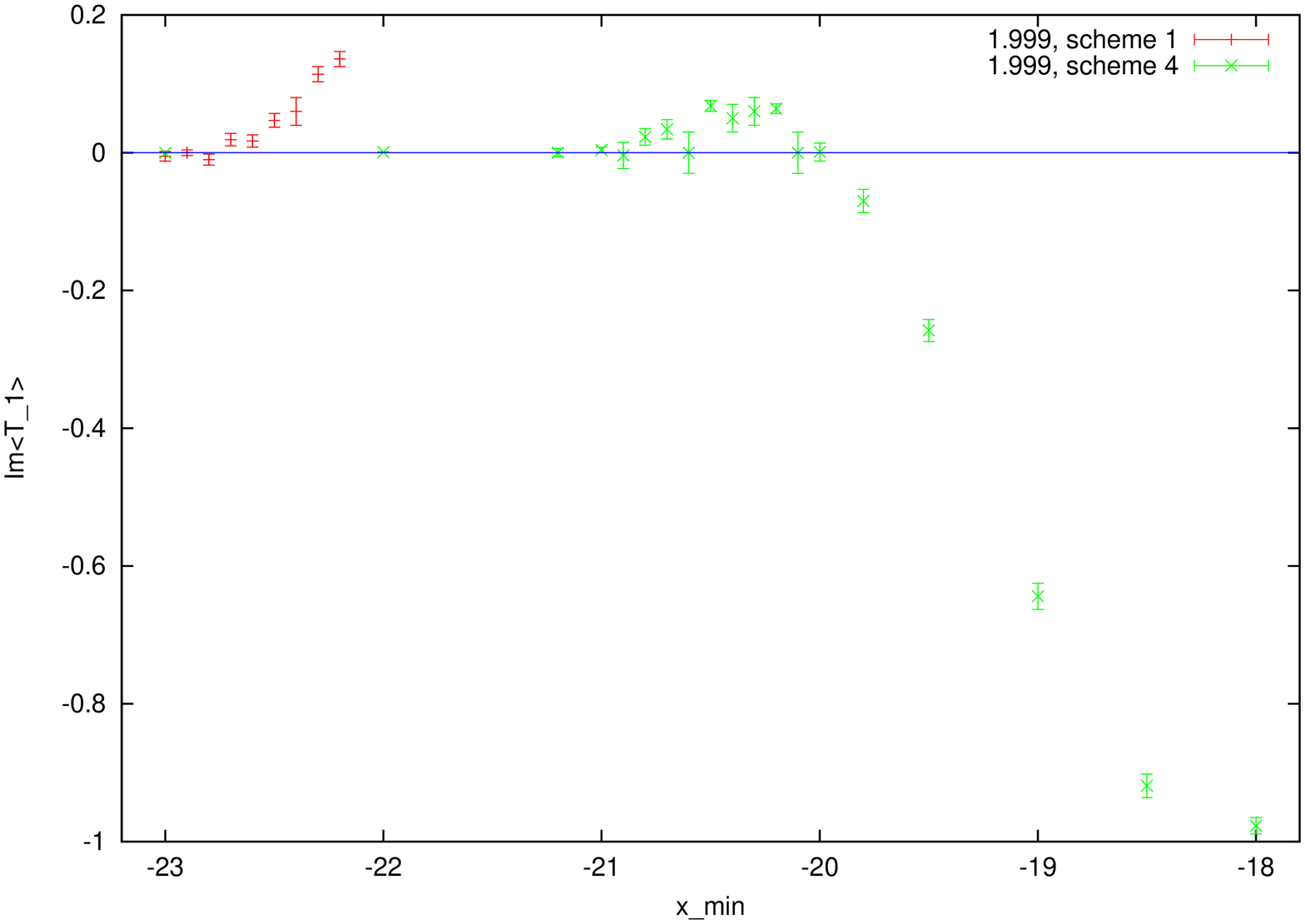,scale=0.43,angle=-90,clip}  
  \caption{$\Im\langle T_1 \rangle$ as a function of $x_{min}$ for $L=20$,
    $l_{max}=1.999$ for scheme 1 (red) and scheme 4 (green).}
  \label{scheme1t1fig}
\end{figure}

We plot the results for $\Im \langle S \rangle$ in
Fig.~\ref{scheme1actfig} and for  $\Im \langle T_1 \rangle$ in
Fig.~\ref{scheme1t1fig}, in both cases together with the corresponding
scheme 4 results. In both cases it is clear that the instability is
much greater, or equivalently appears at smaller $x_{min}$, for scheme
1. The shift in $x$ for equivalent imaginary parts is about 2. This
corresponds to a change in size of the cubic terms of roughly $e^{2V}
\approx 60$.

This is a very significant change, but it can be understood by the following argument. 
Consider first just the terms in the action quadratic in $A$
and $\mathcal{B}$, and the case where each field has only one mode:
\begin{equation}
S=-\frac{1}{2}\left( m_1^2+\left( \frac{\pi}{L}\right) \right) A^2 -
\frac{1}{\alpr} \mathcal{B}^2 -\frac{V}{\sqrt{2\alpr}}A\mathcal{B}.
\end{equation}
Completing the square, and taking $\alpr=1$, one finds this has an
almost massless mode in the direction
$\mathcal{B}=-\frac{V}{2\sqrt{2}}A$, with mass only
$\frac{\pi}{L}$. Indeed, in the large-volume limit this mode becomes
massless --- presumably this is due to the fact that scheme 1 includes
gauge degrees of freedom.

By itself this small but still positive mass would not lead to an
instability. However, it does when the cubic terms are included. The
reason is that there will be large fluctuations of this mode, giving
large values of the fields $A$ and $\mathcal{B}$. These will then lead
to the cubic terms of the form $AAT$ and $T\mathcal{B}\mathcal{B}$ being large,
much larger than they would be if this light mode was not
present. Since some of these cubic terms are unstable, these
instabilities will be much larger than they would be without the
light mode. As a partial check on this mechanism, we have observed
that in our simulations the modes 
$A_1$ and $\mathcal{B}_1$ indeed fluctuate together, in exactly the direction
$\mathcal{B}=-\frac{V}{2\sqrt{2}}A$.
Hence it appears that trying to included gauge degrees of freedom like
this will lead to problems, and we did not continue further with
simulations of scheme~1.

\subsection{Scheme 2}
\label{scheme2sec}

In this section we briefly discuss our results for scheme 2. This is the
same as scheme 4 except that the Grassmann-odd fields $B$ and $C$ are
not present. For $0 \le l_{max} < 1$ it is the same as scheme 4.

\begin{figure}
  \centering
  \epsfig{file=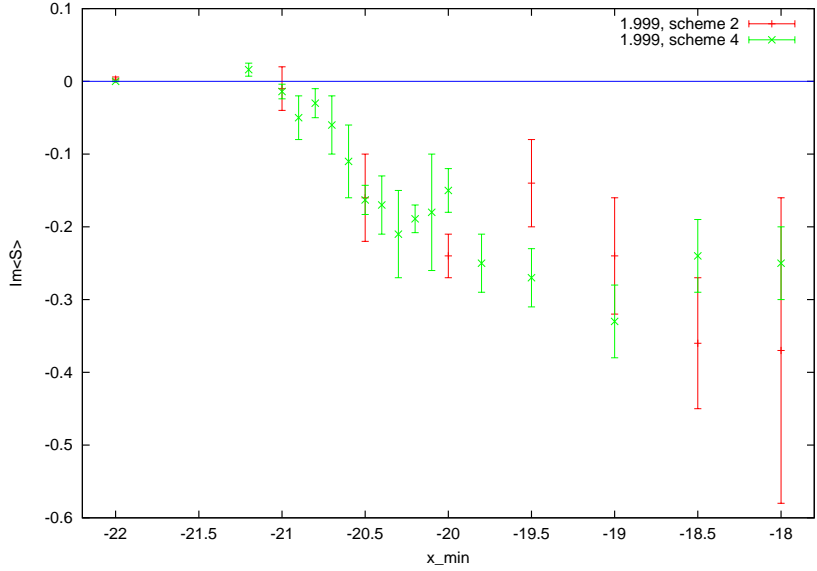,scale=0.43,angle=-90,clip}  
  \caption{$\Im\langle S \rangle$ as a function of $x_{min}$ for $L=20$,
    $l_{max}=1.999$ for scheme 2 (red) and scheme 4 (green).}
  \label{scheme2actfig}
\end{figure}

\begin{figure}
  \centering
  \epsfig{file=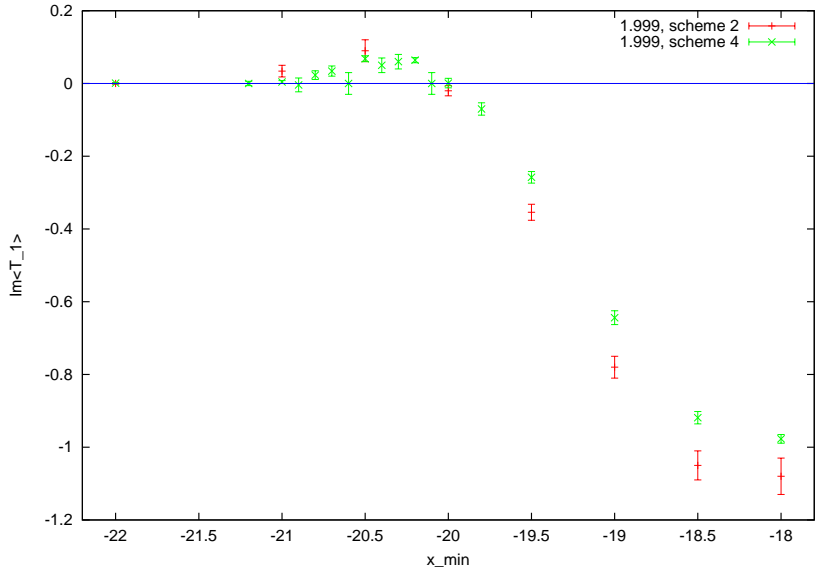,scale=0.43,angle=-90,clip}  
  \caption{$\Im\langle T_1 \rangle$ as a function of $x_{min}$ for $L=20$,
    $l_{max}=1.999$ for scheme 2 (red) and scheme 4 (green).}
  \label{scheme2t1fig}
\end{figure}

Again, we have only done simulations for a few sets of parameters, to compare the
results with those for scheme 4. Specifically, we have scanned
in $x_{min}$ for
$\alpr=1$, $L=20$, $V=-\sqrt\frac{25}{6\alpr}$,
and $l_{max}=1.999$. We have results for several
$x_{min}$ from $-22$ to $-18$.
These are plotted together with the
corresponding scheme 4 results in Figs.~\ref{scheme2actfig}
and~\ref{scheme2t1fig} for $\Im \langle S \rangle$
and $\Im \langle T_1 \rangle$ respectively.

We see that the results are very similar, with scheme 2 being perhaps slightly
more unstable. This includes a region where the instability is strong
so it is not
just a weak-coupling phenomenon.
Furthermore,
the other observables are also very similar between the two schemes.

In addition to the above scan, we also found only small differences
between schemes 2 and 4 for $L=10$ at several values of $V$.
Thus it appears that there is not much difference
between schemes 2 and 4, at least up to $l_{max}=2$.

\section{Discussion}
\label{sec:conc}

In this work we performed the first quantum non-perturbative study
of a string field theory. Our aim was to estimate the feasibility of the
lattice approach.
In principle, a lattice string field theory
could
be used to examine the validity of
a given string field
theory,
as well as to enable a
numerical study of various non-perturbative aspects of string theory.
One could hope to identify known as well as unknown solitons, to
measure the mass of non-BPS
objects\footnote{Mass shifts in string theory can be studied even without an explicit
string field theoretical formulation. However,
such a formulation would make the search more systematic~\cite{Pius:2013sca,Pius:2014iaa}.
Moreover, by studying string field theory on a lattice we could also identify mass shifts
due to non-perturbative effects.},
and to examine various dualities as well as other
conjectures. It might also be useful for identifying and calculating
generalizations of Ellwood invariants~\cite{Ellwood:2008jh} (see
also~\cite{Hashimoto:2001sm,Gaiotto:2001ji,Kawano:2008ry,Kiermaier:2008qu,Kudrna:2012re}).

At this stage, however, our examination was of a much more preliminary nature.
We identified the technical and fundamental obstructions towards a lattice
approach in string field theory, suggested possible resolutions and
examined lattice simulations in order to check whether the advocated
methodology could work for the simplest possible model.

While some of our results seem to suggest that our approach does make sense,
others point towards further obstacles that following studies will have to face.
In particular we found out that the lattice approach in the case of a linear dilaton theory
must sample deep into the strong interaction regime. However,
such sampling is very expensive computationally and one must look for
a resolution of this problem. Possible directions include the examination
of theories without linear dilatons, as well as the introduction of
a Liouville wall that can potentially free us from the need to sample
the strong coupling region.
Another source of high computational cost is the fact that the action is complex.
It could be interesting to find other ways of dealing with the original action
that do not involve analytical continuation, or to find some other way to trade the
analytically continued action by another, real, action.
An interesting approach could be to work not with straight lines in the complex
plane, but with Lefschetz Thimbles, following Witten's suggestions on the
proper definition of the action of some Chern-Simons theories~\cite{Witten:2010cx}.
String field theory looks, at least superficially, very similar to Chern-Simons theory
and the idea of using Lefschetz Thimbles was already implemented in lattice
field theory, see e.g.~\cite{Cristoforetti:2012su,Aarts:2013fpa,Fujii:2013sra}.
It could also be useful to find a way to implement the Langevin method for
a lattice string field theory.

A related issue is that of boundary conditions. First, the linear dilaton
prevented us from using periodic boundary conditions and then it turned out
that the obtained functions tend to concentrate at the rightmost part of the
working segment. Again, working with different backgrounds might be
useful in order to avoid this whole state of affairs. However, working
with such a background might be very difficult for reasons mentioned in the
introduction.
The study of the theory with a Liouville wall, on the other hand, seems to be
relatively simple and would probably enable us to resolve some of the difficulties
we are facing. It would also be interesting to see whether the current
framework is sensitive to some sort of a modification of the boundary conditions.

There could have been other reasons to object the feasibility of our approach.
To begin with, we try to approximate an infinite number of fields by a truncation
that takes into account only modes from a finite number of fields. Furthermore,
we approximate a non-local action using an expansion which retains only a
finite number of derivatives.
While such an approach is a standard one for the description of low energy physics,
it is well known that it might be inadequate for a complete description of such a
theory~\cite{Eliezer:1989cr,Moeller:2002vx}. However, it is also
known that a classical level truncation approach is often very accurate and
useful in string field theory. Moreover, string field theory is expected to
behave better than other non-local theories. On the other hand, we are dealing
with a situation that is more subtle than standard level truncation computations
due to the presence of the linear dilaton background, the fact that we work with
more general component fields than in the usual case and, most importantly, the
quantum nature of the analysis. We prefer to follow the footsteps of the original
level truncation papers and examine these questions
experimentally~\cite{Kostelecky:1988ta,Kostelecky:1990nt}.

With the introduction of higher levels new problems might be encountered.
One complication is related to the fact that in string field theory some of the
auxiliary fields have kinetic terms with wrong signs.
This problem could be avoided by explicitly integrating out all the auxiliary fields
prior to any numerical analysis.
Another problem is the appearance of imaginary
interaction terms. Such terms will become real after the analytical continuation
and will, therefore, lead to instabilities. It seems to us that the origin of these
terms is the breakdown of momentum conservation by the lattice together with the
fact that this momentum conservation has an imaginary part in a linear dilaton
background. It might be possible that the freedom of adding trivial terms,
discussed in section~\ref{sec:Adding_triv_terms}, could be used for setting all
the imaginary parts to zero. Exploring alternative theories and other boundary
conditions, as discussed above, might be useful also in this context.

Another important issue, which we did not address in this work, is renormalization,
namely comparing results at different levels. A related question is that of
renormalizability. In standard field theory one of the advantages of the lattice
approach is that it respects the gauge symmetry and is therefore not expected
to imperil the renormalizability of the theory. In string field theory, on the other
hand, gauge symmetry mixes different levels and is therefore broken by the
lattice. While future works will have to study higher levels, they will also
have to address the issues of gauge symmetry breaking, renormalizability and
renormalization schemes. A possible approach that might be helpful in the context
of breaking the gauge symmetry, as well as for other reasons, could be to
construct a lattice in the continuous $\kappa$-basis of string field
theory~\cite{Rastelli:2001hh,Okuyama:2002yr,Douglas:2002jm,Erler:2002nr,Fuchs:2002zz,Fuchs:2002wk,Fuchs:2003wu,Erler:2003eq,Fuchs:2006an,Fuchs:2006gs,Bonora:2007tm,Bonora:2009he,Bonora:2009hf}.
We leave this approach to future work.

Finally, one could have objected the idea of using open string field theory
for such a study on the ground that it is not expected to describe the closed
string moduli~\cite{Moeller:2010mh,Munster:2011ij,Muenster:2013}.
A related objection is that the theory obeys only the classical master
equation, while the quantum equation is divergent~\cite{Thorn:1988hm}.
This observation suggests that the theory is not really consistent at
the quantum level. The difficulties with the quantum master equation
most probably originate from the somewhat singular nature of Witten's
star product. They are probably common also to other
formulations based on this product, see e.g.~\cite{Kroyter:2010rk,Kroyter:2012ni}.

The singularity of the star product seems to be related to the fact that it
is used in order to describe solely the open string sector.
Indeed, one can consider continuous families of
open-closed string field theories and Witten's theory, which is obtained as
a singular limit of such an interpolation, is the only theory of the family
whose master equation has quantum singularities~\cite{Zwiebach:1992bw}.
Problems with the quantum master equation suggest that the gauge symmetry might
be broken at the quantum level. However, in the case at hand it is known
that at least in the Siegel gauge open string field theory leads to the
correct covering of moduli spaces and to correct expressions for all
amplitudes~\cite{Giddings:1986wp,Zwiebach:1990az}. World-sheet open
string theory is known to be renormalizable. Hence, as long as one
uses the gauge fixed scheme we mostly used, the theory before the introduction
of the lattice should be consistent quantum mechanically. Nonetheless, it would
be very useful in principle to have a lattice formulation of closed string
field theory~\cite{Zwiebach:1992ie}, in which closed string moduli could
also be varied. However, numerical analysis of this
theory is extremely complicated already at the classical
level~\cite{Yang:2005rx,Moeller:2006cv,Moeller:2006cw,Moeller:2007mu}.

To summarize, while lattice string field theory could be a useful framework,
there are many obstacles on the way, some of which we dealt with in this work
and others which still lie ahead. It seems that a gauge fixed approach is
the most promising one. But it remains to be seen which theories can be
studied and what are the most adequate backgrounds and boundary conditions.
We hope that future studies will clarify these points.

\section*{Acknowledgements}

We benefited from discussing the matters presented in this
paper with Gert Aarts, Ted Erler, Udi Fuchs, Carlo Maccaferri, Yaron Oz, Leonardo Rastelli,
Martin Schnabl and Barton Zwiebach.
We would further like to thank Udi Fuchs and Carlo Maccaferri for valuable comments on
a draft of the manuscript.
The calculations for this work were, in part, performed on the University of Cambridge
HPCs as a component of the DiRAC facility jointly funded by STFC and the
Large Facilities Capital Fund of BIS.

\newpage

\bibliography{bib}

\vfill\eject

\end{document}